\documentclass{aa}  
\usepackage{graphicx}
\usepackage{txfonts}
\begin{document}

\title{Stellar populations of
early-type galaxies in different environments III}
\subtitle{Line-strength gradients}

\author{P. S\'{a}nchez--Bl\'{a}zquez\inst{1,2}
\and
J.\ Gorgas\inst{2}
\and
N.\ Cardiel\inst{2,3}
}
\institute{
Laboratoire d'Astrophysique, \'Ecole Polytechnique F\'ed\'erale de Lausanne (EPF
L), Observatoire, 1290 Sauverny, Switzerland\\
\and
Dpto.\ de Astrof\'{\i}sica, Fac. de Ciencias F\'{\i}sicas, Universidad
Complutense de Madrid, E-28040 Madrid, Spain \\
\and
Calar Alto Observatory, CAHA, Apartado 511, E-04004 Almer\'{\i}a, Spain}

\date{Received September 15, 1996; accepted March 16, 1997}

\abstract
{}
{This is the third paper of a series devoted to study the stellar content
of early-type galaxies. The goal of the series is to set constraints on the 
evolutionary status of these objects.}
{We present line-strength gradients for 22 spectral indices measured in a
sample of 82 early-type galaxies in different environments, including the
high-density core of the Coma cluster, the Virgo cluster, poor groups, and
isolated field galaxies.  Using new evolutionary population synthesis
models we derive age and metallicity gradients, and compare the
mean values with the predictions of different galaxy formation models. We
explore the behaviour of individual chemical species by deriving the
metallicity gradient with different indicators.}
{We find that the strength of the metallicity gradient inferred 
from stellar population models depends on the specific Lick index 
employed. In particular, metallicity gradients obtained with CN$_2$ and Ca4668 
 combined with H$\beta$ are steeper than when  measured 
using Ca4227 or Fe4383. The correlation of the metallicity gradients
with other parameters also depends on the specific index employed.
If the metallicity gradient is obtained using CN$_2$ and Mgb then 
it correlates with the central age of the galaxies. On the contrary, 
if Fe4383 or Ca4227 are used, the metallicity gradient correlates with 
the velocity dispersion gradient.}
{This may suggests that  several mechanism have helped to 
set the age and metallicity gradients in early-type galaxies. 
While we do not find any correlation between the metallicity gradient and
the central velocity dispersion for galaxies in low-density environments,
we find a marginal correlation between the metallicity gradient and the
mass for galaxies in the centre of the Coma cluster. We also find a
trend for which galaxies in denser environments show a steeper metallicity
gradient than galaxies in less dense environments. We interpret these
results in light of the different mechanisms proposed to explain the
observed changes between galaxies as a function of environment.}
{}

\titlerunning{Line-strength gradients in early-type galaxies}
\maketitle
\section{Introduction}

This is the third paper in a series devoted to studying the properties of
the stellar populations in early-type galaxies as a function of their
local environment. In S\'anchez-Bl\'azquez et al.\ (2006a, hereafter
Paper~I) we presented central Lick/IDS index measurements for a sample of
98 early-type galaxies and described their relation to the velocity
dispersion. In S\'anchez-Bl\'azquez et al.\ (2006b, hereafter Paper~II) we
compared the indices with the stellar population synthesis models of
Vazdekis et al.\ (2006, in preparation; hereafter V06) to derive ages and
metallicities.   These models  are an improved  
version of those described by Vazdekis et al.\ (2003) 
strengthened by the addition of a new stellar library
(S\'anchez-Bl\'azquez et al. 2006, in preparation).
The synthetic spectra cover a spectral range of   
$\lambda\lambda 3500-7500$ \AA~ at resolution 2.3 \AA. 
The previous papers concentrated on the analysis of the
central regions of the galaxies. In this paper we investigate the
behaviour of these properties as a function of galactocentric radius. 

It is well known that early-type galaxies show a variation of their
stellar population properties with radius.  The first evidence of this
phenomenon was observed in the colours of bulges and elliptical galaxies
(see the review by Kormendy \& Djorgovski 1989) and indicated that the
central regions tend to be redder than the outer parts. Other authors have
measured optical surface brightness profiles for large sets of local
ellipticals (e.g. Franx, Illingworth \& Heckman 1989; Peletier et al.\
1990), reaching the same conclusion. The first study of gradients in the
strength of absorption features was performed by McClure (1969). In a
sample of 7 galaxies, McClure found that the C($41-42$) index (a measure
of the CN$\lambda$4216 band strength) was stronger in the centres of the
galaxies than at a distance of 1--1.5 kpc, which he interpreted as a
metallicity difference between the two regions. Subsequent work in the
field has explored the strength of absorption feature
gradients using a broad range of line indices (Spinrad et al.\ 1971;
Spinrad, Smith \& Taylor 1972; Welch \& Forrester 1972; Joly \& Andrillat
1973; Oke \& Schwarzschild 1975; Cohen 1979;  Efstathiou \& Gorgas 1985;
Couture \& Hardy 1988; Peletier 1989; Thomsen \& Baum 1989; Gorgas,
Efstathiou \& Arag\'on-Salamanca 1990; Boroson \& Thompson 1991;  Bender
\& Surma 1992; Davidge 1992; Davies, Sadler \& Peletier 1993; Carollo,
Danziger \& Buson 1993; Gonz\'alez 1993; Fisher, Franx \& Illingworth
1995; Gorgas et al.\ 1997; Cardiel, Gorgas \& Arag\'on-Salamanca 1998a;
Mehlert et al.\ 2003).  In general, these studies indicate the existence
of intense gradients in CN$\lambda 3883$ and CN$\lambda 4216$, less
pronounced gradients in the Mg$\lambda 5176$, G band, NaD, Ca H\&K
features, and in some lines of Fe~{\sc i}, and a weak or null gradient in
H$\beta$, MgH, TiO, Ca~{\sc i} and in the calcium triplet in the
near-infrared. Most of these studies have suggested that the existence of
gradients in the metallic spectral features is a consequence of a
decreasing metallicity with increasing galactocentric radius (e.g.\ Mc
Clure 1969;  Cohen 1979; Davies et al.\ 1993; Kobayashi \& Arimoto 1999;
Mehlert et al.\ 2003). Some, however, have argued that, apart from a
variation of metallicity with radius, there is also a radial variation in
the luminosity-weighted mean age of the stellar populations, with the
central regions being younger than the outer regions (Gorgas et al.\ 1990;
Munn 1992; Gonz\'alez 1993; Gonz\'alez \& Gorgas 1996). 

How the physical properties of galaxies vary with radius can prove
invaluable for constraining the processes of galaxy formation and
evolution.  For example, metallicity gradients are a measure of the
quantity, velocity, and duration of gas dissipation. Likewise, age and
metallicity gradients contain  information concerning the
relative importance of interactions during galaxy formation. 

In the classical models of monolithic collapse (Eggen et al.\ 1962; Larson
1974a; Carlberg 1984; Arimoto \& Yoshii 1987; Gibson 1997), stars form in
essentially all regions during the collapse and remain in their orbits
with little inward migration, whereas the gas dissipates inwards, being
continuously enriched by the evolving stars. 
In this way, the stars formed at the centres of galaxies are predicted to
be more metal-rich than those born in the outer regions.  Supernova-driven
galactic winds (Mathews \& Baker 1971; Larson 1974b;  Arimoto \& Yoshii
1987; Gibson 1997), initiated when the energy injected into the
interstellar medium (ISM) by supernovae matches that of its binding
energy, act to evacuate the galaxy of gas, thereby eliminating the fuel
necessary for star formation. The external parts of the galaxy (with a
shallower potential well) develop winds before the central regions, where
the star formation and, therefore, the chemical enrichment continue for
longer. The monolithic collapse models, therefore, predict very steep
metallicity gradients as both processes -- the dissipation of gas toward
the central parts of the galaxy and the different timescales for the
occurrence of the galactic winds -- act in the direction of steepening any
nascent metallicity gradient.\footnote{Galactic winds \it can \rm produce
metallicity gradients without invoking dissipation (e.g.\ Franx \&
Illingworth 1990; Martinelli, Matteucci \& Colafrancesco 1998), with the
local metallicity coupled directly to the local potential well depth and
essentially independent of the galaxy formation collapse physics.}
 
Simulations of galaxy mergers within the concordant hierarchical
clustering cold dark matter framework (e.g.\ Cole et al.\ 1994; Baugh,
Cole \& Frenk 1996; Kauffmann 1996; Kauffmann \& Charlot 1998) offer
somewhat contradictory predictions as to the radial variation of stellar
properties in early-type galaxies -- while some (e.g.\ White 1980; Bekki
\& Shioya 1999) suggest mergers lead of a flattening of metallicity
gradients, others  (e.g.\ van Albada 1982) argue that the gradients
are affected only moderately by the mergers, as the violent relaxation
preserves the position of the stars in the local potential. This apparent
dichotomy between the simulations is driven in part by the sensitivity of
the outcome to the fraction of gaseous versus stellar mass present in the
progenitor galaxies. Broadly speaking, the predicted gradients are steeper
if the progenitor galaxies have a large fraction of their pre-merger
baryonic mass in the form of gas. Furthermore, numerical simulations
suggest that during the merger a significant fraction of this gas migrates
inward toward the central regions of the merging galaxies, resulting in
increased central star formation (Barnes \& Hernquist 1991).  Mihos \&
Hernquist (1994) showed that the observed gradients in elliptical galaxies
may be a consequence of the occurrence of secondary bursts of star
formation triggered by these mergers. 
 
One of the keys  to dermining the physical 
underlying the formation and evolution 
of galaxies is to study of the relations between the gradients
and other fundamental (global) properties of galaxies. For instance,
dissipational collapse models predict a strong positive correlation
between metallicity gradient and galactic mass.  The empirical evidence
for such putative correlations remains contentious -- e.g. Gorgas et al.\
(1990, G90 hereafter) did not find any relation between Mg$_2$ gradient
and the rotation or total luminosity of a sample of early-type galaxies,
although they found some evidence of a positive correlation between Mg$_2$
gradient and central velocity dispersion (a probability of 95\% in a
non-parametric Spearman test).  Conversely, Franx \& Illingworth (1990)
found a correlation between colour gradient and local escape velocity (later
confirmed by other authors, as Davies et~al. 1993),
arguing that the aforementioned galactic winds were the dominant mechanism
controlling the metal content of early-type galaxies. Davidge (1992) analysed 12 bright
ellipticals, searching for correlations between Mg$_2$ gradients and the
central velocity dispersion ($\sigma_0$), the total luminosity, the shape
of the isophotes, the fine structure parameter (Schweizer et al.\ 1990),
and the anisotropy parameter (v/$\sigma$)*, which defines the degree of
rotation in a galaxy (Binney 1978). Davidge found weak correlations
between the Mg$_2$ gradient and (v/$\sigma$)* and $\sigma_0$, and an
absence of correlation with any of the other parameters. Carollo et al.\
(1993) carried out an exhaustive study of the gradients of several
spectral features (Mg$_1$, Mg$_2$, NaD, TiO$_1$, TiO$_2$, and Fe5270) in a
sample of 42 galaxies. Carollo et~al. found a tendency for the slope of
the gradients in the Mg$_2$ index to increase with the mass of the galaxy,
but only for galaxies with masses below 10$^{11}$ M$_{\odot}$. Carollo \&
Danziger (1994) studied the line-strength gradients in Mg$_2$ and
$\langle$Fe$\rangle$ for five early-type galaxies, confirming the
dependency between metallicity and local potential well depth as a
function of galactocentric radius, albeit with significant scatter. 
Gonz\'alez \& Gorgas (1996) surveyed the Mg$_2$ gradient literature
available at the time, discovering that galaxies with steeper Mg$_2$
gradients also possess stronger central Mg$_2$. They proposed a scenario
in which star formation episodes in the centre of the galaxies are
responsible for the correlation, pointing out that its existence implies
that the global mass-metallicity relation is flatter than the
mass-metallicity relation inferred from the central values. Recently,
Mehlert et al.\ (2003) found a correlation between the gradients of some
spectral features and the velocity dispersion gradient (which can be
considered a measure of the potential well depth gradient).  In contrast
with Gonz\'alez \& Gorgas, Mehlert et~al. did not find a correlation
between the line-strength index gradients and the central values, or
between the index gradients and the central velocity dispersion. 

The results to date are, to some extent, contradictory and therefore
incapable of providing unequivocal support to any particular galaxy
formation scenario. This contradictory nature is driven, in part, by the
requisite high signal-to-noise data and associated care in data reduction
necessary to extract reliable gradients.  Further, very few absorption
features (mainly Mg$_2$) have been systematically explored with a suitably
large sample of ellipticals.  Our work has been designed specifically to
address these shortcomings which have plagued the interpretation of the
extant data, using a sample of 82 galaxies populating a range of local
environmental conditions. 

In Section~2, we describe the measurement of the line-strength gradients
within our dataset.  In Section~3, we compare these gradients with
synthesis models to derive explicit age and metallicity gradients. 
Sections~4 to 7 analyse the stellar population gradients for a sample
drawn from low density environments. In particular, Section~4 discusses
the mean age and metallicity gradients, while in Section~5 we explore the
existence of possible variations of the chemical abundance ratios with
radius.  In Section~6, we explore the existence of putative correlations
between metallicity gradient and other galactic parameters. In Section~7,
we use the stellar population global parameters, derived with the help of
the gradients, to study whether the trends found in Papers~I and II hold
only for the central parts, or can be extended to the entire galaxy. 
Section~8 studies the mean gradients in the sample of galaxies drawn from
the high-density core of the Coma cluster. Finally, in Section~9, we
summarise our findings and conclusions. 

\section{Measurement of gradients} 
\label{section.measurement}
The parent sample from which  galaxies are drawn consists of 98
early-type systems, of which 37 belong to the Coma cluster (high-density
environment galaxies, hereafter HDEGs), while the rest are drawn from the
field, small groups, and the Virgo cluster (low-density environment
galaxies, hereafter LDEGs).  Details of the sample, the observations, and
associated data reduction, can be found in Paper~I.  From each fully
reduced galaxy frame, a final frame was created by extracting spectra
along the slit, binning in the spatial direction to guarantee a minimum
signal-to-noise ratio per \AA~(S/N) of 20 in the spectral region of the
H$\beta$ index, ensuring a maximum relative error for this index (assuming
a typical value of 1.5~\AA) of 20\% (see Cardiel et al.\ 1998b). Those
galaxies which did not have a minimum of 4 spectra along the radius
meeting this criterion were eliminated from the final sample, which left a
total of 82 galaxies of sufficient quality to measure gradients -- 21
HDEGs and 61 LDEGs. 

An equivalent radius $\bar{r}(i)$ was assigned to each binned spectrum as
\begin{equation}
\bar{r}(i)=\frac{\int_{i} \mu_{i}(r)rdr}{\int_{i}\mu_{i}(r)dr},
\end{equation}
where
\begin{equation}
\mu_{i}(r)=k\; 10^{b_{i}r}.
\end{equation}
The value of $b_{i}$ for each spectrum was calculated by fitting the number of
counts in each row of the detector (including all the rows in each spectrum
plus the neighbours) to the relation
\begin{equation}
\log N = b_{i}\;r+c,
\end{equation}
where $N$ is the number of counts, $r$ the radius in arc-seconds, and $c$ a
constant. We projected the gradients over the major axis, multiplying the mean
radius by a factor \mbox{$f=\sqrt{\cos^2\theta + (\frac{a}{b} \sin^2\theta})$},
where $\theta$ is the difference between the major axis and the position angle
of the slit, and $a$ and $b$ represent the major and minor axis of the galaxy
respectively. The $a$, $b$ and $\theta$ values for each galaxy can be found in
Table~2 of Paper~I. 

Radial velocities and velocity dispersions were measured for each spectrum
as a function of galactocentric radius as described in Paper~I.  The
rotation curve and the velocity dispersion profile were used to measure
the Lick/IDS indices at the resolution defined by this system, and to
correct them for velocity broadening (again, as described in Paper~I). 
Instead of using the individual values of the velocity dispersion for each
spectrum measured at each radial ``bin'', we fitted a smooth curve to the
$\sigma$ profile. This minimises the errors in the derived velocity
dispersion, due to the lower signal-to-noise ratios in the external parts
of the galaxies. Residual [OIII] emission profiles were also estimated as
described in Paper~I. These values were used to correct the H$\beta$ index
for emission using the the relation derived by Trager et al.\ (2000a) --
$\Delta$H$\beta$=0.6 [OIII], where $\Delta$H$\beta$ is the correction to
the index, and [OIII] the equivalent width of the [OIII]$\lambda 5007$
emission line. The Mgb and Fe5015 indices were also corrected from
emission, as in the central spectra. The final kinematic, line-strength
and emission profiles, together with their associated errors, are
presented in S\'anchez-Bl\'azquez (2004). 

We measured 19 Lick/IDS indices defined by Trager et al.\ (1998) for all the
galaxies except for those observed in Run~3, for which only 15 indices could be
measured. With the aim of reducing the random errors in the external parts of
the galaxies, we have defined the following composite indices:
\begin{eqnarray}
{\rm HA}      &=& \frac{{\rm H}\gamma_A+{\rm H}\delta_A}{2}\\
{\rm Balmer}  &=& \frac{{\rm H}\gamma_A+{\rm H}\delta_A+{\rm H}\beta}{3}
\end{eqnarray}
To obtain the final values for the line strength indices we applied the same
corrections as for the central indices (see Paper~I).  Finally, all indices
were transformed into magnitudes, as described in Paper~I.

\begin{figure*}
\resizebox{0.32\hsize}{!}{%
\includegraphics[angle=-90]{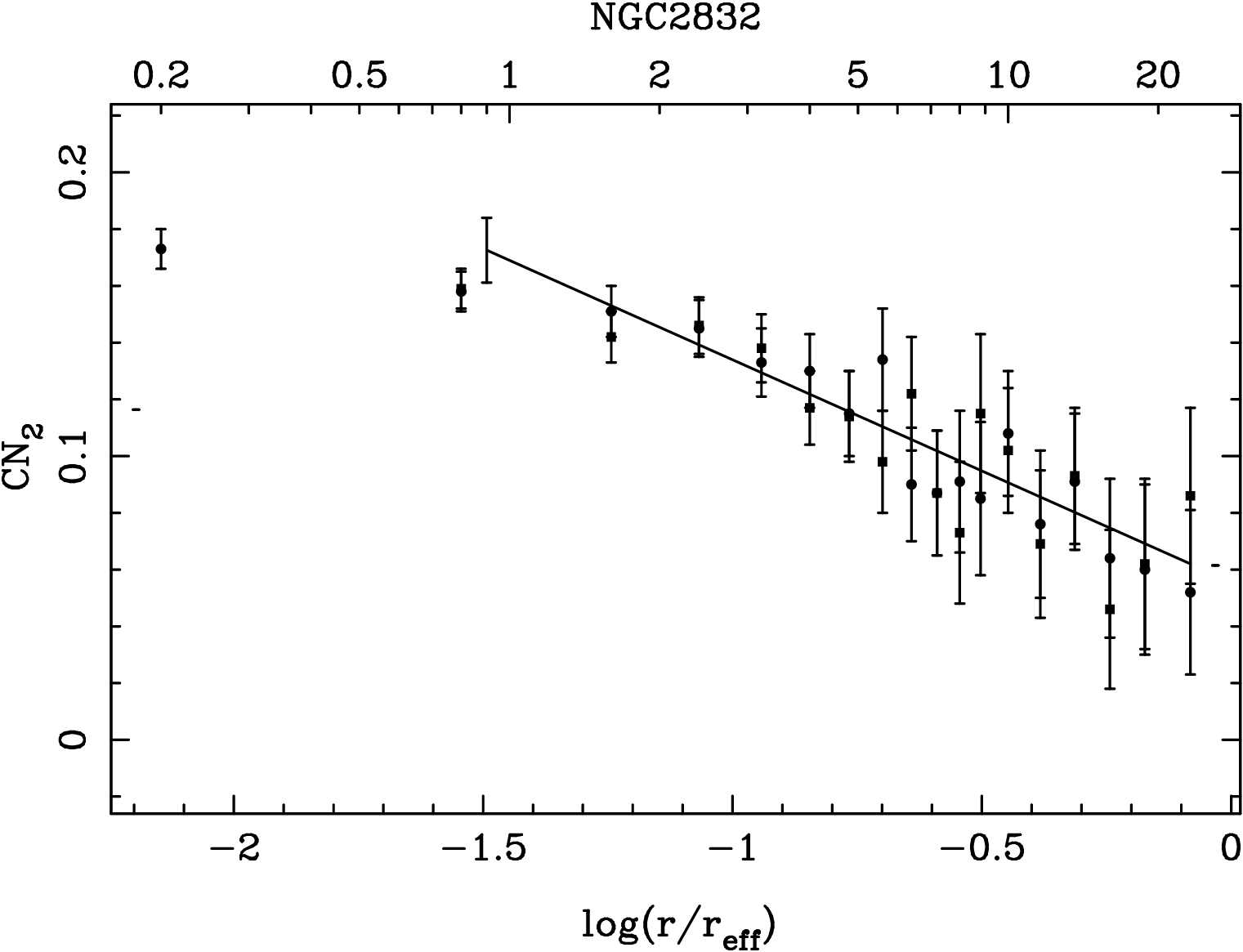}}
\hfill
\resizebox{0.32\hsize}{!}{%
\includegraphics[angle=-90]{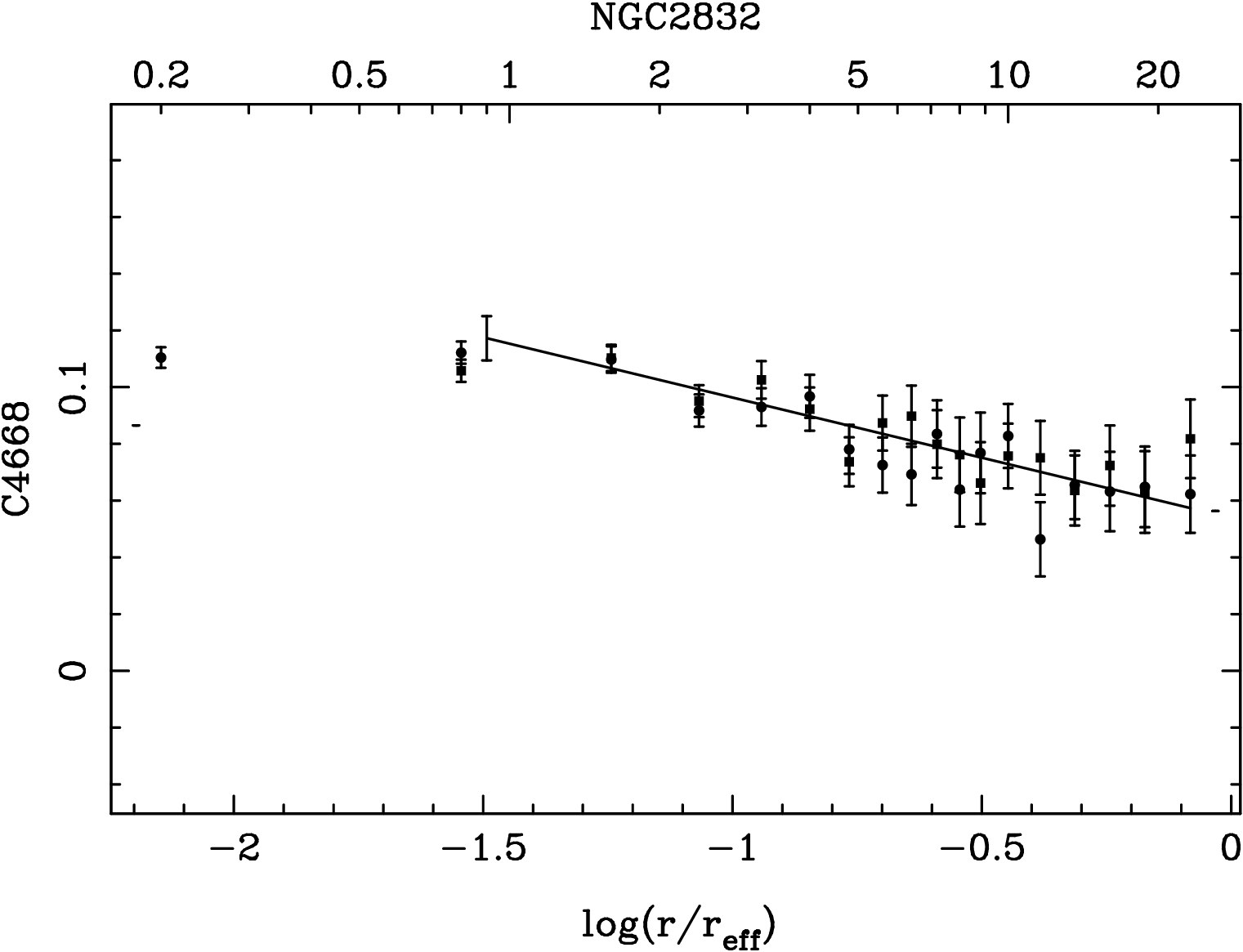}} 
\hfill
\resizebox{0.32\hsize}{!}{%
\includegraphics[angle=-90]{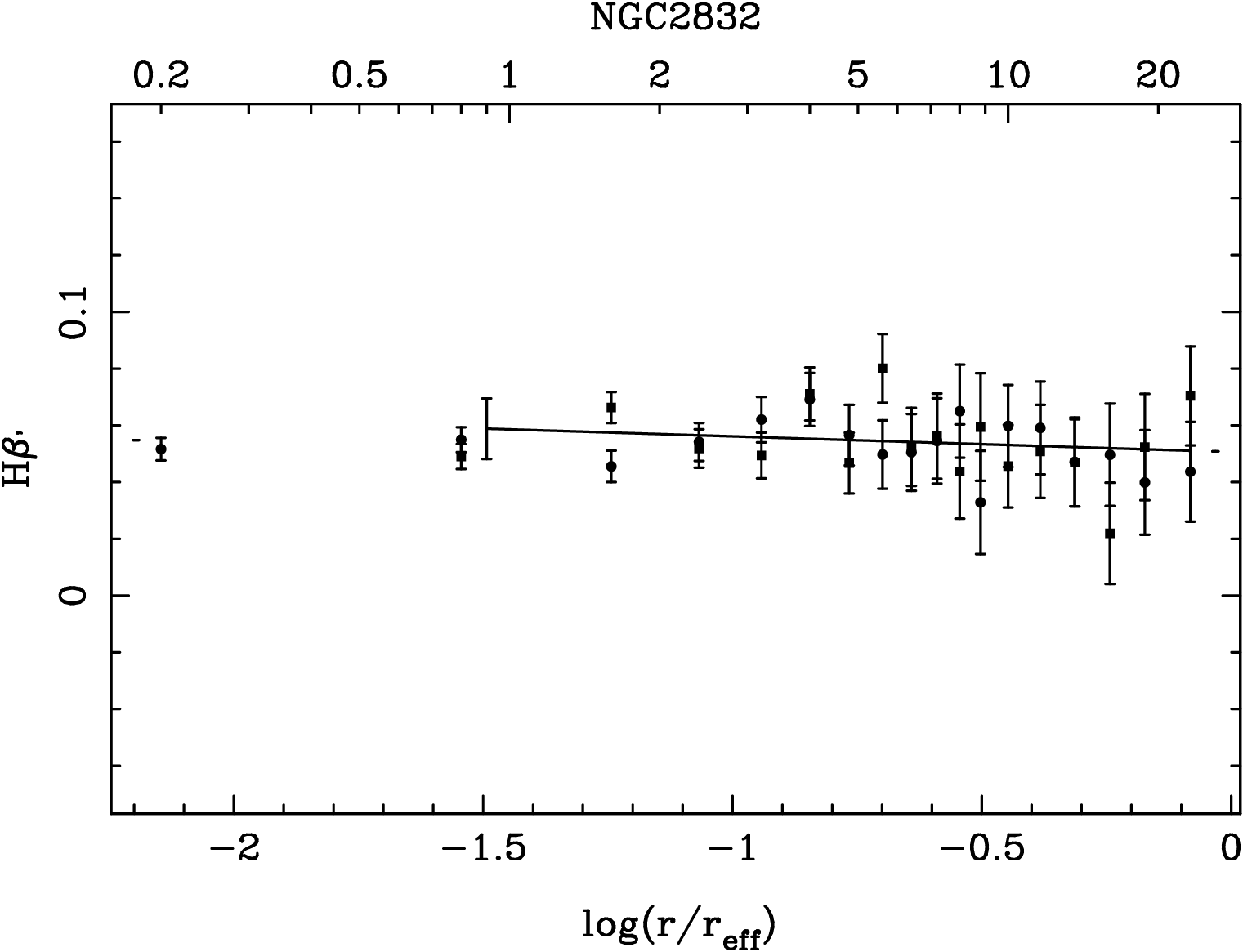}}
\caption{Variation of the indices CN$_2$, C4668 and H$\beta$ as a function of
galactocentric radius
for the galaxy NGC~2832. The upper axes show the radial distance from the
centre of the galaxy in arcseconds. 
The solid line represents a linear fit to the
data, as described in the text. The data within 0$^{\prime\prime}$.9
of the galaxy's centre were
excluded from the fit to minimise the effects of seeing on the fitted 
gradients.\label{example.gradients}}
\end{figure*}

Although the behaviour of the indices as a function of galactocentric
radius is not perfectly linear for all 82 galaxies, with the aim of
obtaining a quantitative measurement of their values, we performed a
linear fit, weighted by the errors over all the indices, to the relation
$I'=c+d\log \frac{r}{a_{e}}$, where $I'$ represents the index expressed in
magnitudes, $a_{e}$ is the effective radius projected over the major axis,
i.e.  $a_{e}=r_{\rm eff}/\sqrt{\frac{a}{b}}$ ($r_{\rm eff}$ being the
effective radius), and $d$ the index gradient, hereafter denoted by
$grad$. This projection can be done under the assumption that the contours
of constant $I'$ coincide with the isophotes, which has been confirmed by
several earlier studies (e.g. Davies et al.\ 1993).   To avoid the
uncertainties and radial flattening due to seeing effects, the data within
the central area corresponding to the seeing were excluded from the fit.
The seeing was estimated directly from the observations on the 4 different 
runs.
Fig.~\ref{example.gradients} shows some examples of the line-strength
gradients for the galaxy NGC~2832. 

\begin{figure*}
\resizebox{0.18\hsize}{!}{\includegraphics[angle=-90,bb=88 147 591 631]{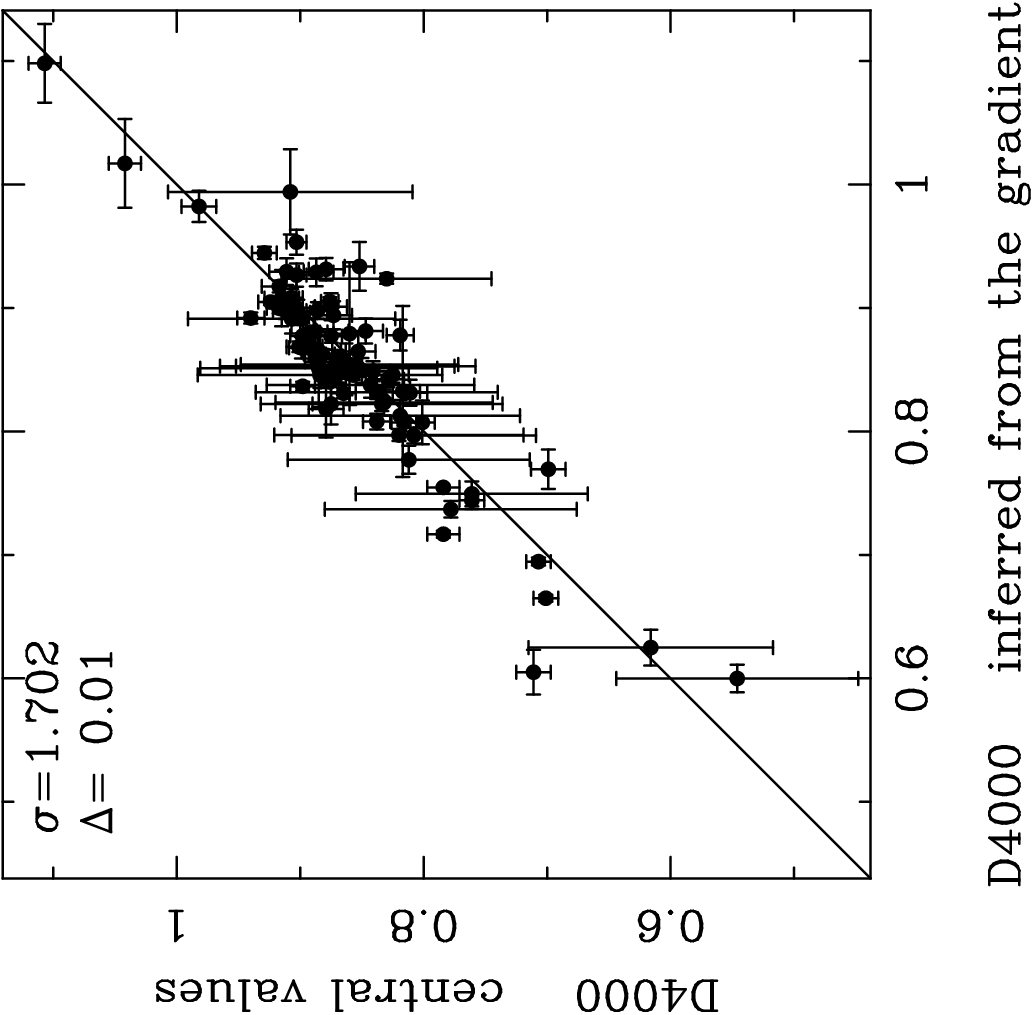}}
\hfill
\resizebox{0.18\hsize}{!}{\includegraphics[angle=-90,bb=88 147 591 631]{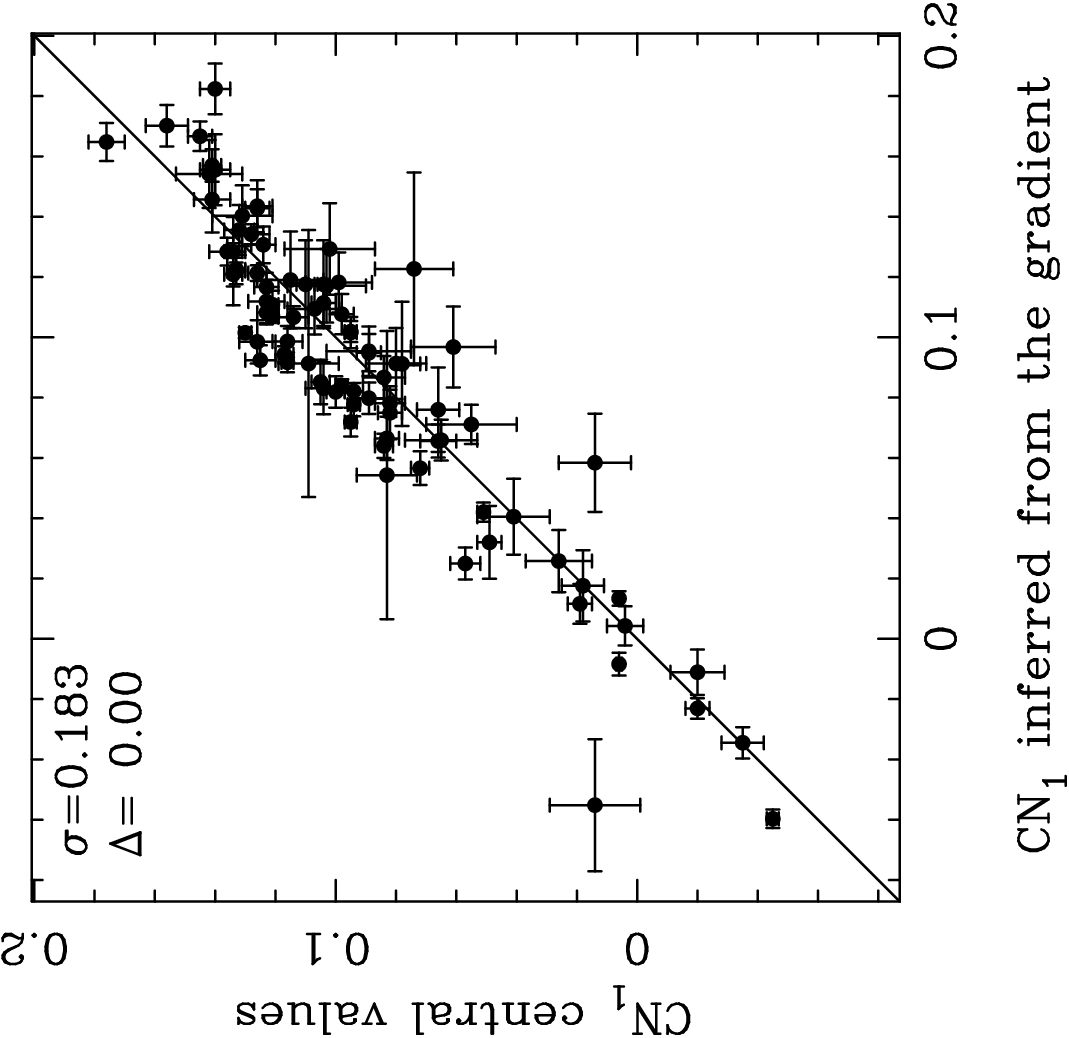}}
\hfill
\resizebox{0.18\hsize}{!}{\includegraphics[angle=-90,bb=88 147 591 631]{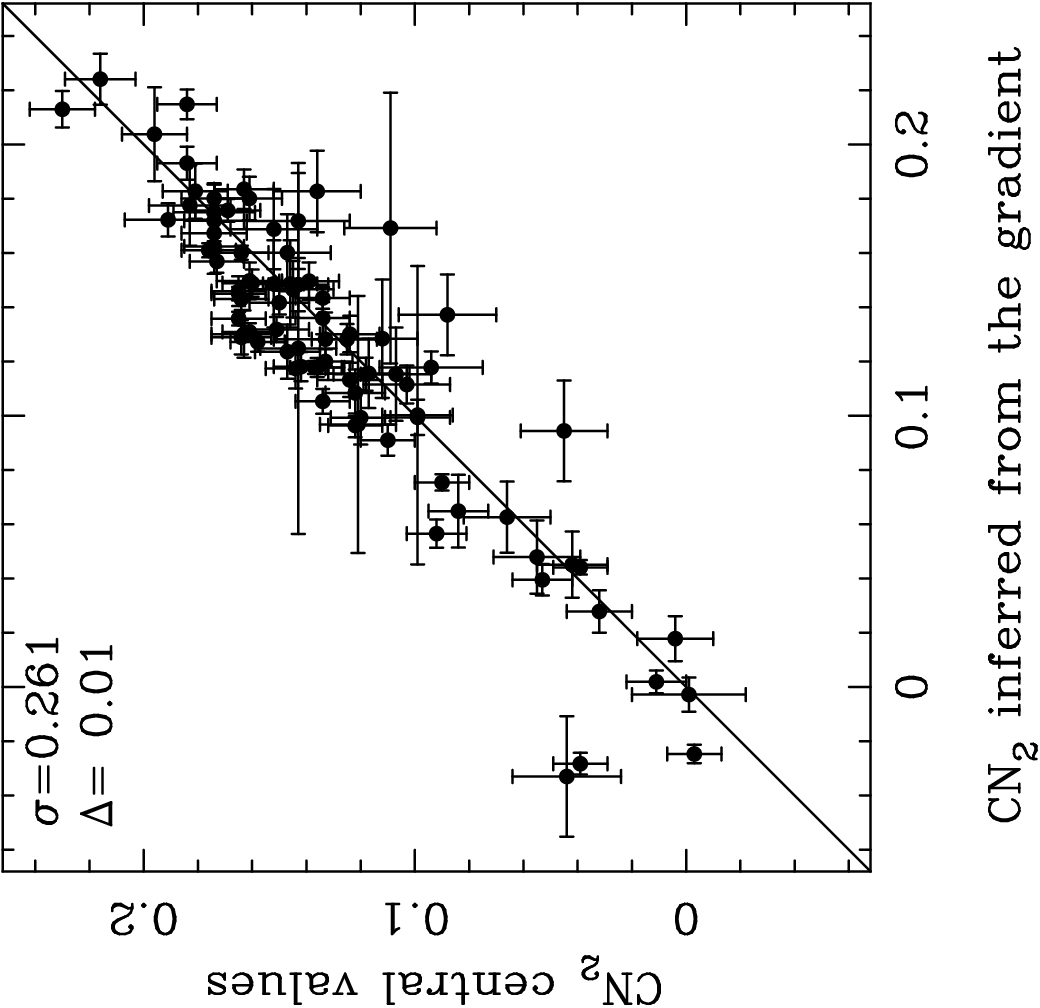}}
\hfill
\resizebox{0.18\hsize}{!}{\includegraphics[angle=-90,bb=88 147 591 631]{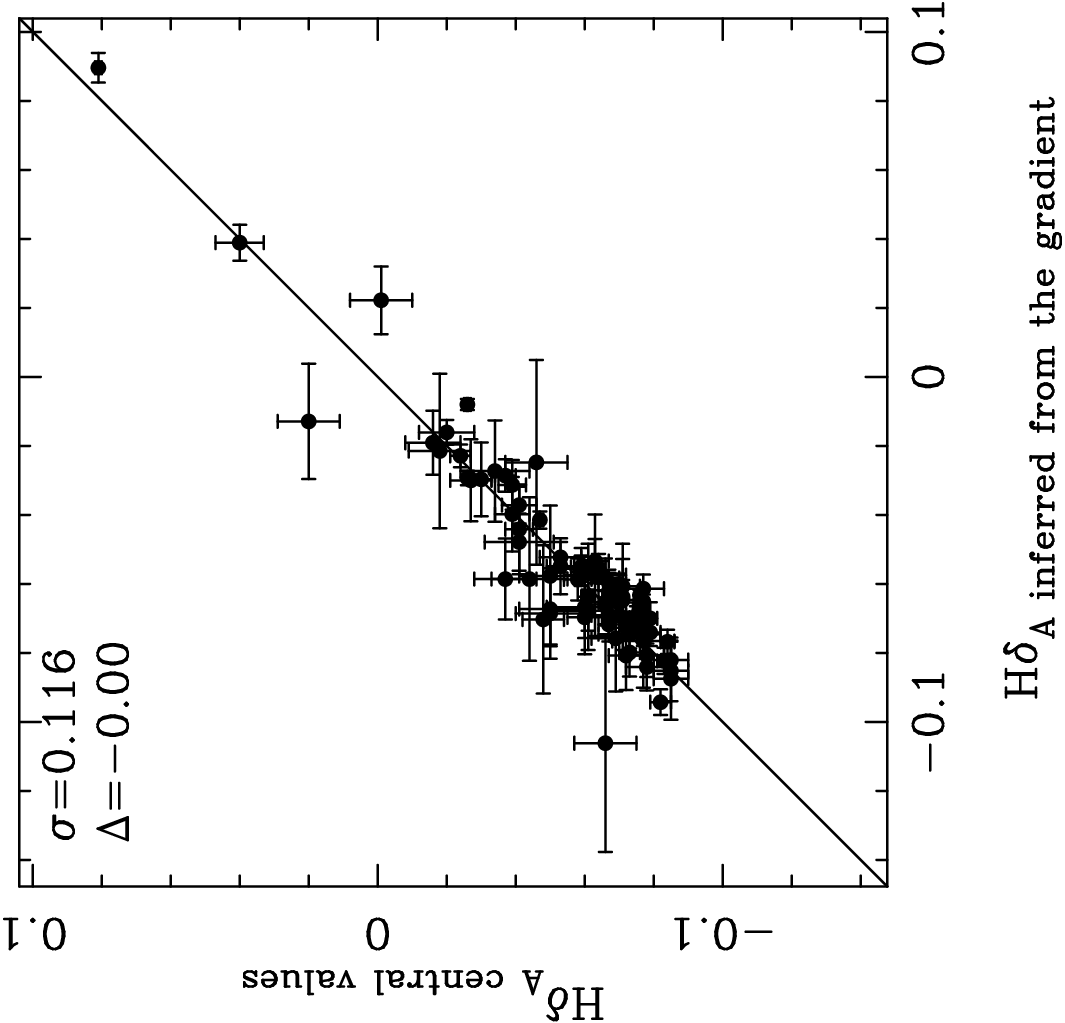}}
\hfill
\resizebox{0.18\hsize}{!}{\includegraphics[angle=-90,bb=88 147 591 631]{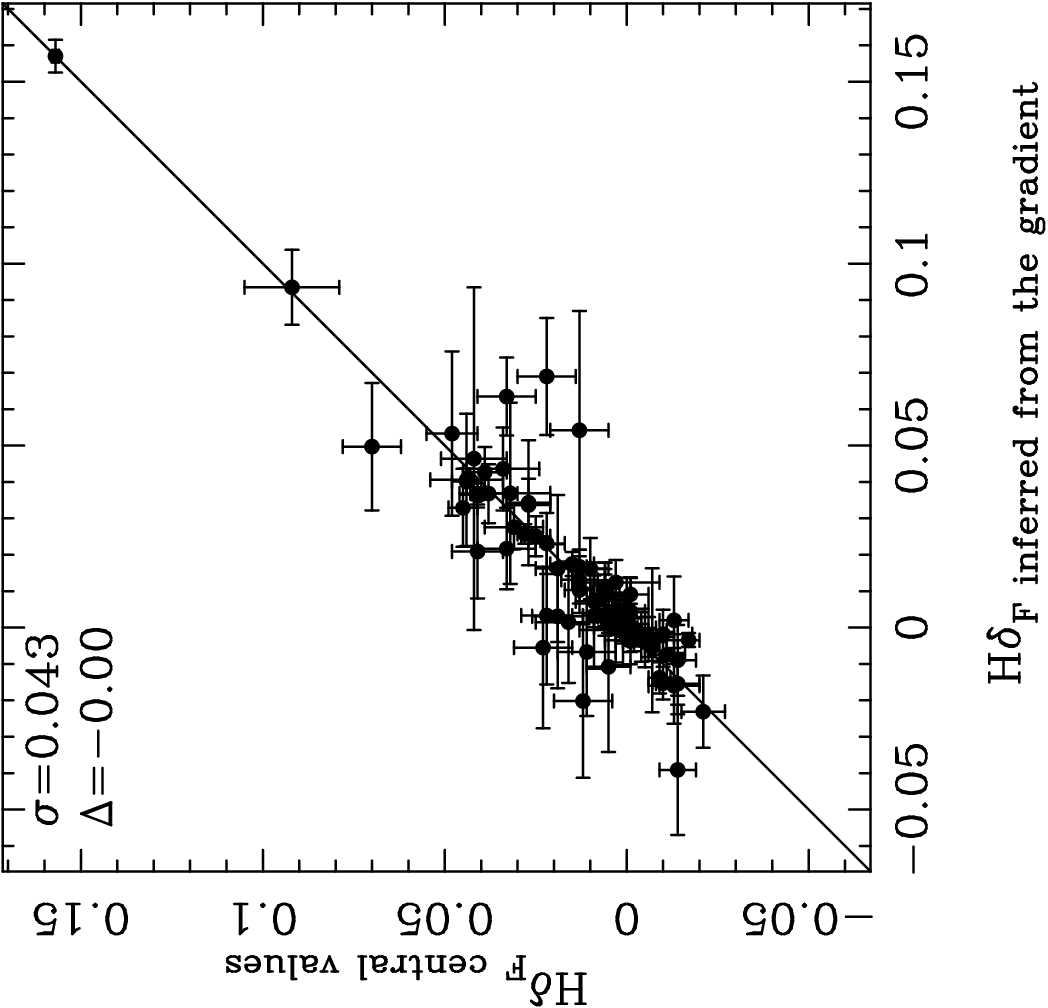}}

\vspace{4mm}

\resizebox{0.18\hsize}{!}{\includegraphics[angle=-90,bb=88 147 591 631]{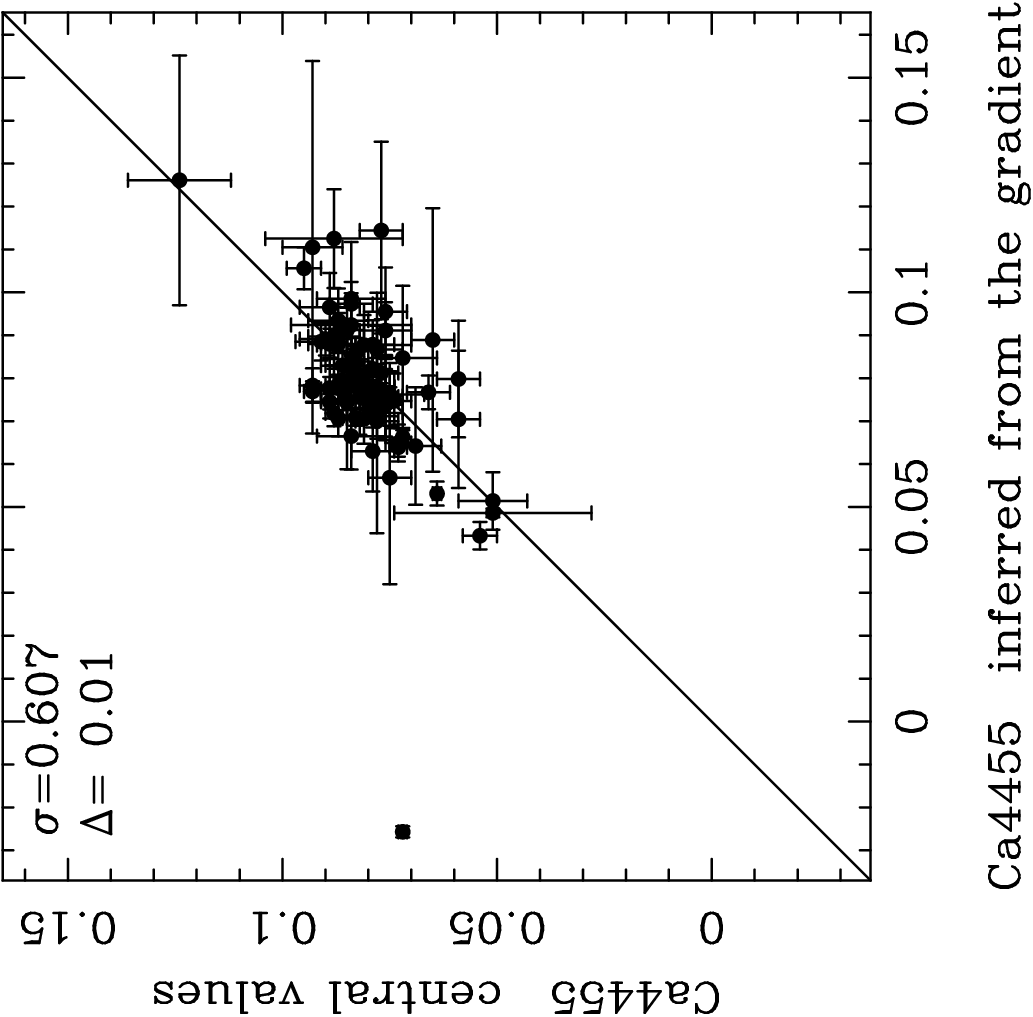}}
\hfill
\resizebox{0.18\hsize}{!}{\includegraphics[angle=-90,bb=88 147 591 631]{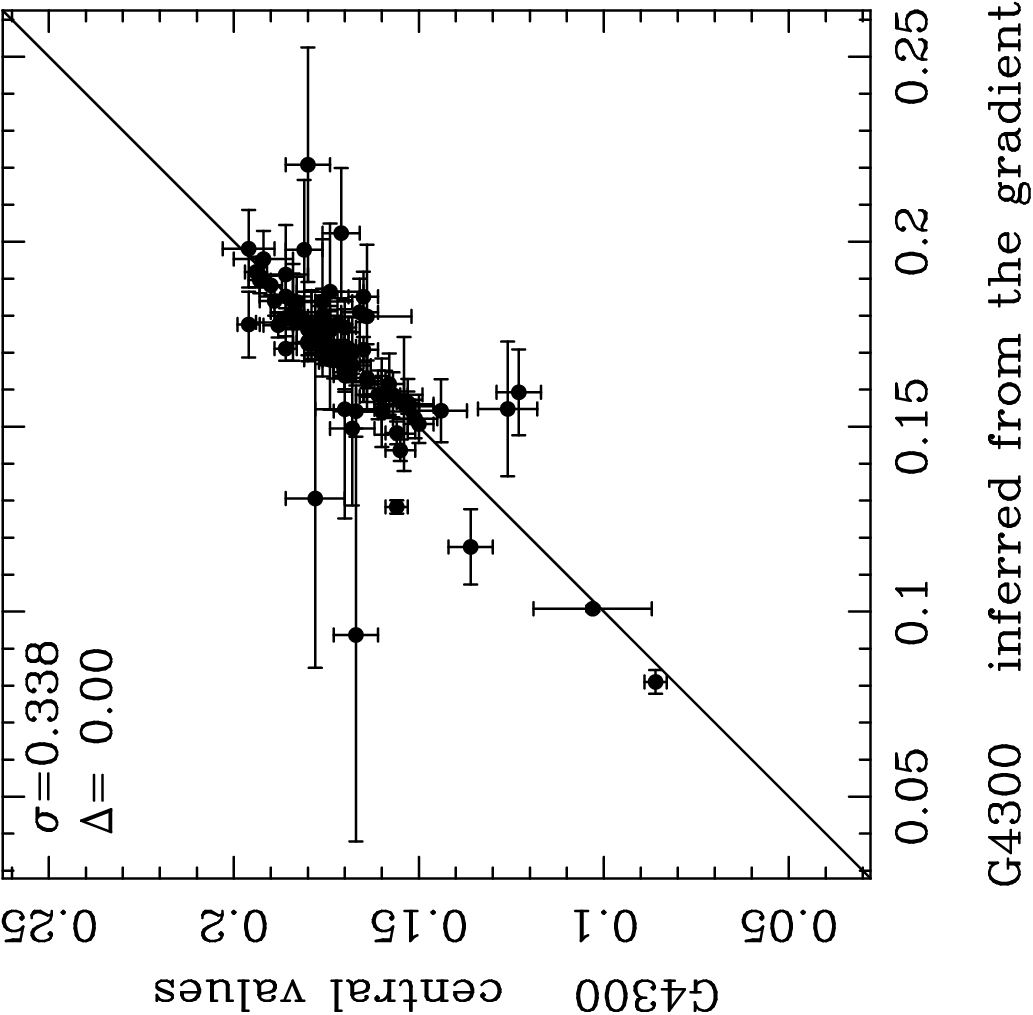}}
\hfill
\resizebox{0.18\hsize}{!}{\includegraphics[angle=-90,bb=88 147 591 631]{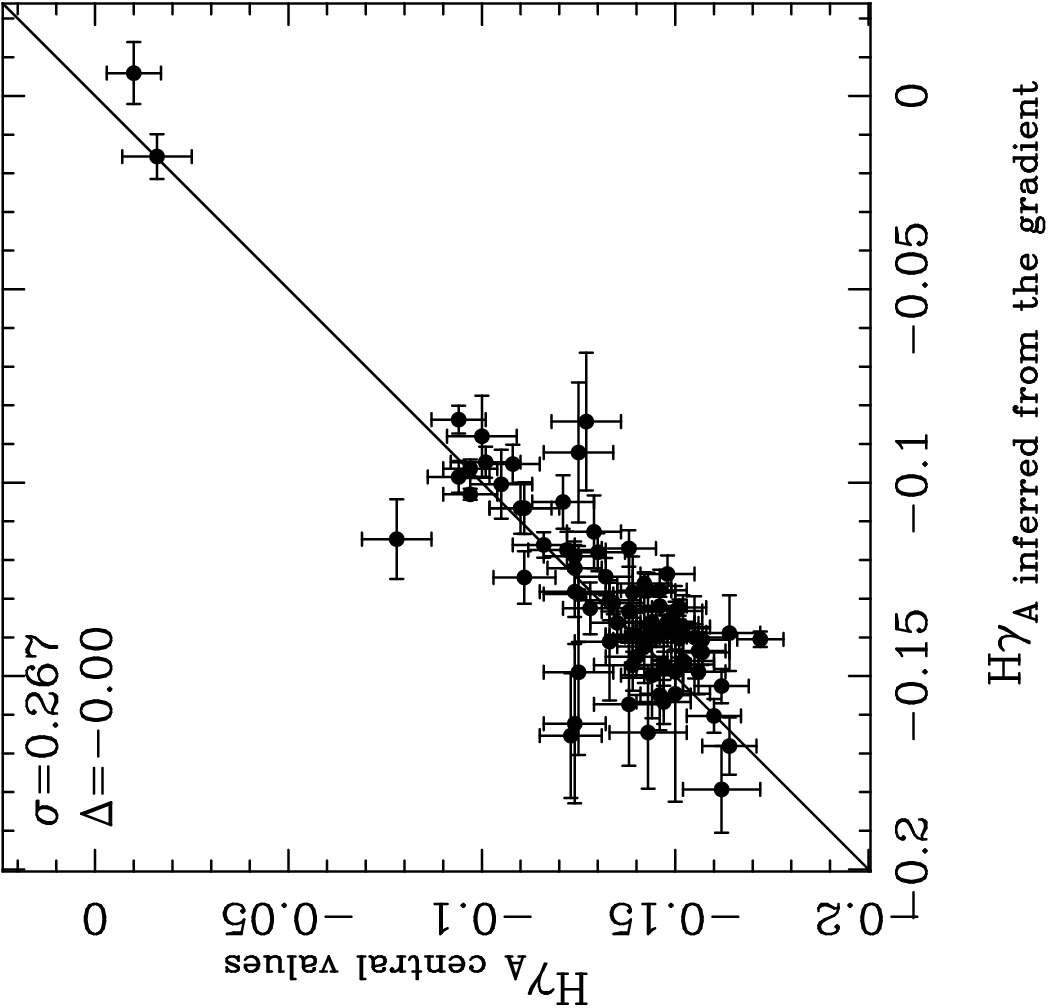}}
\hfill
\resizebox{0.18\hsize}{!}{\includegraphics[angle=-90,bb=88 147 591 631]{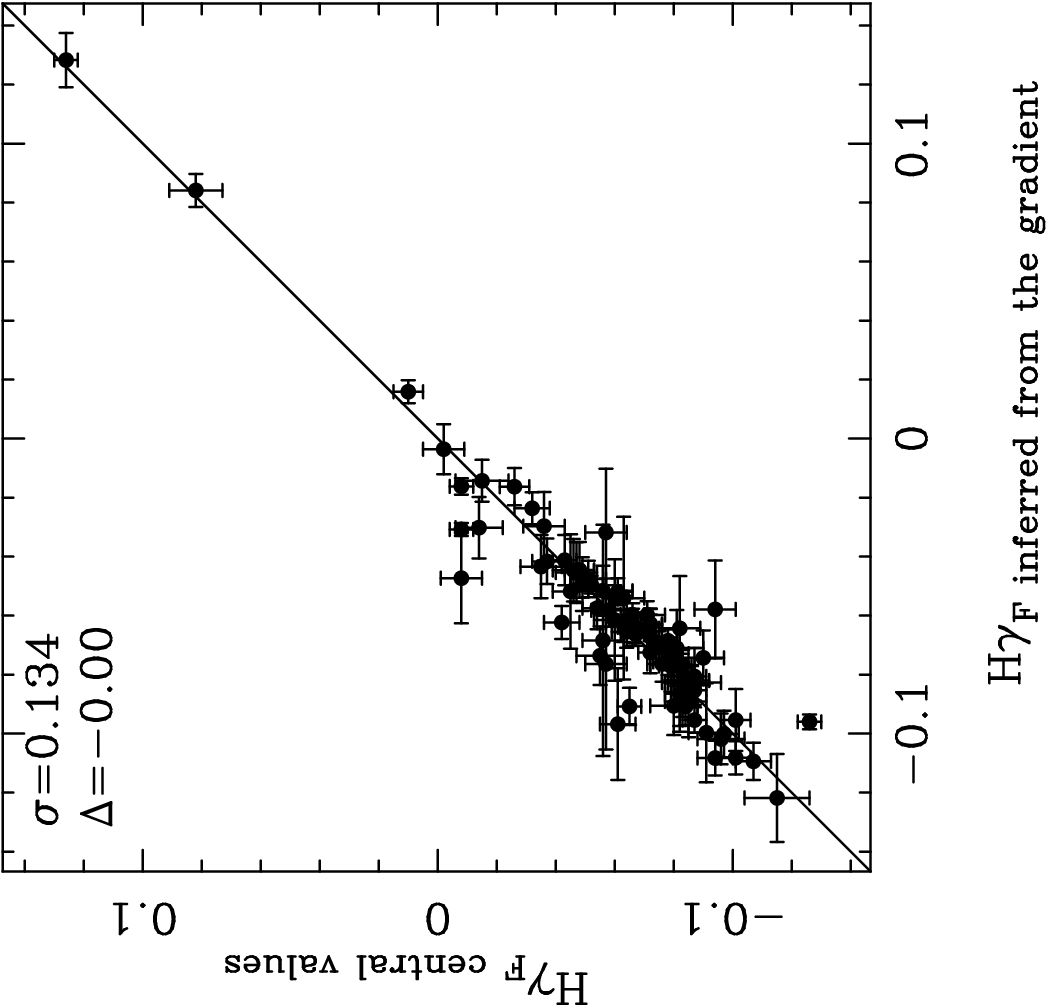}}
\hfill
\resizebox{0.18\hsize}{!}{\includegraphics[angle=-90,bb=88 147 591 631]{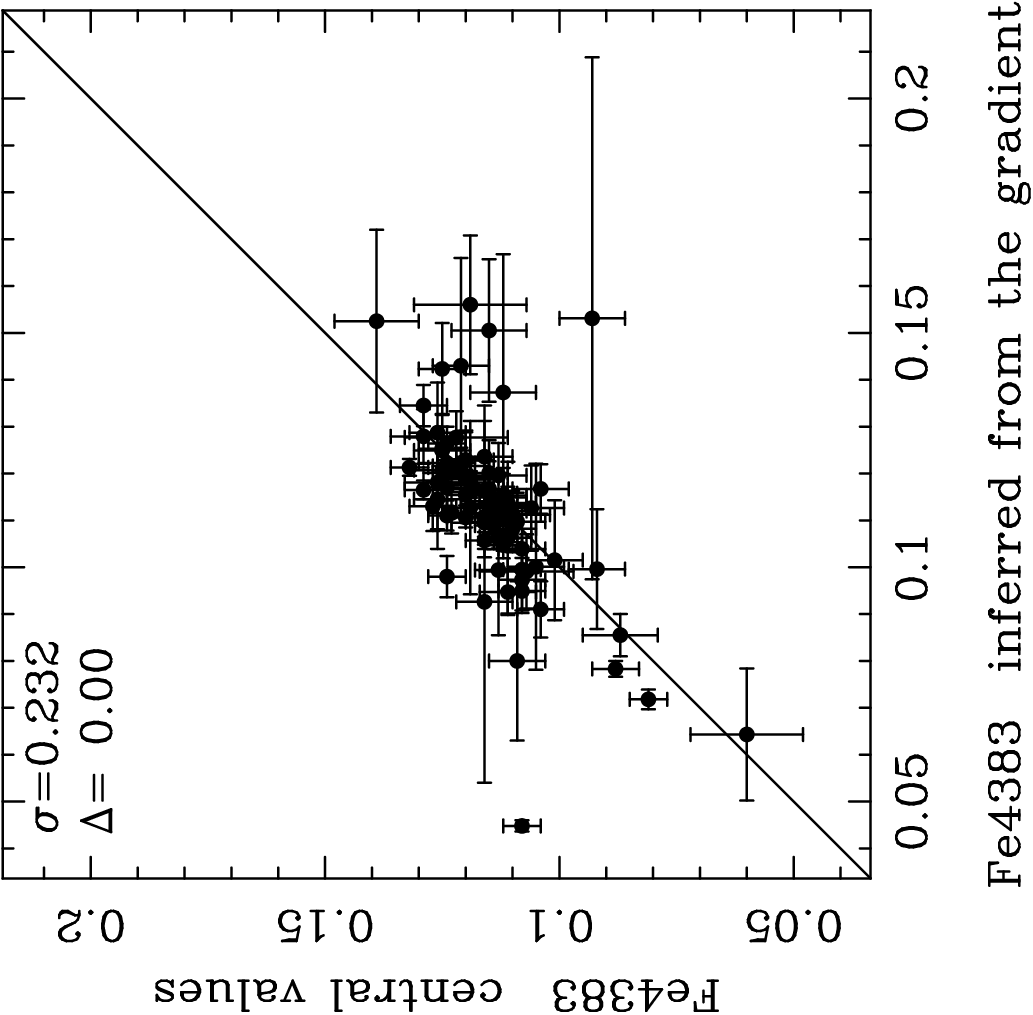}}

\vspace{4mm}

\resizebox{0.18\hsize}{!}{\includegraphics[angle=-90,bb=88 147 591 631]{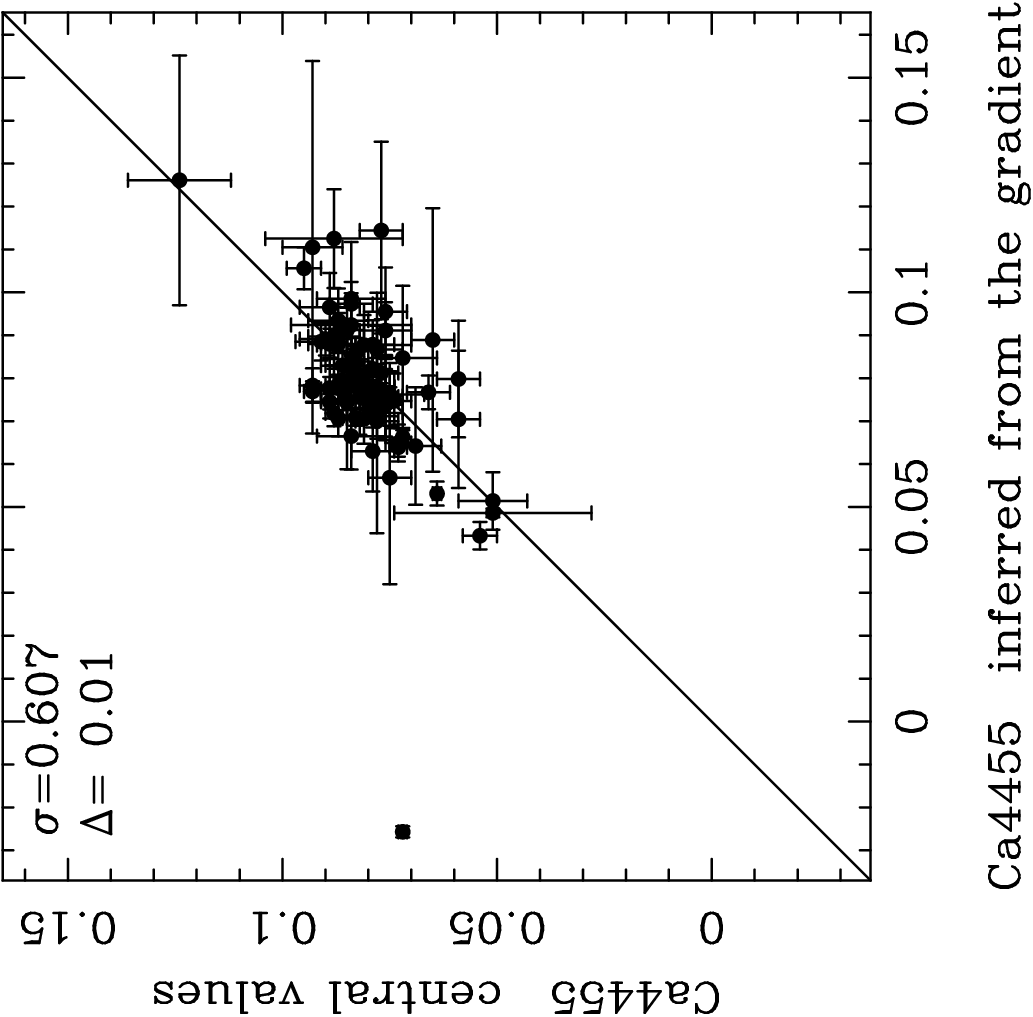}}
\hfill
\resizebox{0.18\hsize}{!}{\includegraphics[angle=-90,bb=88 147 591 631]{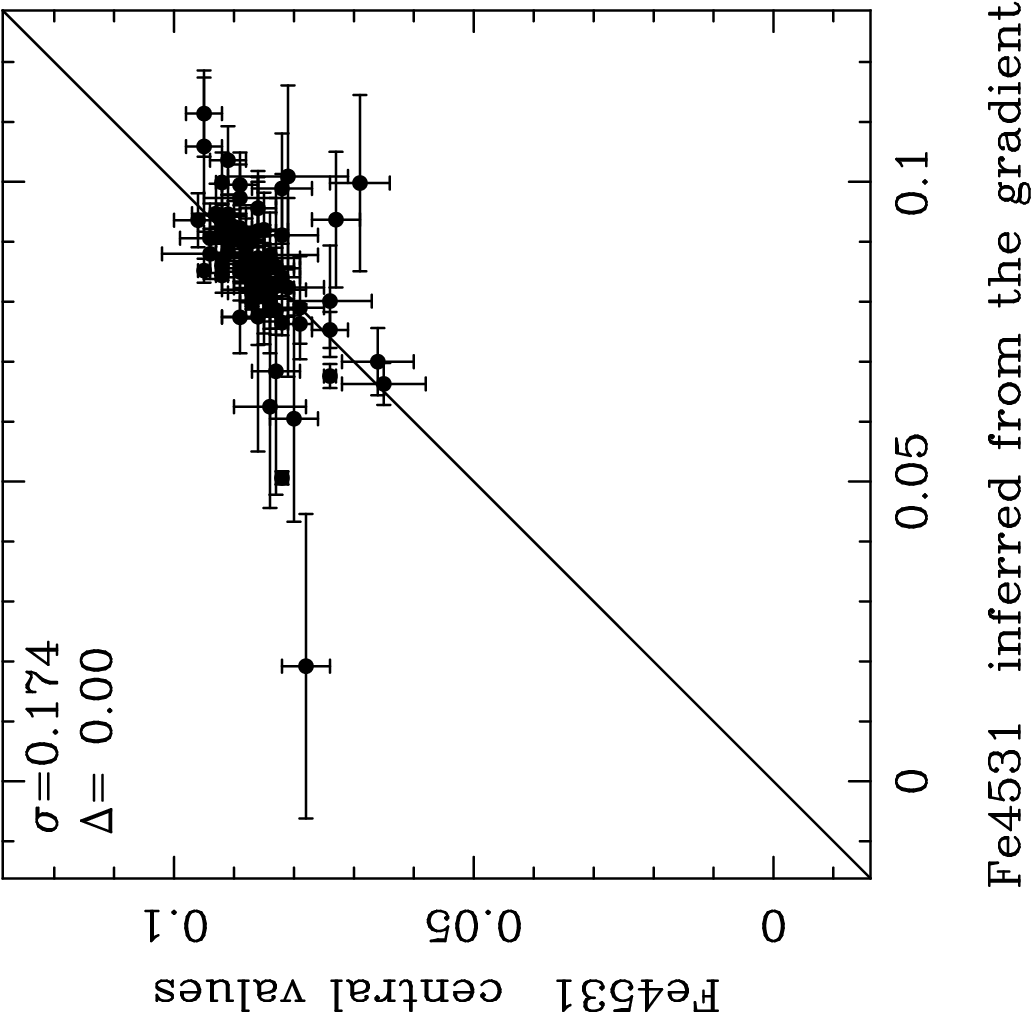}}
\hfill
\resizebox{0.18\hsize}{!}{\includegraphics[angle=-90,bb=88 147 591 631]{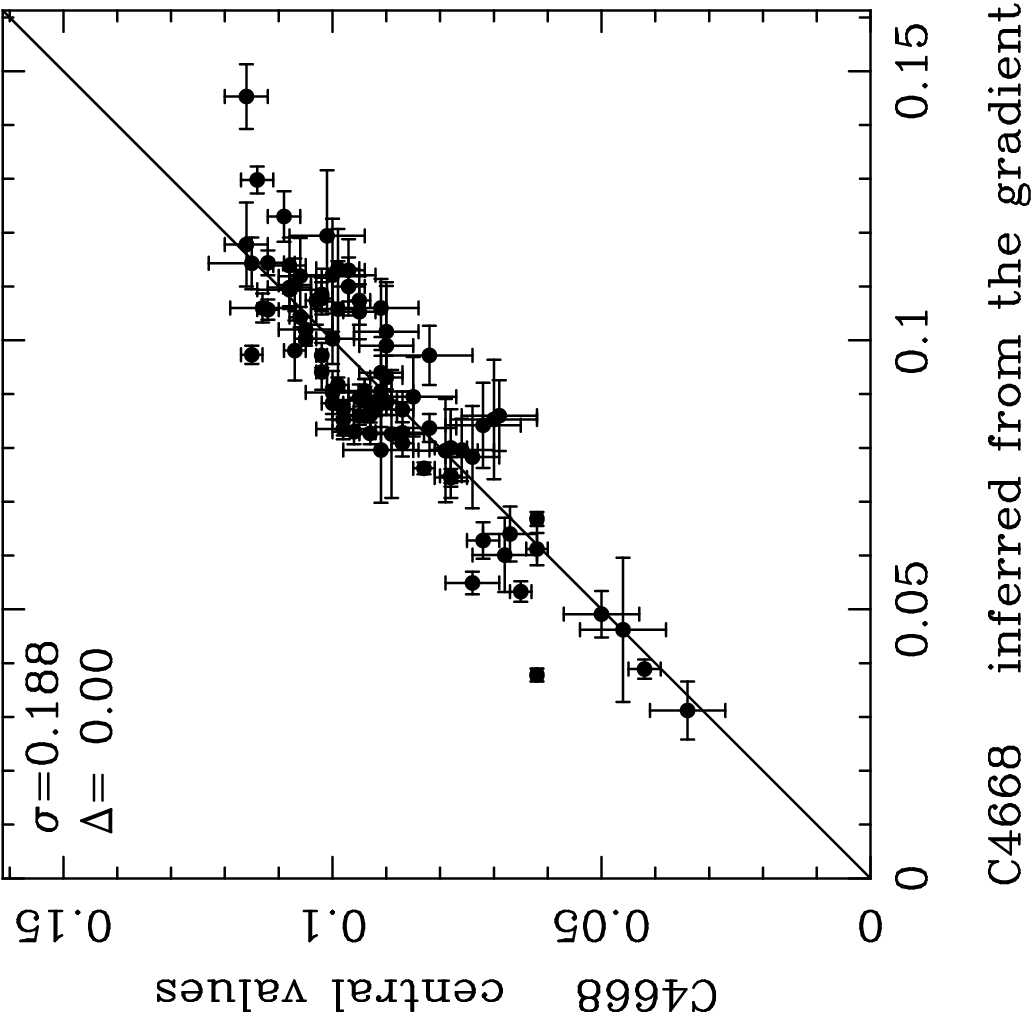}}
\hfill
\resizebox{0.18\hsize}{!}{\includegraphics[angle=-90,bb=88 147 591 631]{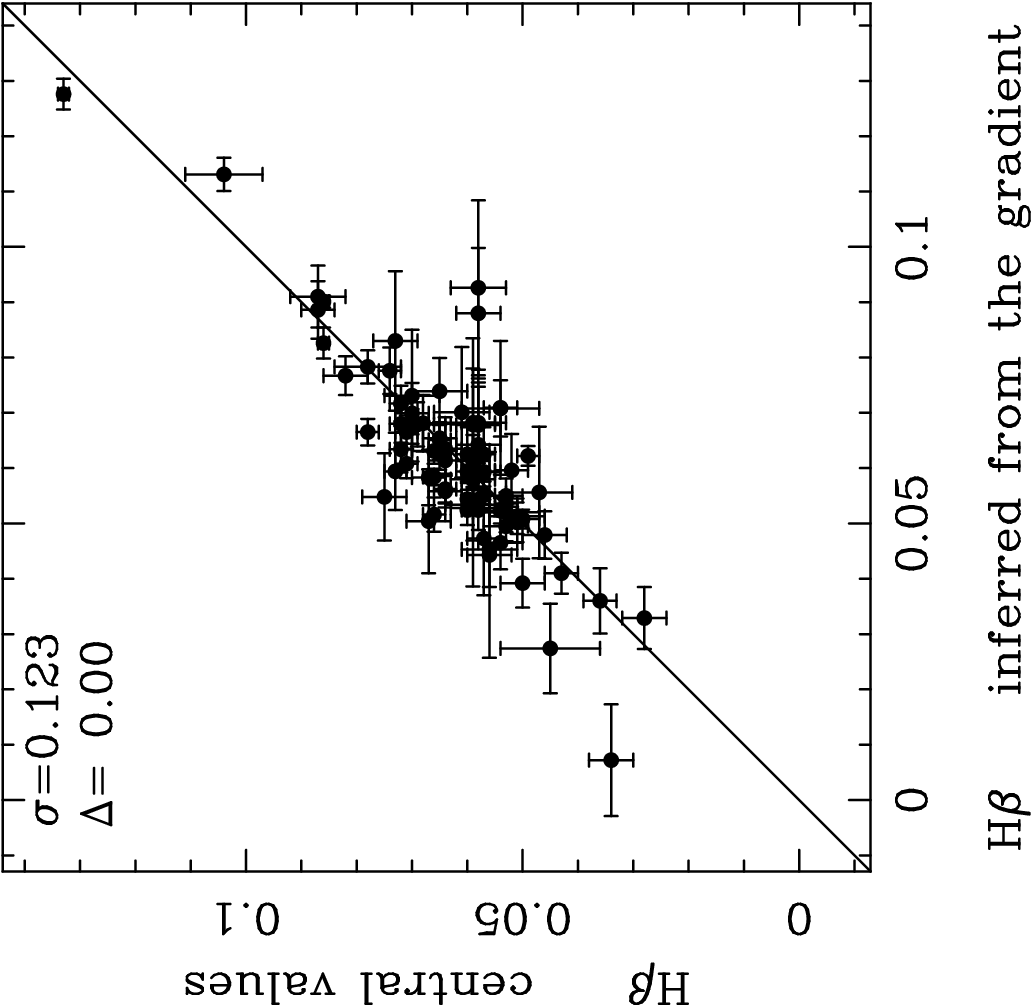}}
\hfill
\resizebox{0.18\hsize}{!}{\includegraphics[angle=-90,bb=88 147 591 631]{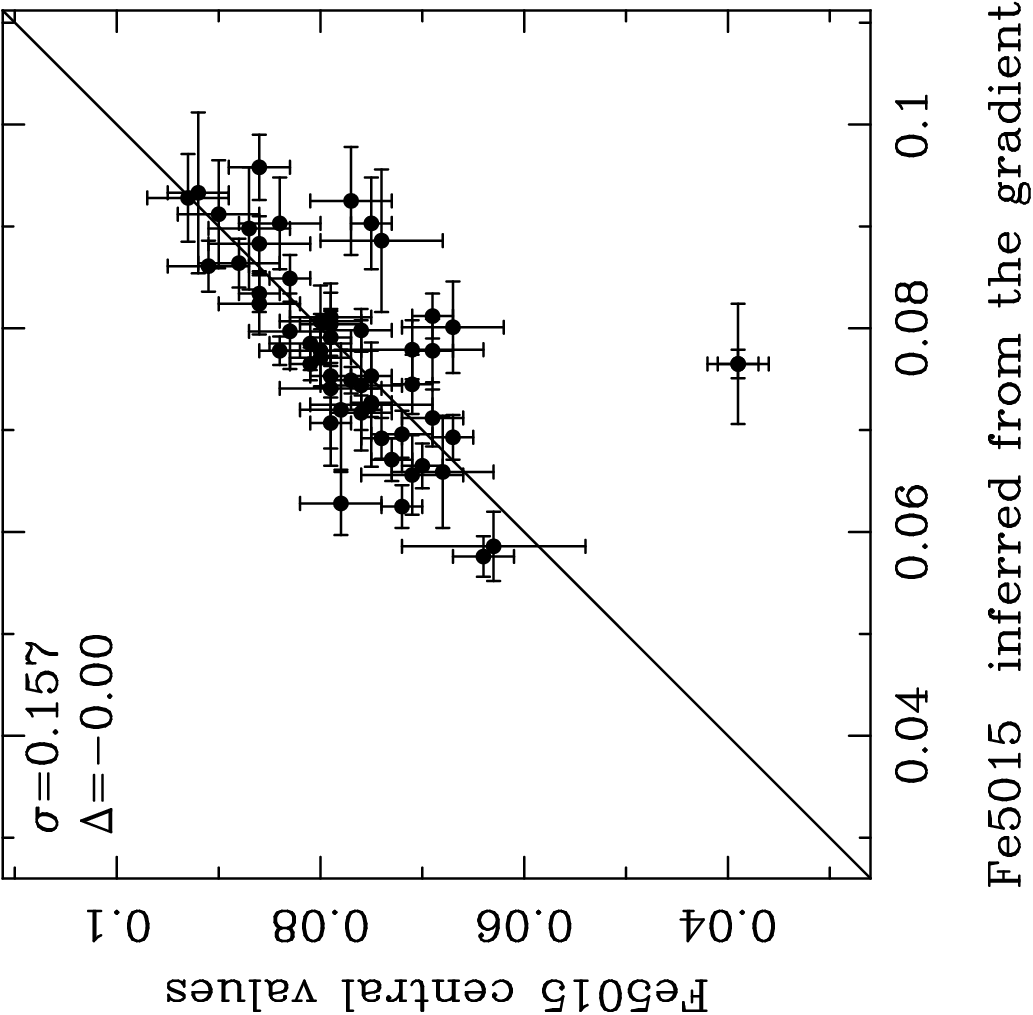}}

\vspace{4mm}

\resizebox{0.18\hsize}{!}{\includegraphics[angle=-90,bb=88 147 591 631]{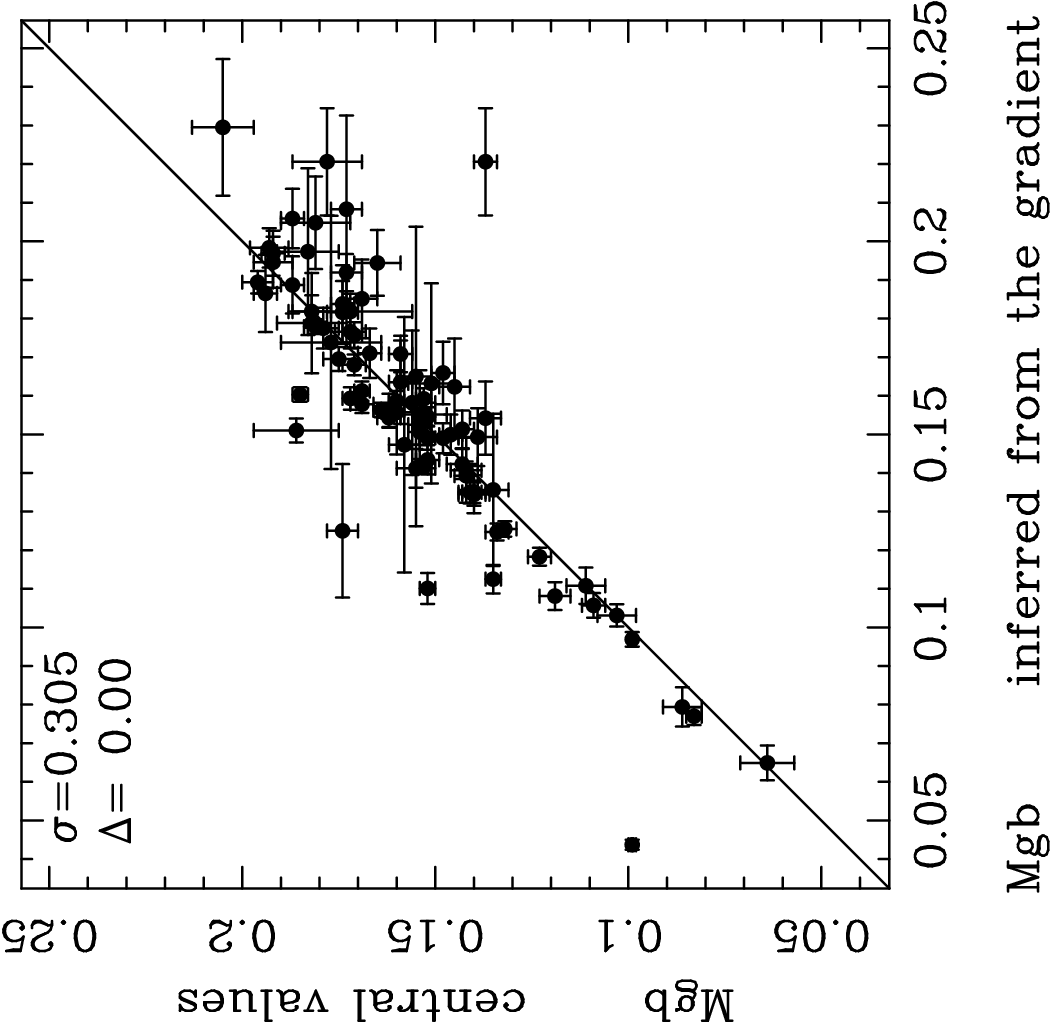}}
\hfill
\resizebox{0.18\hsize}{!}{\includegraphics[angle=-90,bb=88 147 591 631]{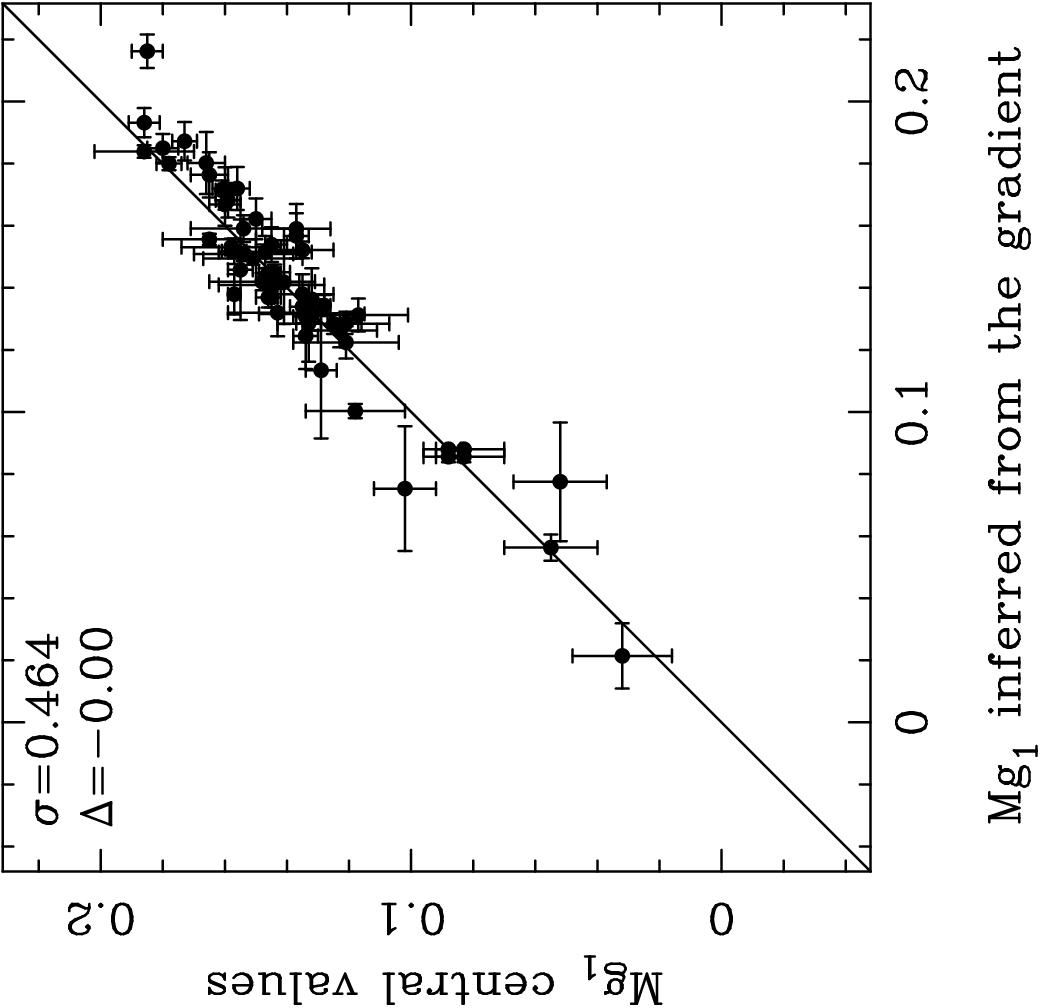}}
\hfill
\resizebox{0.18\hsize}{!}{\includegraphics[angle=-90,bb=88 147 591 631]{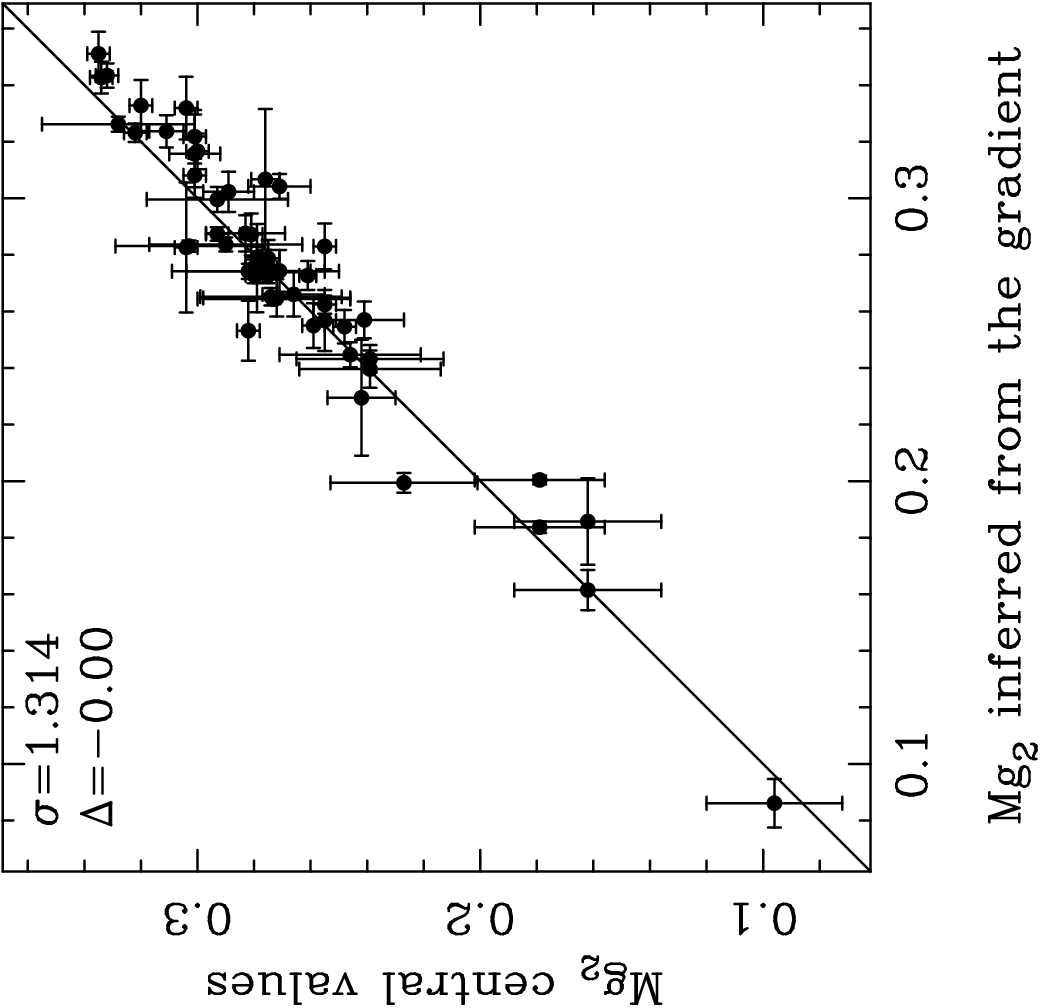}}
\hfill
\resizebox{0.18\hsize}{!}{\includegraphics[angle=-90,bb=88 147 591 631]{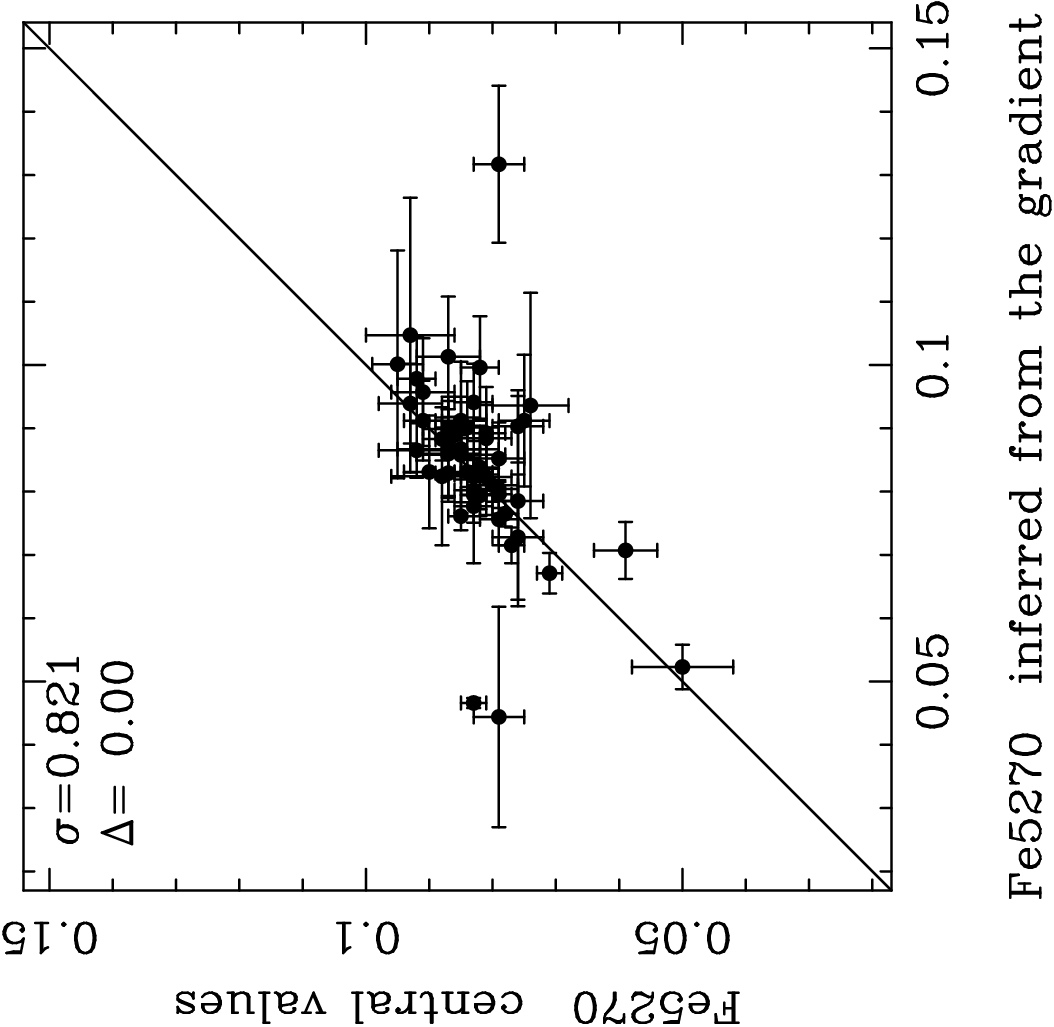}}
\hfill
\resizebox{0.18\hsize}{!}{\includegraphics[angle=-90,bb=88 147 591 631]{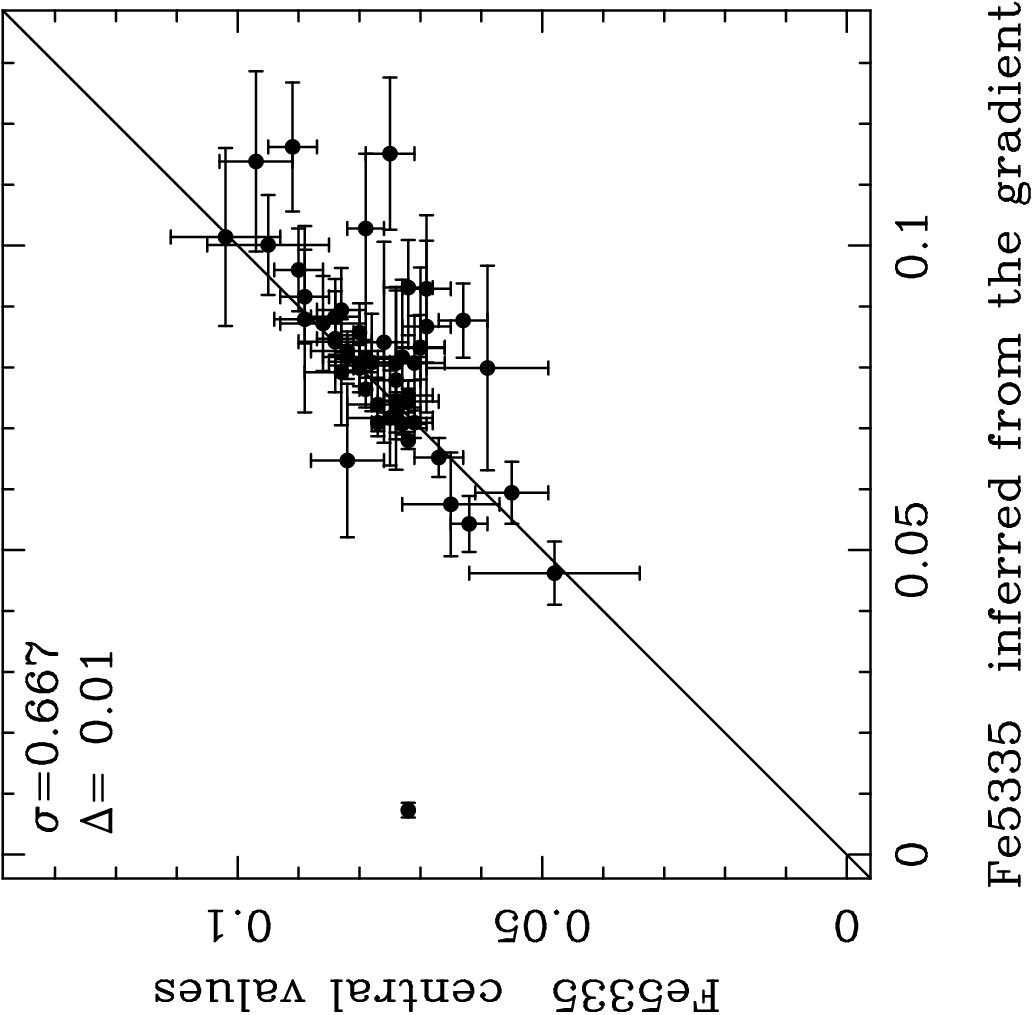}}
\caption{Comparison of the Lick indices measured in the central spectra
(integrated within an equivalent aperture of 4$^{\prime\prime}$ 
at redshift $z=0.016$ -- i.e., $\sim$1.3~kpc) and
the indices obtained from the gradients integrating over the same aperture;
$\sigma$ indicates the root mean square and $\Delta$ the mean offset between
both measurements. The solid line indicates the 1:1 correspondence. In all
cases, the mean offsets are not statistically significant.\label{fig.linearity}}
\end{figure*}

In order to check the assumption of linearity for the gradients we have
compared the central values derived in Paper~I (integrated within an
equivalent aperture of 4$^{\prime\prime}$ at redshift $z=0.016$ -- i.e.,
$\sim$1.3~kpc), with the expected values for this aperture inferred from
the gradients. Fig. \ref{fig.linearity} shows this comparison for all the
indices. Within each panel, the mean offset ($\Delta$) and the root mean
square of the deviations ($\sigma$) are indicated. In general, the
agreement is very good -- the offsets are not statistically significant
for any of the indices and the scatter is compatible with the errors,
lending support to the adopted approximation of a linear fit. 

Table \ref{gradientes.electronic}, available in the electronic edition and
at the web site\footnote{\tt
http://www.ucm.es/info/Astrof/users/pat/pat.html}, lists the gradients in
all the line-strength indices measured for each of the 82 galaxies in our
sample.  A portion of this table is reproduced here to demonstrate its
content and structure. 

\begin{table*}
\caption{Velocity dispersion and line-strength gradients for the sample of 82
galaxies. A portion of the table is shown for guidance regarding its form and
content. For each galaxy and index, the first line shows the measured gradient,
while the second row lists its corresponding formal error. The full table is
available in the electronic edition of this
paper.\label{gradientes.electronic}}
\centering
\begin{tabular}{@{}lrrrrrrrrrr@{}}
\hline\hline
Galaxy  & 
\multicolumn{1}{c}{$\sigma$}    & 
\multicolumn{1}{c}{D4000}       &
\multicolumn{1}{c}{H$\delta_A$} &
\multicolumn{1}{c}{H$\delta_F$} &
\multicolumn{1}{c}{CN$_2$}      &
\multicolumn{1}{c}{Ca4227}      & 
\multicolumn{1}{c}{G4300}       &
\multicolumn{1}{c}{H$\gamma_A$} &
\multicolumn{1}{c}{H$\gamma_F$} &
\multicolumn{1}{r@{}}{Fe4383}      \\
\hline
NGC 221 &$-0.0877$  &$-0.0520$&$-0.0014$  &$0.0008$   &$-0.0247$&$-0.0110$&$0.0104$&$0.0018$   &$0.0079$   &$0.0025$\\
        &$ 0.0126$  &$ 0.0058$&$ 0.0023$  &$0.0026$   &$ 0.0057$&$ 0.0035$&$0.0025$&$0.0020$   &$0.0032$   &$0.0017$\\
        &           &         &          &            &         &         &        &           &          &         \\
        & 
\multicolumn{1}{c}{Ca4455}     &
\multicolumn{1}{c}{Fe4531}     &  
\multicolumn{1}{c}{C4668}      &  
\multicolumn{1}{c}{H$\beta$}   &  
\multicolumn{1}{c}{Fe5015}     &  
\multicolumn{1}{c}{Mg$_1$}     & 
\multicolumn{1}{c}{Mg$_2$}     & 
\multicolumn{1}{c}{Mgb}        & 
\multicolumn{1}{c}{Fe5270}     &  
\multicolumn{1}{r@{}}{Fe5335}     \\ 
\cline{2-11}
        &$0.0027$   &$-0.0030$&$-0.0037$  &$-0.0021$  &$0.0052$ &$-0.0007$&$0.0092$&$ -0.0017$ &$ -0.0038$ &$  0.0020$\\
        &$0.0019$   &$ 0.0016$&$ 0.0017$  &$ 0.0016$&$0.0019$&$ 0.0011$&$  0.0022$&$  0.0018$&$  0.0011$&$  0.0017$\\
\hline
\end{tabular}
\end{table*}

\section{Gradients in age and metallicity}
\label{sec-age.metallicity.gradients}

\begin{figure}
\resizebox{1.0\hsize}{!}{%
\includegraphics[angle=-90]{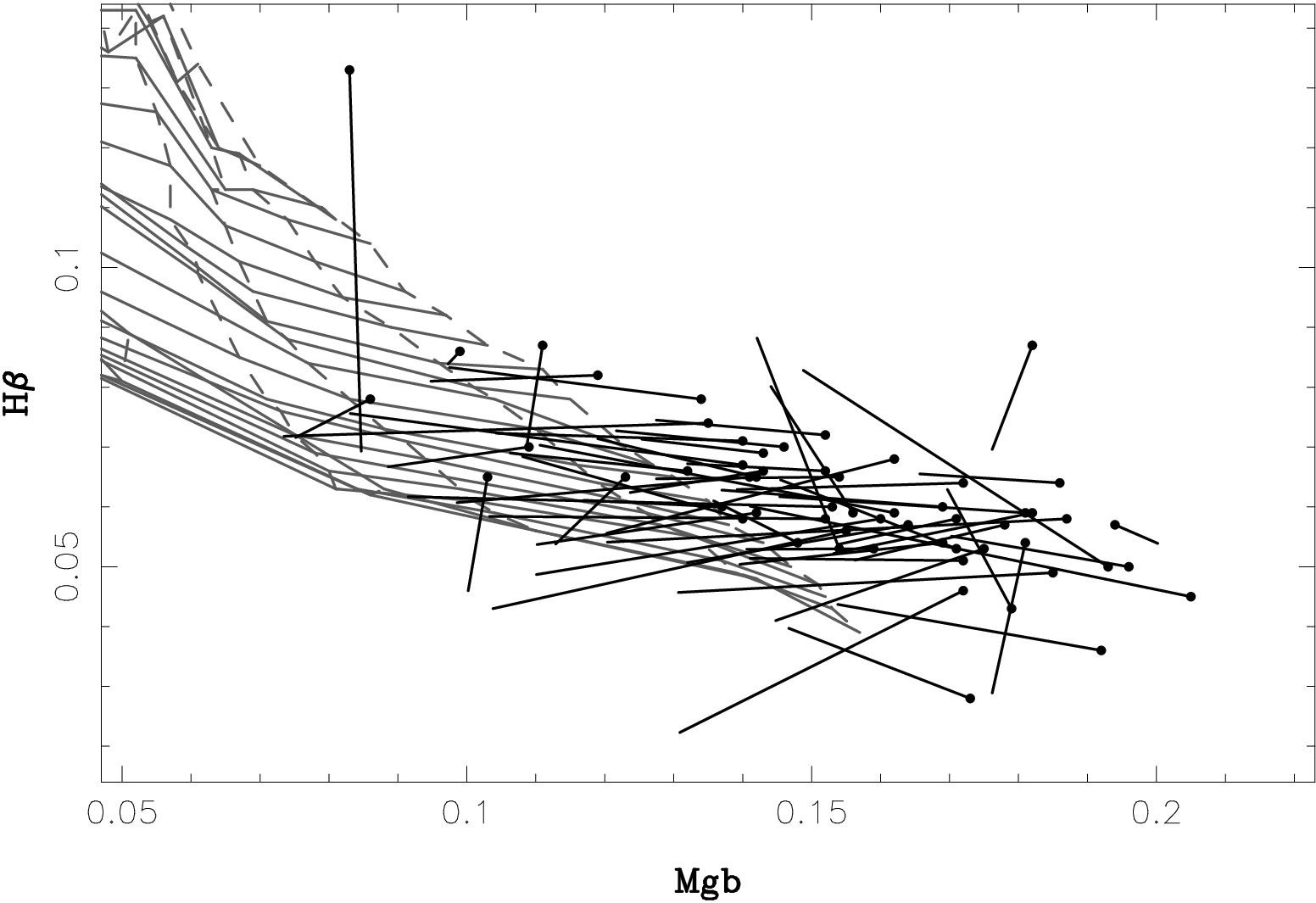}}

\vspace{4mm}

\resizebox{1.0\hsize}{!}{%
\includegraphics[angle=-90]{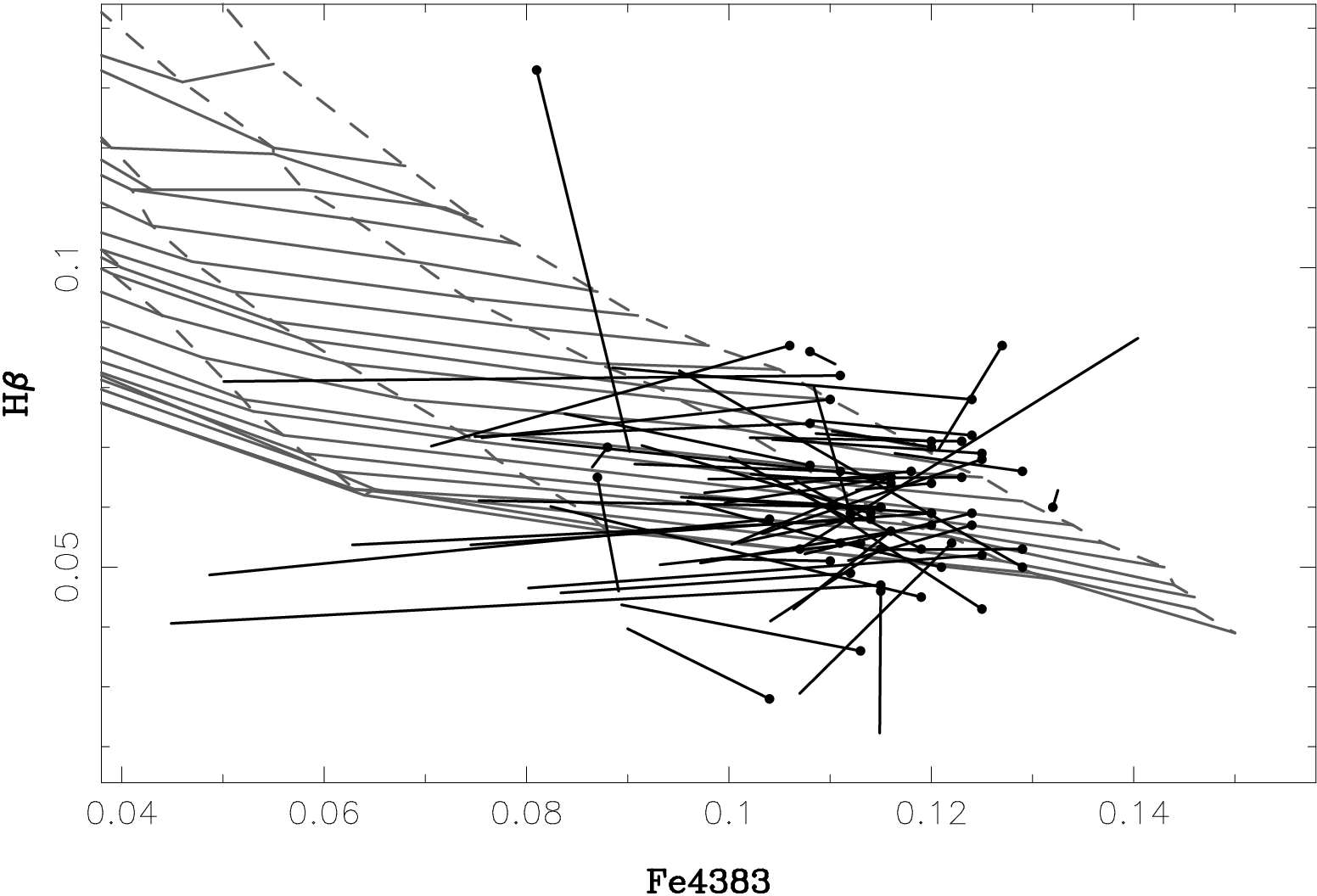}}
\caption{Index--index diagrams using the models of V06. Solid lines represent
the predictions for populations of constant age ({\it top to the bottom}, 1.41, 2.00, 
2.82, 3.98, 5.62, 7.98, 11.22, 15.85 Gyr), while dashed lines show
populations of constant metallicity ({\it left to right}, [M/H]=$-0.68$, 
$-0.38$, 0.0, +0.2). The solid circles in the figure indicate
the position of the centre of the galaxies, while the lines connect with the
measurements at one effective radius. \label{index-index}}
\end{figure}

Figure~\ref{index-index} shows the line-strength gradients in Fe4383, Mgb
and H$\beta$, presented in index--index diagrams that combine both the
central values and the line-strengths at one effective radius. Each
central measurement (solid circle) is connected to its corresponding
measurement in the outer region. Overplotted are the stellar population
models of V06. As can be seen, most of the lines tend to be nearly
parallel (although, admittedly, not entirely) to the iso-age lines taken
from the theoretical grid, which indicates that the gradients are mostly
due to variations in metallicity. However, there are several galaxies
which show significant variations in both age and metallicity with radius. 
We return to this point towards the end of this Section. 

To quantify this behaviour, we have transformed the line-strength
gradients into age and metallicity gradients using the model predictions
of V06, who make use of the code of Vazdekis (1999) updated with a new and
improved stellar library (MILES, S\'anchez-Bl\'azquez et al.\ 2006; these
models will be presented in the forthcoming paper V06). 

With an aim to minimising the effects of low signal-to-noise data, we have
designed a method that makes simultaneous use of 10 different indices. The
selected indices are: H$\delta_A$, CN$_2$, Ca4227, G4300, H$\gamma_A$,
Fe4383, H$\beta$, Fe5015, HA, and Balmer.   We checked that the results
did not depend on the particular choice of indices. None of the Mg indices were
employed, as they could not been measured in all the galaxies from Run~3 
due to the wavelength converage of the spectra.
In order to derive single
estimates of age and metallicity gradients, we have followed a procedure
which is divided into two steps: (i) obtaining individual errors for the
age and metallicity gradients for all the galaxies in each index-index
diagram, and (ii) deriving the final values of the age and metallicity
gradients, along with their associated errors. 

\subsection{Obtaining the errors in the age and metallicity gradients}

\begin{figure}
\resizebox{1.0\hsize}{!}{\includegraphics[angle=-90]{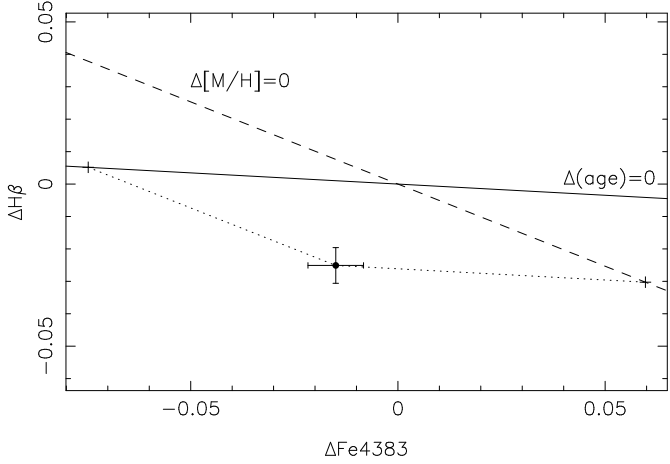}}
\caption{Variations in the H$\beta$ and Fe4383 line-strength indices expected
through changes in metallicity at a constant age of 10~Gyr (solid line), and by
variations in the age at constant solar metallicity (dashed line). The point
represents the values of the gradients for the galaxy NGC~4842A with their
associated 
error bars. The dotted lines indicate the projections of the values over the
lines of constant age and constant metallicity. \label{Deltaindex}}
\end{figure}

We first calculated the slopes of the constant age and metallicity lines
in the models. The slopes of these lines change with the absolute values
of age and metallicity but, in the region in which the galaxies are
located, the values are almost constant. The derived slopes define a new
diagram that we call ``$\Delta$index--$\Delta$index'' -- in the associated
Fig.~\ref{Deltaindex}, the dashed line represents the expected gradients
in the indices if these were due, exclusively, to a variation in age
(assuming a constant solar metallicity). On the other hand, the solid
lines show the predicted gradients if the only parameter changing with
radius was the metallicity (assuming a constant age of 10~Gyr).
Fig.~\ref{Deltaindex} also shows the gradients in H$\beta$ and Fe4383
measured in the galaxy NGC~4842A. To derive the age and metallicity
gradients, we project the position of the galaxy over the lines
$\Delta$age=0 and $\Delta$[M/H]=0, as shown by the dotted lines in the
figure, and interpolate the projection over these lines. 
 
\begin{figure}
\resizebox{1.00\hsize}{!}{\includegraphics[angle=-90]{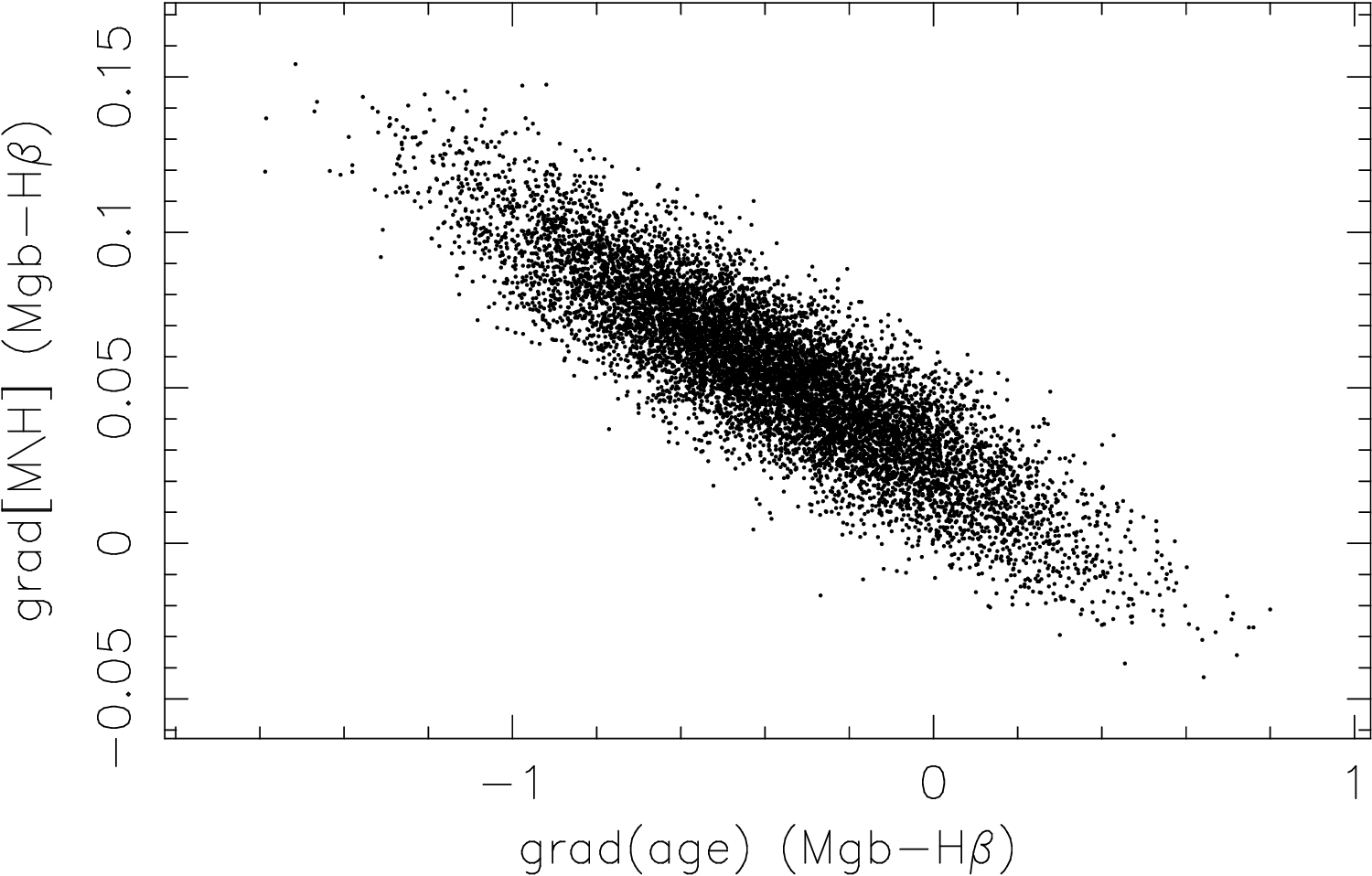}}

\vspace{3mm}

\resizebox{1.00\hsize}{!}{\includegraphics[angle=-90]{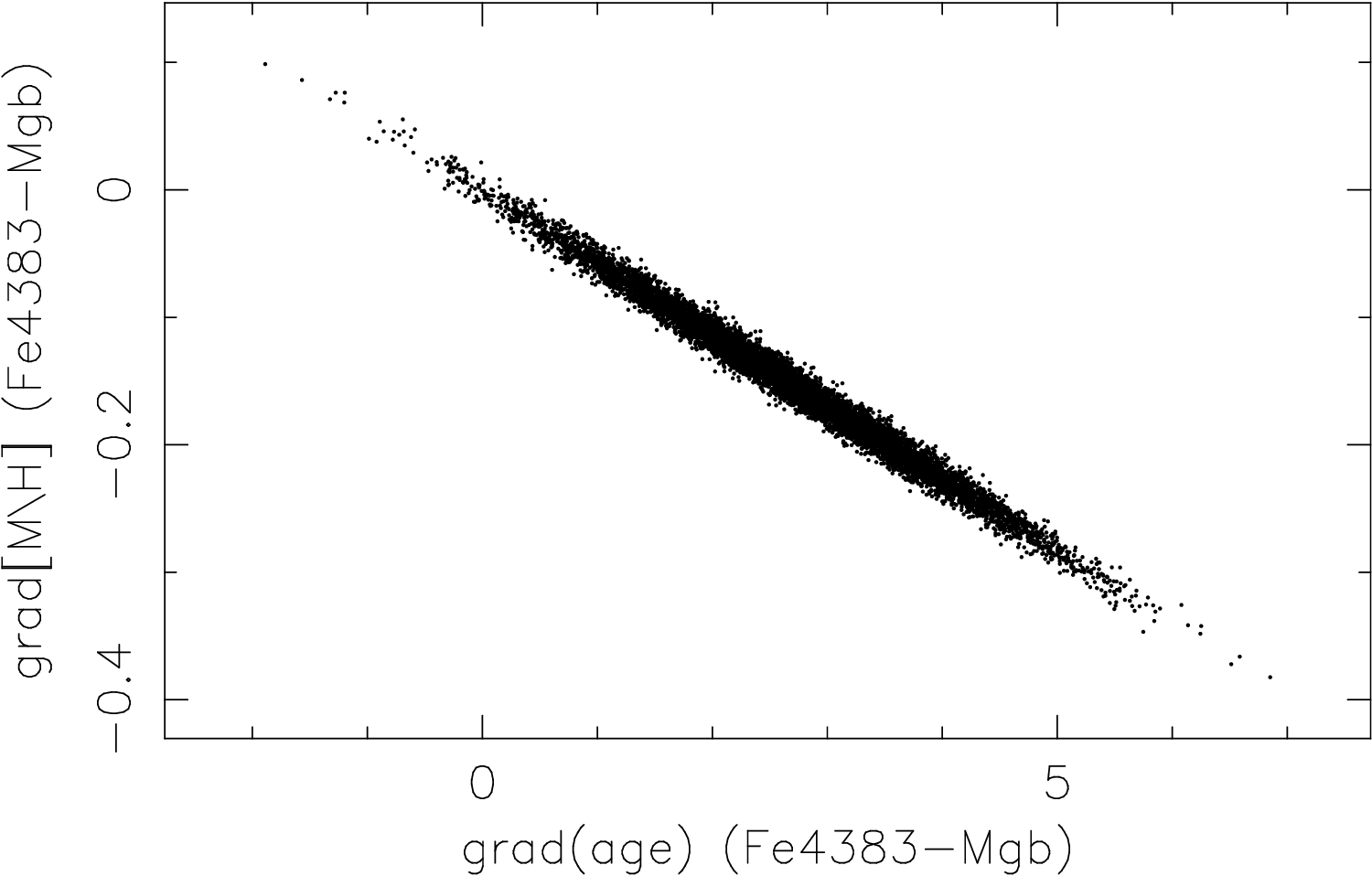}}
\caption{Age and metallicity gradients obtained for the 10$^4$ Monte Carlo
simulations performed in the diagrams $\Delta$H$\beta$--$\Delta$Mgb (top panel)
and $\Delta$Mgb--$\Delta$Fe4383 (bottom panel) for NGC~4842A.  The correlation
of the errors is larger when the age and metallicity are inferred from the
latter.
\label{age.meta.simulaciones}}
\end{figure}

For each galaxy, and in every $\Delta$index--$\Delta$index diagram, we
performed 10$^4$ Monte Carlo simulations in which each point was perturbed
with our errors, following a Gaussian probability distribution.  For each
simulation, an age and metallicity gradient was obtained. This process was
repeated with all possible paired combinations of indices. As an
illustration, Fig.~\ref{age.meta.simulaciones} shows the values of age and
metallicity gradients derived for NGC~4842A, using the
$\Delta$H$\beta$--$\Delta$Mgb and $\Delta$Mgb--$\Delta$Fe4383 diagrams. As
can be seen, due to the non-orthogonality of the index-index diagrams,
there is an artificial anti-correlation between the age and the
metallicity gradients.  To obtain the error in the age and metallicity
gradient for each galaxy, in every diagram, we projected these ellipses
over the $x$- and \mbox{$y$-axes} and calculated the standard deviation of
the resultant Gaussian. We denote the typical errors by
$\sigma$(grad[age])$_{i}$ and $\sigma$(grad[M/H])$_{i}$, where $i$ refers
to each index--index diagram. 
 
\subsection{Obtaining the age and metallicity gradients}

Once we have obtained the typical errors associated with the inferred age
and metallicity gradients for each index-index combination, we again
performed (for each galaxy) 10$^4$ Monte Carlo simulations, into which
Gaussian noise was added to each line-strength index. Simulations were
performed for each index independently, as opposed to each iindex-index
diagram (note that once an index value is simulated, its particular value
is fed simultaneously to all the possible diagrams taking into account the
fact that not all of the different diagrams are independent). 

We then measured the age and metallicity gradients of all the values obtained
in all the simulations $j$, using all the different diagrams $i$. The mean age
and metallicity gradient of simulation $j$ was obtained as an average of
the age and metallicity gradients obtained for this particular simulation using
all the different diagrams $i$, namely
\begin{equation}
   {\rm grad[age]}_{j}=
    \frac{\displaystyle\sum_{i=1}^{n_{\rm com}}
             \frac{{\rm grad[age]}_{i,j}}{\sigma({\rm grad[age]})_{i}^2} }
         {\displaystyle\sum_{i=1}^{n_{\rm com}}
             \frac{1}{\sigma({\rm grad[age]})_{i}^2}},
\end{equation}
and
\begin{equation}
   {\rm grad}[{\rm M}/{\rm H}]_{j}=
   \frac{\displaystyle\sum_{i=1}^{n_{\rm com}}
            \frac{{\rm grad}[{\rm M}/{\rm H}]_{i,j}}
                 {\sigma({\rm grad}[{\rm M}/{\rm H}])_{i}^2} }
        {\displaystyle\sum_{i=1}^{n_{\rm com}}
            \frac{1}{\sigma({\rm grad} [{\rm M}/{\rm H}])_{i}^2}},
\end{equation}   
where grad[age]$_{i,j}$ represents the age gradient obtained with the $i^{th}$
diagram in the simulation $j$, $\sigma$(grad[age])$_{i}$ the uncertainty of the
age gradient in $i$-th diagram (obtained in the first step of the process),
grad[M/H]$_{i,j}$ the metallicity gradient in simulation $j$ (using the
$i^{th}$ diagram), and $\sigma$(grad[M/H])$_{i}$ the error in the metallicity
gradient in this diagram; $n_{\rm com}$ indicates the total number of
diagrams, which is the number of possible index pairs ($n_{\rm com}$=45).

Finally, the age and metallicity gradients for each galaxy were calculated as
the mean values of all the simulations, i.e.
\begin{equation}
{\rm grad[age]}_{\rm final}=\sum_{j=1}^{n_{\rm sim}}
   \frac{{\rm grad}[{\rm age}_{j}]}{n_{\rm sim}},
\end{equation}
and
\begin{equation}
{\rm grad[M/H]}_{\rm final}=\sum_{j=1}^{n_{\rm sim}}
   \frac{{\rm grad}[{\rm M}/{\rm H}]_{j}}{n_{\rm sim}},
\end{equation}
being $n_{\rm sim}$ the number of simulations performed for each line-strength
index gradient.  The final errors in the age and metallicity gradients were
obtained as the standard deviations of the values obtained for each simulation,
that is
\begin{equation}
\sigma_{(\rm age)}=\sqrt{\frac{\displaystyle\sum_{j=1}^{n_{\rm sim}}
  \left({\rm grad[age]}_{j}-{\rm grad[age]}_{\rm final}\right)^2}
    {n_{\rm sim}-1}},
\end{equation}
and
\begin{equation}
\sigma_{(\rm [M/H])}=\sqrt{\frac{\displaystyle\sum_{j=1}^{n_{\rm sim}}
  \left({\rm grad[M/H]}_{j}-{\rm grad[M/H]}_{\rm final}\right)^2}
    {n_{\rm sim}-1}}.
\end{equation}

This process was performed for all 82 galaxies in our sample. The final
values for the age and metallicity gradients obtained in this way,
together with the associated errors, are listed in Table
\ref{tab-gradients}. 

\begin{figure}
\resizebox{0.8\hsize}{!}{%
\includegraphics[angle=-90]{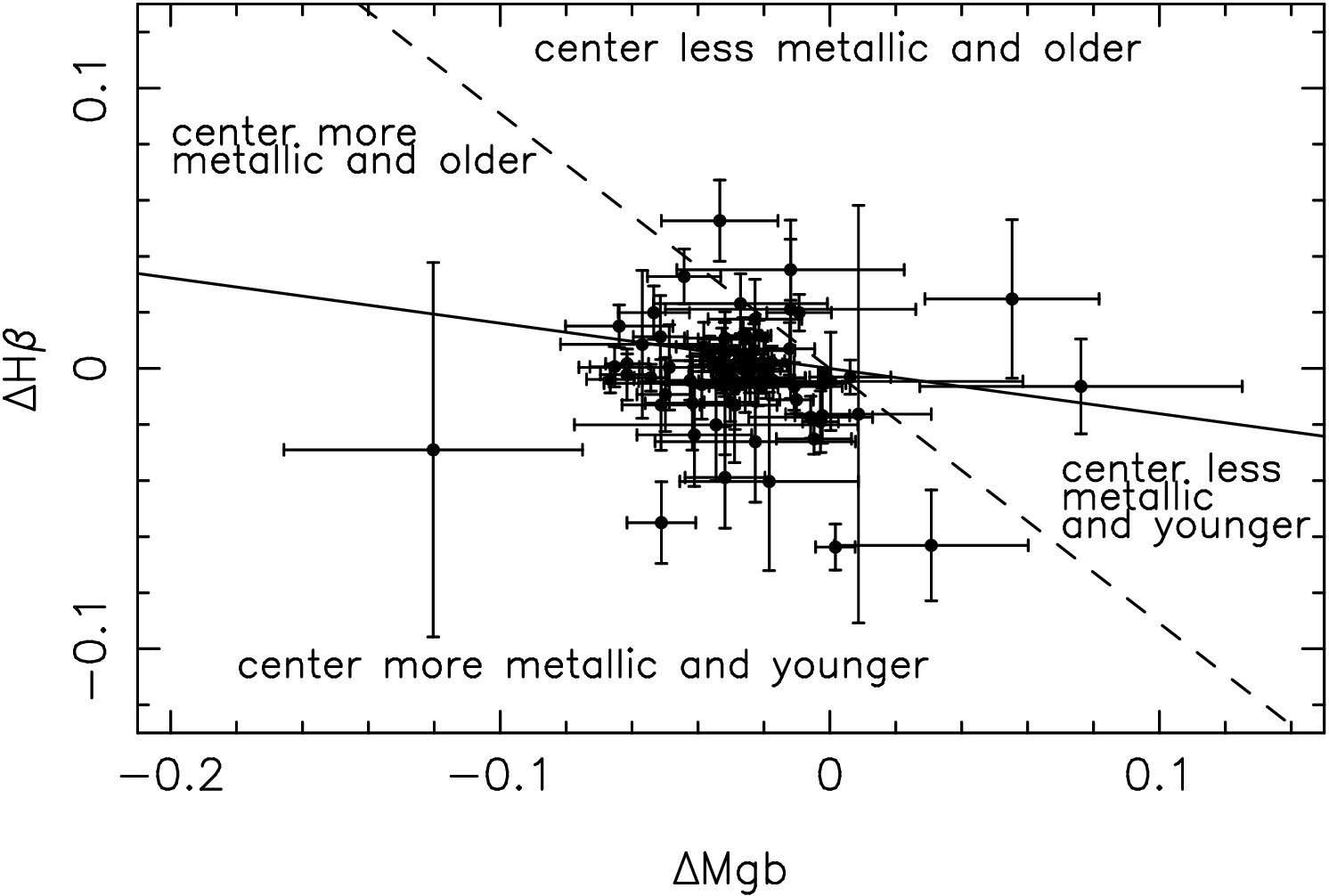}}

\vspace{4mm}

\resizebox{0.8\hsize}{!}{%
\includegraphics[angle=-90]{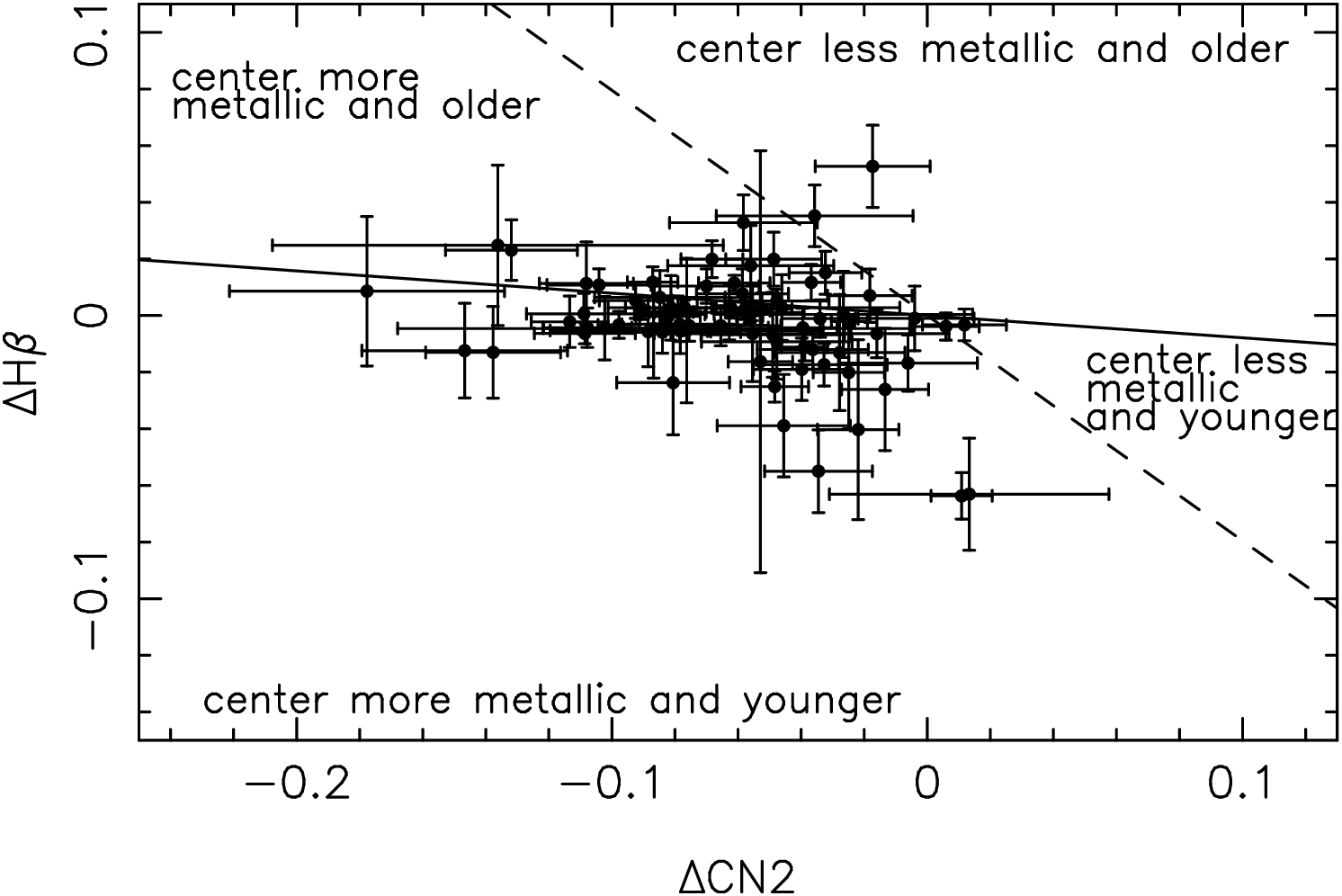}}

\vspace{4mm}

\resizebox{0.8\hsize}{!}{%
\includegraphics[angle=-90]{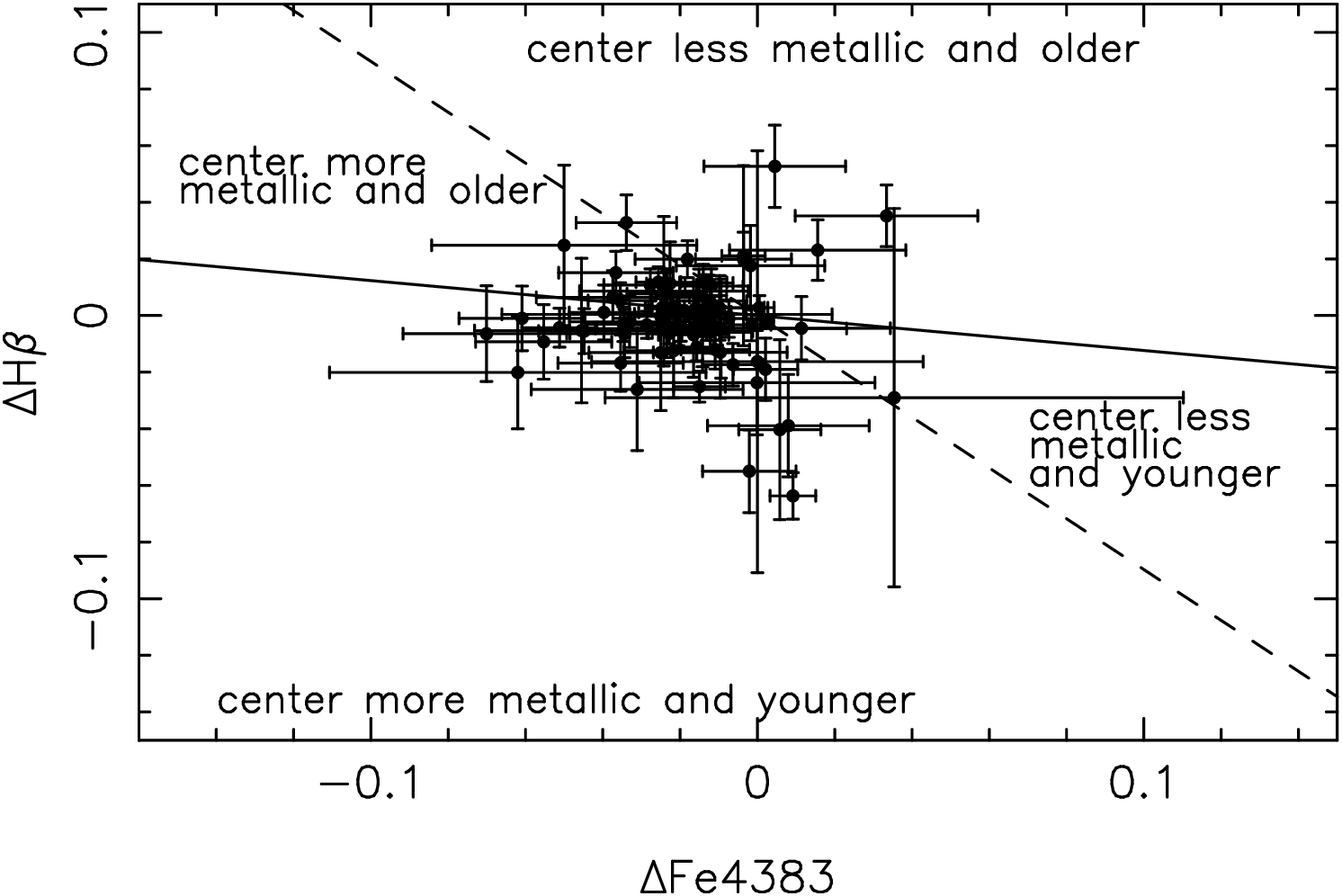}}
\caption{$\Delta$index--$\Delta$index diagrams combining the H$\beta$ gradient
with Mgb, CN$_2$, and Fe4383. Solid lines indicate the slope of the
predictions at constant age (10~Gyr); the dashed line shows the slope of the
simple stellar population at solar metallicity. The solid circles 
represent the values of the
line-strength gradients for the sample of galaxies in low-density environments.
\label{index_uno}}
\end{figure}

Figure~\ref{index_uno} shows three different $\Delta$index--$\Delta$index
diagrams in which we have over-plotted the galaxies in low-density
environments. The lines of constant age and metallicity divide these
diagrams into four distinct regimes. These regimes indicate the position
of the galaxies with positive/negative differences in age and metallicity
between the central and the external parts, as labeled in the panels. As
can be seen, most of the galaxies are situated in the regime consistent
with their centres being younger and more metal-rich than their outer
regions. The bulk of the galaxies, however, are further away from the line
$\Delta$[M/H]=0 (dashed line) than from the line $\Delta$age=0 (solid
line), suggesting that the gradients in the line-strength indices are due
(mostly) to variations of metallicity with radius. 

\begin{table*}
\caption{Age and metallicity gradients derived with the method of
Sec.~\ref{sec-age.metallicity.gradients} (second and third columns), and
metallicity gradients derived in several index--index diagrams using different
metallicity indicators (as indicated in the column headers) combined with
H$\beta$. The second row for each galaxy indicates the associated
errors in the gradients.
\label{tab-gradients}}
\centering
\begin{tabular}{@{}l rrrrrrr@{}}
\hline\hline
        &                               &                                 & CN$_2$&C4668&Fe4383&Mgb&Ca4227\\      
 Galaxy & \multicolumn{1}{c}{grad(age)}&\multicolumn{1}{c}{grad($[$M/H$]$)}&\multicolumn{1}{c}{grad($[$M/H$])$}
 &\multicolumn{1}{c}{grad($[$M/H$])$}&\multicolumn{1}{c}{grad($[$M/H$])$}&\multicolumn{1}{c}{grad($[$M/H$])$}&\multicolumn{1}{c@{}}{grad($[$M/H$])$}\\
        &$\pm \sigma$                   &  $\pm \sigma$                  &  $\pm \sigma$    & $\pm \sigma$  & $\pm \sigma$  & $\pm \sigma$
        &$\pm \sigma$\\
   \hline
  NGC 221     &$ 0.0151$&$ -0.0273$  &  $-0.170$&$-0.246$ &$ 0.023$&$-0.171$&$ 0.082$\\
              &  0.0180 &   0.0226   &  0.241  &  0.139  &  0.222 &  0.259 &  0.373\\
  NGC 315     &$ 0.0850$&$ -0.2745$  &  $ 0.010$&$-0.689$ &$-0.214$&$-1.266$&$-0.209$ \\
              &  0.0486 &   0.0651   &  0.112  &  0.089  &  0.095 &  0.148 &  0.151\\
  NGC 507     &$-0.6291$&$  0.5756$  &  $ 0.076$&$ 0.451$ &$ 1.004$&$ 0.453$&$ 0.935$\\
              &  0.1508 &   0.2140   &  0.309  &  0.324  &  0.354 &  0.611 &  0.664\\
  NGC 584     &$ 0.0814$&$ -0.1937$  &  $-0.673$&$-0.510$ &$-0.341$&$-0.414$&$-0.294$\\
              &  0.0868 &  0.1185    &  0.196  &  0.114  &  0.145 &  0.242 &  0.355\\
  NGC 636     &$ 0.0779$&$ -0.3627$  &  $-0.740$&$-0.804$ &$-0.427$&$-0.291$&$ 0.117$ \\
              &  0.1148 &  0.1592    &  0.220  &  0.200  &  0.300 &  0.274 &  0.382 \\
  NGC 821     &$ 0.0563$&$ -0.8237$  &  $-1.560$&$-1.058$ &$-0.493$&$-0.981$&$-1.057$\\
              &  0.1157 &  0.9981    &  0.351  &  0.235  &  0.293 &  0.391 &  0.478\\
  NGC 1453    &$-0.2652$&$ -0.0141$  &  $-0.170$&$-0.246$ &$ 0.023$&$-0.171$&$ 0.082$\\
              &  0.1036 &  0.1259    &  0.241  &  0.139  &  0.222 &  0.259 &  0.373\\
  NGC 1600    &$ 0.2293$&$ -0.0732$  &  $-0.765$&$-0.521$ &$-0.592$&$-0.561$&$-0.897$\\
              &  0.0504 &  0.0708    &  0.116  &  0.072  &  0.121 &  0.216 &  0.203\\
  NGC 1700    &$ 0.0803$&$ -0.3432$  &  $-0.750$&$-0.693$ &$-0.096$&$-0.384$&$ 0.534$ \\
              &  0.1116 &  0.1349    &  0.186  &  0.183  &  0.237 &  0.274 &  0.508\\
  NGC 2300    &$ 0.1531$&$ -0.2416$  &  $-0.553$&$-0.902$ &$-0.335$&$-0.644$&$-0.049$\\
              &  0.1284 &  0.1582    &  0.270  &  0.156  &  0.252 &  0.353 &  0.508\\
  NGC 2329    &$ 0.2413$&$ -0.3977$  &  $-0.610$&$-0.510$ &$-1.081$&$ 1.243$&$-0.014$\\
              &  0.1731 &  0.1840    &  0.346  &  0.284  &  0.377 &  0.884 &  0.478\\
  NGC 2693    &$-0.1917$&$ -0.0072$  &  $-0.420$&$-0.220$ &$ 0.049$&$ 0.210$&$-0.168$\\
              &  0.0786 &  0.0997    &  0.116  &  0.138  &  0.206 &  0.205 &  0.265\\
  NGC 2694    &$-0.1551$&$ -0.0214$  &  $ 0.119$&$-0.531$ &$ 0.270$&$ 0.186$&$-0.882$\\
              &  0.2540 &  0.2572    &  1.146  &  0.218  &  0.467 &  0.853 &  1.057\\
  NGC 2778    &$ 0.0644$&$ -0.3501$  &  $-1.041$&$-0.733$ &$-0.284$&$-1.158$&$-0.233$\\
              &  0.0879 &  0.1155    &  0.186  &  0.125  &  0.178 &  0.182 &  0.249\\
  NGC 2832    &$ 0.1436$&$ -0.3521$  &  $-0.819$&$-0.747$ &$-0.712$&$ 4.801$&$-1.200$\\
              &  0.0696 &  0.1046    &  0.120  &  0.102  &  0.196 &  1.205 &  0.297\\
  NGC 3115    &$ 0.2518$&$ -0.4864$  &  $-1.121$&$-0.691$ &$-0.416$&$-0.692$&$-0.367$\\
              &  0.0501 &  0.0845    &  0.154  &  0.070  &  0.090 &  0.103 &  0.129\\
  NGC 3377    &$ 0.0334$&$ -0.3600$  &  $-1.118$&$-0.960$ &$-0.498$&$-1.143$&$-0.229$\\
              &  0.0679 &  0.1147    &  0.152  &  0.103  &  0.241 &  0.214 &  0.222\\

 \hline
\end{tabular}
\end{table*}

\addtocounter{table}{-1}
\begin{table*}
\caption{\it Continued.}
\centering
\begin{tabular}{@{}l rrrrrrr@{}}
\hline\hline
        &                               &                                 & CN$_2$&C4668&Fe4383&Mgb&Ca4227\\      
 Galaxy & \multicolumn{1}{c}{grad($\log$
 age)}&\multicolumn{1}{c}{grad($[$M/H$]$)}&\multicolumn{1}{c}{grad($[$M/H$]$)}
 &\multicolumn{1}{c}{grad($[$M/H$]$)}&\multicolumn{1}{c}{grad($[$M/H$]$)}&\multicolumn{1}{c}{grad($[$M/H$]$)}&\multicolumn{1}{c@{}}{grad($[$M/H$]$)}\\
        &$\pm \sigma$                   &  $\pm \sigma$                  &  $\pm \sigma$    & $\pm \sigma$  & $\pm \sigma$  & $\pm \sigma$
        &$\pm \sigma$\\
   \hline
  NGC 3379    &$ 0.1055$&$ -0.2279$  &  $-0.515$&$-0.419$ &$-0.239$&$-1.069$&$-0.262$\\
              &  0.0347 &  0.0528    &  0.066  &  0.051  &  0.072 &  0.128 &  0.098\\
  NGC 3605    &$ 0.0574$&$ -0.3890$  &  $-0.050$&$-0.385$ &$-0.869$&$-0.452$&$ 0.235$\\
              &  0.1046 &  0.1486    &  0.218  &  0.191  &  0.278 &  0.288 &  0.348\\
  NGC 3608    &$ 0.0595$&$ -0.3767$  &  $-0.906$&$-0.575$ &$-0.145$&$-0.706$&$ 0.104$\\
              &  0.1043 &  0.1180    &  0.216  &  0.209  &  0.230 &  0.306 &  0.334\\
  NGC 3641    &$ 0.0067$&$ -0.3821$  &  $-1.608$&$-0.547$ &$-0.209$&$-0.856$&$-0.261$\\
              &  0.2529 &  0.2634    &  0.501  &  0.359  &  0.491 &  0.636 &  0.731\\
  NGC 3665    &$ 0.0616$&$ -0.1746$  &  $-0.092$&$-0.383$ &$-0.105$&$-0.086$&$ 0.284$\\
              &  0.0716 &  0.0941    &  0.166  &  0.095  &  0.163 &  0.215 &  0.285\\
  NGC 3818    &$ 0.1865$&$ -0.3875$  &  $-1.089$&$-0.874$ &$-0.295$&$-0.840$&$-0.216$ \\
              &  0.0728 &  0.1087    &  0.152  &  0.106  &  0.147 &  0.172 &  0.203\\
  NGC 4261    &$ 0.1011$&$ -0.2411$  &  $-0.538$&$-0.350$ &$-0.171$&$-0.335$&$-0.119$\\
              &  0.0453 &  0.0597    &  0.075  &  0.068  &  0.091 &  0.132 &  0.184\\
  NGC 4278    &$ 0.1895$&$ -0.4505$  &  $-0.981$&$-0.623$ &$-0.451$&$-1.034$&$-0.397$\\
              &  0.0408 &  0.0647    &  0.081  &  0.056  &  0.095 &  0.121 &  0.132 \\
  NGC 4365    &$ 0.1087$&$ -0.2385$  &  $-0.567$&$-0.459$ &$-0.305$&$-0.586$&$-0.247$\\
              &  0.0378 &  0.0542    &  0.069  &  0.050  &  0.077 &  0.091 &  0.121 \\
  NGC 4374    &$-0.1112$&$ -0.0716$  &  $-0.452$&$-0.258$ &$-0.000$&$-0.239$&$ 0.154$\\
              &  0.0421 &  0.0668    &  0.109  &  0.077  &  0.092 &  0.126 &  0.158\\
  NGC 4415    &$ 0.1048$&$ -0.0759$  &  $ 0.074$&$-0.038$ &$-0.071$&$-0.428$&$-0.581$ \\
              &  0.0625 &  0.0658    &  0.136  &  0.080  &  0.104 &  0.243 &  0.183\\
  NGC 4431    &$ 0.1725$&$ -0.1285$  &  $-0.610$&$-0.216$ &$-0.260$&$-0.409$&$-0.548$\\
              &  0.0928 &  0.1048    &  0.189  &  0.145  &  0.194 &  0.219 &  0.423\\
  NGC 4464    &$-0.1041$&$ -0.0150$  &  $-0.550$&$-0.206$ &$-0.009$&$-0.374$&$ 0.213$\\
              &  0.0796 &  0.0931    &  0.169  &  0.141  &  0.141 &  0.178 &  0.226\\
  NGC 4467    &$ 0.0916$&$ -0.5234$  &  $-0.798$&$-1.202$ &$-0.719$&$-0.669$&$-0.303$\\
              &  0.2661 &  0.2767    &  0.492  &  0.504  &  0.526 &  0.782 &  0.719\\
  NGC 4472    &$ 0.1062$&$ -0.3919$  &  $-0.543$&$-0.472$ &$-0.220$&$-0.331$&$-0.506$\\
              &  0.0252 &  0.0460    &  0.047  &  0.035  &  0.056 &  0.072 &  0.091\\
  NGC 4478    &$ 0.1997$&$ -0.2858$  &  $-0.508$&$-0.428$ &$-0.391$&$-0.393$&$-0.454$\\
              &  0.0508 &  0.0667    &  0.087  &  0.056  &  0.090 &  0.122 &  0.135\\
  NGC 4486B   &$ 0.4092$&$ -0.3663$  &  $-1.059$&$-0.891$ &$-0.362$&$-1.184$&$-0.337$\\
              &  0.2165 &  0.2441    &  0.260  &  0.200  &  0.483 &  0.442 &  0.576\\
  NGC 4489    &$ 0.3077$&$ -0.4354$  &  $-0.259$&$-0.700$ &$-0.752$&$-0.359$&$-0.483$\\
              &  0.1042 &  0.1423    &  0.228  &  0.164  &  0.260 &  0.228 &  0.492\\
  NGC 4552    &$ 0.0526$&$ -0.2075$  &  $-0.782$&$-0.584$ &$-0.158$&$-0.546$&$-0.091$\\
              &  0.0674 &  0.0939    &  0.127  &  0.076  &  0.151 &  0.144 &  0.287 \\

  \hline
\end{tabular}
\end{table*}

\addtocounter{table}{-1}
\begin{table*}
\caption{\it Continued.}
\centering
\begin{tabular}{@{}l rrrrrrr@{}}
\hline\hline
        &                               &                                 & CN$_2$&C4668&Fe4383&Mgb&Ca4227\\      
 Galaxy &
 \multicolumn{1}{c}{grad(age)}&\multicolumn{1}{c}{grad($[$M/H$]$)}&\multicolumn{1}{c}{grad($[$M/H$]$)}
 &\multicolumn{1}{c}{grad($[$M/H$]$)}&\multicolumn{1}{c}{grad($[$M/H$]$)}&\multicolumn{1}{c}{grad($[$M/H$]$)}&\multicolumn{1}{c@{}}{grad($[$M/H$]$)}\\
        &$\pm \sigma$                   &  $\pm \sigma$                  &  $\pm \sigma$    & $\pm \sigma$  & $\pm \sigma$  & $\pm \sigma$
        &$\pm \sigma$\\
   \hline
  NGC 4564    &$ 0.1010$&$ -0.3264$  &  $-0.842$&$-0.740$ &$ 0.050$&$-0.520$&$-0.613$\\
              &  0.0453 &  0.0635    &  0.107  &  0.059  &  0.080 &  0.109 &  0.131\\
  NGC 4594    &$ 0.0938$&$ -0.3880$  &  $-1.481$&$-0.929$ &$-0.332$&$-1.162$&$-0.281$\\
              &  0.1078 &  0.1491    &  0.272  &  0.142  &  0.335 &  0.353 &  0.496\\
  NGC 4621    &$ 0.1645$&$ -0.4335$  &  $-0.770$&$-0.702$ &$-0.332$&$-0.594$&$-0.274$\\
              &  0.0461 &  0.0718    &  0.097  &  0.067  &  0.114 &  0.143 &  0.184\\
  NGC 4636    &$ 0.2275$&$ -0.3357$  &  $-0.493$&$-0.446$ &$-0.335$&$-0.768$&$-0.215$\\
              &  0.0766 &  0.0981    &  0.178  &  0.120  &  0.158 &  0.182 &  0.289\\
  NGC 4649    &$ 0.0305$&$ -0.2151$  &  $-0.398$&$-0.465$ &$-0.095$&$-0.364$&$-0.215$\\
              &  0.0624 &  0.0845    &  0.159  &  0.105  &  0.142 &  0.137 &  0.261 \\
  NGC 4673    &$-0.1593$&$ -0.1049$  &  $-0.231$&$-0.540$ &$ 0.252$&$-0.580$&$-0.415$ \\
              &  0.0960 &  0.1239    &  0.176  &  0.175  &  0.213 &  0.249 &  0.339\\
  NGC 4692    &$ 0.1630$&$ -0.3817$  &  $-0.920$&$-0.737$ &$-0.286$&$-0.802$&$-0.490$\\
              &  0.1103 &  0.1285    &  0.233  &  0.204  &  0.251 &  0.305 &  0.349\\
  NGC 4697    &$ 0.0425$&$ -0.1537$  &  $-0.706$&$-0.502$ &$-0.131$&$-0.403$&$-0.113$ \\
              &  0.0248 &  0.0423    &  0.046  &  0.040  &  0.055 &  0.071 &  0.086\\
  NGC 4742    &$ 0.7591$&$ -0.6467$  &  $-0.655$&$-0.771$ &$-0.840$&$-1.174$&$-1.053$ \\
              &  0.0804 &  0.0949    &  0.129  &  0.098  &  0.143 &  0.182 &  0.207\\
  NGC 4839    &$ 0.2800$&$ -0.3535$  &  $-0.440$&$-0.460$ &$-0.833$&$-0.898$&$ 0.056$\\
              &  0.1708 &  0.1540    &  0.278  &  0.220  &  0.488 &  0.647 &  0.693\\
  NGC 4842A   &$-0.2147$&$  0.0637$  &  $-1.015$&$ 0.098$ &$-0.324$&$ 1.459$&$ 1.735$\\
              &  0.2617 &  0.2364    &  0.721  &  0.295  &  0.619 &  0.676 &  0.709\\
  NGC 4864    &$ 0.4900$&$ -0.5792$  &  $-0.479$&$-0.697$ &$-1.175$&$-0.998$&$-0.692$ \\
              &  0.1762 &  0.1618    &  0.251  &  0.244  &  0.722 &  0.812 &  0.563\\
  NGC 4865    &$ 0.0672$&$ -0.2587$  &  $-0.410$&$-0.359$ &$-0.143$&$-0.532$&$ 0.106$\\
              &  0.1232 &  0.1416    &  0.370  &  0.175  &  0.261 &  0.339 &  0.368\\
  NGC 4874    &$ 0.5845$&$ -0.7181$  &  $-0.988$&$-1.078$ &$-0.866$&$-1.955$&$-1.657$\\
              &  0.1029 &  0.1382    &  0.225  &  0.136  &  0.265 &  0.316 &  0.411\\
  NGC 4875    &$ 0.1777$&$ -0.3231$  &  $-0.704$&$-0.869$ &$-0.248$&$-0.152$&$-0.582$\\
              &  0.1390 &  0.2141    &  0.852  &  0.588  &  1.212 &  1.398 &  1.876\\
  NGC 4889    &$ 0.2578$&$ -0.4363$  &  $-0.793$&$-0.665$ &$-0.534$&$ 0.052$&$ 0.136$ \\
              &  0.0711 &  0.1120    &  0.182  &  0.109  &  0.163 &  0.237 &  0.307\\
  NGC 4908    &$ 0.0131$&$ -0.2354$  &  $-0.251$&$-0.629$ &$-0.322$&$-0.864$&$ 0.613$\\
              &  0.2231 &  0.2812    &  0.419  &  0.604  &  0.618 &  0.549 &  0.794\\
  NGC 5638    &$ 0.0766$&$ -0.3273$  &  $-0.774$&$-0.515$ &$-0.140$&$-0.614$&$ 0.038$\\
              &  0.0524 &  0.0673    &  0.097  &  0.061  &  0.103 &  0.147 &  0.187\\
  NGC 5796    &$ 0.1234$&$ -0.2664$  &  $-0.340$&$-0.400$ &$-0.120$&$-0.608$&$-0.371$ \\
              &  0.0744 &  0.0938    &  0.133  &  0.091  &  0.148 &  0.208 &  0.216\\

 \hline
\end{tabular}
\end{table*}

\addtocounter{table}{-1}
\begin{table*}
\caption{\it Continued.}
\centering
\begin{tabular}{@{}l rrrrrrr@{}}
\hline\hline
        &                               &                                 & CN$_2$&C4668&Fe4383&Mgb&Ca4227\\      
 Galaxy &
 \multicolumn{1}{c}{grad(age)}&\multicolumn{1}{c}{grad($[$M/H$]$)}&\multicolumn{1}{c}{grad($[$M/H$]$)}
 &\multicolumn{1}{c}{grad($[$M/H$]$)}&\multicolumn{1}{c}{grad($[$M/H$]$)}&\multicolumn{1}{c}{grad($[$M/H$]$)}&\multicolumn{1}{c@{}}{grad($[$M/H$]$)}\\
        &$\pm \sigma$                   &  $\pm \sigma$                  &  $\pm \sigma$    & $\pm \sigma$  & $\pm \sigma$  & $\pm \sigma$
        &$\pm \sigma$\\
   \hline
  NGC 5812    &$ 0.1436$&$ -0.3077$  &  $-0.795$&$-0.567$ &$-0.355$&$-0.480$&$-0.434$\\
              &  0.0645 &  0.0881    &  0.117  &  0.079  &  0.127 &  0.172 &  0.214\\
  NGC 5813    &$ 0.0150$&$ -0.0826$  &  $-0.747$&$-0.272$ &$-0.266$&$-0.335$&$-0.011$\\
              &  0.0527 &  0.0634    &  0.105  &  0.078  &  0.113 &  0.156 &  0.202 \\
  NGC 5831    &$ 0.0299$&$ -0.2412$  &  $-0.827$&$-0.608$ &$-0.417$&$-0.553$&$-0.401$\\
              &  0.0682 &  0.0980    &  0.127  &  0.123  &  0.142 &  0.149 &  0.230\\
  NGC 5845    &$ 0.0536$&$ -0.2156$  &  $-0.591$&$-0.482$ &$-0.540$&$-0.337$&$-0.304$\\
              &  0.1005 &  0.1117    &  0.244  &  0.150  &  0.180 &  0.216 &  0.287\\
  NGC 5846    &$ 0.1258$&$ -0.2327$  &  $-0.432$&$-0.363$ &$ -0.786$&$-0.389$&$ 0.369$\\
              &  0.0554 &  0.0764    &  0.130  &   0.096 & 0.165 &  0.161 &  0.236\\
  NGC 5846A   &$ 0.1829$&$ -0.2979$  &  $-0.570$&$-0.848$ &$ -0.917 $&$-1.069$&$-0.451$\\
              &  0.1110 &  0.1620    &  0.265  &   0.156 & 0.305 &  0.278 &  0.488\\
  NGC 6127    &$ 0.2307$&$ -0.4596$  &  $-0.879$&$-0.667$ &$ -0.261 $&$-0.500$&$-0.258$\\
              &  0.0855 &  0.1293    &  0.240  &   0.118 & 0.179 &  0.375 &  0.344\\
  NGC 6166    &$-0.1318$&$ -0.0738$  &  $-0.131$&$ -0.346$ &$ -0.284 $&$-0.858$&$-0.329$\\
              &  0.0920 &  0.1051    &  0.137   &   0.178  & 0.228 &  0.310 &  0.271\\
  NGC 6411    &$ 0.0348$&$ -0.1593$  &  $-0.477$&$ -0.475$ &$ -0.255 $&$-0.272$&$-0.166$\\
              &  0.0528 &  0.0745    &  0.107   &   0.098  & 0.122 &  0.154 &  0.181\\
  NGC 6482    &$ 0.0031$&$ -0.1937$  &  $-0.472$&$ -0.263$ &$ -0.214 $&$-0.538$&$-0.318$\\
              &  0.0877 &  0.0951    &  0.170   &   0.117  & 0.185 &  0.214 &  0.321\\
  NGC 6702    &$ 0.3652$&$ -0.5531$  &  $-0.522$&$ -0.666$ &$ -0.353 $&$-0.433$&$-0.168$\\
              &  0.0854 &  0.1245    &  0.204   &   0.140  & 0.150 &  0.355 &  0.343\\
  NGC 6703    &$ 0.0898$&$ -0.2319$  &  $-0.526$&$ -0.510$ &$ -0.186 $&$-0.391$&$ 0.112$\\
              &  0.0423 &  0.0684    &  0.102   &   0.059  & 0.089 &  0.093 &  0.132\\
  NGC 7052    &$-0.0011$&$ -0.1360$  &  $-0.215$ &$-0.240$ &$ -0.018 $&$-0.250$&$ 0.309$  \\
              &  0.0596 &  0.0763    &  0.112    &   0.105 & 0.123 &  0.153 &  0.196\\
  NGC 7332    &$-0.0207$&$ -0.0314$  &  $-0.698$ &$-0.724$ &$ -0.180 $&$-0.162$&$ 0.217$\\
              &  0.0460 &  0.0768    &  0.096    &   0.075 & 0.110 &  0.113 &  0.157\\
  IC 767      &$ 0.3838$&$ -1.1858$  &  $-0.625$ &$-1.015$ &$ -1.211 $&$-0.642$&$-1.250$\\
              &  0.1290 &  0.2090    &  0.459    &   0.579 & 1.359 &  0.619 &  0.683\\
  IC 794      &$ 0.2383$&$ -0.5238$  &  $-0.230$ &$-0.831$ &$ -0.580 $&$-0.314$&$ 0.112$\\
              &  0.1220 &  0.1673    &  0.229    &   0.240 & 0.236 &  0.293 &  0.472\\
  IC 832      &$-0.1432$&$  0.0902$  &  $-0.329$ & 0.191   &$  0.242 $&$-0.073$&$ 0.169$\\
              &  0.1404 &  0.1483    &  0.291    &   0.182 & 0.330 &  0.351 &  0.710\\

\hline
\end{tabular}
\end{table*}
\addtocounter{table}{-1}
\begin{table*}
\caption{\it Continued.}
\centering
\begin{tabular}{@{}l rrrrrrr@{}}
\hline\hline
        &                               &                                 & CN$_2$&C4668&Fe4383&Mgb&Ca4227\\      
 Galaxy &
 \multicolumn{1}{c}{grad(age)}&\multicolumn{1}{c}{grad($[$M/H$]$)}&\multicolumn{1}{c}{grad($[$M/H$]$)}
 &\multicolumn{1}{c}{grad($[$M/H$]$)}&\multicolumn{1}{c}{grad($[$M/H$]$)}&\multicolumn{1}{c}{grad($[$M/H$]$)}&\multicolumn{1}{c@{}}{grad($[$M/H$]$)}\\
        &$\pm \sigma$                   &  $\pm \sigma$                  &  $\pm \sigma$    & $\pm \sigma$  & $\pm \sigma$  & $\pm \sigma$
        &$\pm \sigma$\\
   \hline
  IC 3957     &$ 0.2298$&$ -0.4788$  &  $-0.890$ &$-0.940$ &$ -0.202 $&$-0.082$&$-2.087$\\
              &  0.2385 &  0.2636    &  0.769    &   0.287 & 0.507 &  1.048 &  0.612\\
  IC 3959     &$ 0.0731$&$ -0.1642$  &  $-1.038$ &$-0.077$ &$  0.091 $&$-0.545$&$-0.565$\\
              &  0.1344 &  0.1655    &  0.214    &   0.260 & 0.349 &  0.316 &  0.510\\
  IC 3963     &$ 0.3360$&$ -0.5703$  &  $-0.424$ &$-0.864$ &$ -0.550 $&$-0.765$&$-1.322$\\
              &  0.1340 &  0.1788    &  0.300    &   0.217 & 0.387 &  0.436 &  0.595\\
  IC 3973     &$ 0.4122$&$ -0.3849$  &  $-0.692$ &$-0.435$ &$ -0.532 $&$-1.091$&$-0.595$\\
              &  0.1776 &  0.1643    &  0.374    &   0.285 & 0.483 &  0.735 &  0.820\\
  IC 4042     &$-0.1029$&$ -0.3628$  &  $-0.161$ &$-0.794$ &$  0.919 $&$-0.275$&$ 0.785$\\
              &  0.4661 &  0.6512    &  0.374    &   0.285 & 0.483 &  0.735 &  0.820 \\
  IC 4051     &$-0.0582$&$ -0.3693$  & $-0.873$ &$-0.759$  &$ -0.224 $&$-0.365$&$ 0.251$\\
              &  0.0852 &  0.1261    & 0.165    &   0.208  & 0.184 &  0.234 &  0.411\\
  CGCG 159-41 &$ 0.1629$&$ -0.0679$  & $-0.643$ & 0.156    &$  0.055 $&$-2.702$&$ 0.558$\\
              &  0.5742 &  0.6968    & 1.401    &   1.034  & 1.372 &  1.443 &  1.416\\
  CGCG 159-43 &$ 0.5593$&$ -0.6726$  & $-0.902$&$-0.751$   &$ -0.480 $&$-1.305$&$-0.820$\\
              &  0.1614 &  0.1936    & 0.281    &   0.355  & 0.385 &  0.386 &  0.618\\
  CGCG 159-83 &$-0.4099$&$  0.0939$  & $-0.902$ &$-0.751$  &$ -0.480 $&$-1.305$&$-0.820$\\
              &  0.1310 &  0.1566    & 0.281    &   0.355  & 0.385 &  0.386 &  0.618\\
  CGCG 159-89 &$-0.0815$&$ -0.3979$  & $-0.994$ &$-0.743$  &$  0.570 $&$-0.048$&$ 0.831$\\
              &  0.1410 &  0.1969    &0.229     &  0.134   & 0.351 &  0.484 &  0.513 \\
\end{tabular}
\end{table*}
\section{Mean gradients}
\label{sec.meangradients}

The predicted mean values of the age and metallicity gradients are
dependent upon the merging history of the galaxies. 
 In general, dissipative processes tend to steepen the 
metallicity gradient while major mergers are expected 
to dilute it.
In the simulations of Kobayashi (2004), the
typical gradients for non-merger and merger galaxies are
$\Delta$[Fe/H]/$\Delta \log r \sim -0.45$, $-0.38$, $\Delta$[O/H]$\Delta
\log r \sim -0.25$, $-0.24$, and $\Delta \log Z/\log r \sim -0.30, -0.24$,
respectively.  We now analyse the mean age and metallicity gradients for
the sample of LDEGs.  Table~\ref{mean-gradients} list these values,
together with relevant associated statistics (see below). 

\begin{table*}
\caption{Mean gradients of age, [M/H], and metallicity, derived with different
indices for LDEGs. $\sigma$: typical deviation; $N$: number of galaxies; 
$N_{\rm eff}$: effective number of points, $N_{\rm eff}=\left[\sum
(1/\sigma^2)\right]^2/\sum \left[(1/\sigma^2)^2\right]$; $t$: $t$-statistic to check the
hypothesis ``mean$\ne$0''; $\sigma_{\rm exp}$: typical deviation expected from
errors; $\alpha$: level of significance to reject the hypothesis
``$\sigma=\sigma_{\rm exp}$''. The last column 
contains the $t$-statistic used to
test the hypothesis ``mean grad[M/H] = mean grad[M/H]'' 
(with different indices).
\label{mean-gradients}}
\centering
\begin{tabular}{@{}lrrrrrrrr@{}}
\hline\hline
         & 
\multicolumn{1}{c}{mean} & 
\multicolumn{1}{c}{$\sigma$} & 
$N$ & $N_{\rm eff}$ & $t$ & $\sigma_{\rm exp}$ & \multicolumn{1}{c}{$\alpha$} & $t$\\
\hline
grad($\log$ age)              &$ 0.082\pm 0.015$ & 0.117  & 61  & 19.62&  3.1& 0.032& 1.63E$-13$ &   \\
grad [M/H]                    &$-0.206\pm 0.019$ & 0.151  & 61  & 17.25&  5.7& 0.075&  4.3E$-22$ &   \\
grad [M/H] (CN$_2$-H$\beta$)  &$-0.582\pm 0.032$ & 0.249  & 61  & 23.14& 11.2& 0.111&  2.6E$-31$  & 5.94\\
grad [M/H] (C4668-H$\beta$)   &$-0.459\pm 0.028$ & 0.219  & 61  & 21.78&  9.8& 0.078&  0.000     & 4.26\\
grad [M/H] (Fe4383-H$\beta$)  &$-0.197\pm 0.025$ & 0.197  & 61  & 15.31&  3.9& 0.109&  1.2E$-14$ & 0.14\\
grad [M/H] (Mgb-H$\beta$)     &$-0.407\pm 0.039$ & 0.302  & 61  & 14.53&  5.1& 0.138&  5.3E$-31$  & 2.30\\
grad [M/H] (Ca4227-H$\beta$)  &$-0.238\pm 0.034$ & 0.269  & 61  & 19.18&  3.4& 0.178&  2.9E$-7$  & 0.45\\
\hline
\end{tabular}
\end{table*}

\subsection{Age gradients}

The mean age gradient for the sample of LDEGs is \mbox{$\Delta
\log$(age)/$\log r$=$0.082\pm 0.015$}. A $t$-test indicates that the
probability of this value being different from zero by chance is $<0.1$\%. 
The positive gradient indicates that, on average, the centres of the
galaxies are younger (or at least, they contain a percentage of younger
stars which make the mean luminosity-weighted age lower) than the outer
parts.  The existence of a significant age gradient is difficult to
explain within monolithic collapse scenarios, as the timescales for star
formation are necessarily very short, but it does suggest that secondary
episodes of star formation have occurred recently in the centres of the
galaxies. 

Little work has been done to date on the derivation of age gradients
within early-type galaxies.  Indeed, most of the extant studies of
gradients in the literature {\it assume} the absence of an age gradient in
order to derive the metallicity gradient.  
Munn (1992) was the first author that studied, in a systematic way, the variation
of age with the galactocentric radius.
From a sample of seven early-type galaxies, Munn combined the
CN$\lambda$3883 and CN$\lambda$4216 features with the D4000 index to
compare with the prediction of stellar population models.  Since the age
calibration employed by Munn was necessarily restricted to the assumption
of solar metallicity, his ability to quantify the inferred gradients was
limited.  He did however conclude that in order to explain the dispersion
and trends in the CN--D4000 diagrams, a variation of at least two
parameters with radius was necessary (as suggested already by G90), with
age and metallicity being the obvious candidates.  Using the stellar
population models of Worthey (1994), Gonz\'alez (1993) found a mean
variation in the age of his sample of galaxies of $\sim$ 20\% from the
centre to the effective radius (he also found a variation of $\sim$50\% in
metallicity). Fisher, Franx \& Illingworth (1996) found in their sample of
lenticulars that the centres were slightly younger than the external
parts, although they did not quantify the result. Contrary to these
results, Mehlert et al.\ (2003) found, in a sample of galaxies from the
Coma cluster, a mean age gradient compatible with being null.  The
discrepancy between the above results and those of Mehlert et al.\ (2003)
{\it may} be due to the different environments from which the samples were
drawn. We analyse the gradients for the HDEGs in
Sec.~\ref{sec.comagradientes}. 

Table~\ref{mean-gradients} also shows the values of the rms dispersion
about the mean value ($\sigma$) and the scatter expected from the errors
($\sigma_{\rm exp}$).  We performed a $\chi^2$ test to compare these
values and the results are shown in the 8$^{\rm th}$ column of the table.
There is a real scatter in the age gradients between galaxies that cannot
be explained by the errors, consistent with the earlier results of G90. We
have explored possible correlations between the age gradients and other
parameters of the galaxies, including (but not limited to) the central
velocity dispersion, but we have not found any significant correlations. 

\subsection{Metallicity gradients}

The mean metallicity gradient in our sample of LDEGs is
$\Delta$[M/H]/$\Delta \log r$=$-0.206\pm 0.019$. Earlier studies have
derived comparable gradients using, primarily
$\langle$Fe$\rangle$ and Mg indices. For example, Couture \& Hardy (1988),
using the Mg$_2$ index, found a variation in metallicity with radius of
$\Delta$[M/H]$/\log r=-0.25$. G90 measured mean gradients in their sample
of early-type galaxies of $\Delta$[M/H]$/\log r=-0.23\pm0.09$ and
$\Delta$[M/H]$/\log r=-0.22\pm 0.10$, using the Mg$_2$ and
$\langle$Fe$\rangle$ indices, respectively (later confirmed by Davies
et~al. 1993). To transform the line-strength gradients into metallicity
gradients, each of these studies used the Mould (1978), Burstein (1979),
and Faber et al.\ (1985) calibrations. In all cases, they assumed a null
age gradient.  Fisher et al.\ (1995), through the comparison of the
gradients of Mg$_2$, H$\beta$, and $\langle$Fe$\rangle$, with the stellar
population models of Worthey (1994), obtained a metallicity gradient of
$\Delta$[M/H]$/\log r = -0.25 \pm 0.1$. Finally, Mehlert et al.\ (2003),
using the population synthesis models of Thomas et al.\ (2003), obtained a
metallicity gradient of $\Delta$[M/H]$/\log r=-0.10\pm 0.12$,
significantly more shallow than previous studies had found.  The values
derived in the current study are consistent with the extant literature and
suggest that early-type galaxies exhibit a reduction in metallicity of
$\sim$40\% per decade of variation in radius.  This value is considerably
flatter than the values predicted by the dissipative collapse models. For
example, Larson's hydrodynamical simulations gave $\Delta \log Z/\Delta
\log r \sim -0.35$ (Larson 1974a,b), and $-1.0$ (Larson 1975) and
Carlberg's (1984) N-body simulations gave $\Delta \log Z/\Delta \log r
\sim -0.5$.  It should be stressed though that these particular models
leave significant room for improvement, due to the absence of essential
physical processes, including star formation, thermal feedback from
supernovae, and chemical enrichment.  

Finally, we compared the dispersion about the mean values ($\sigma=0.151$)
and the scatter expected from the errors ($\sigma_{\rm exp}$=0.075) by
performing a $\chi^2$ test. The hypothesis ``$\sigma_{\rm exp}=\sigma$''
can be rejected at a low significance level (see 8$^{\rm th}$ column of
Table~\ref{mean-gradients}).  To investigate the causes of the variation
in the metallicity gradients among galaxies, in
Section~\ref{section.correlation}, we explore putative correlations of the
gradients with other physical galaxy parameters. 

\section{Relative radial abundance ratios}

It is well known that massive early-type galaxies have abundance patterns
that do not match that of the solar neighbourhood (see Paper~II, and
references therein). Although the influence of non-scaled solar abundance
ratio patterns complicates the estimation of mean ages from integrated
light, it does provide an important clue as to the formation and chemical
enrichment histories of galaxies.  In this Section, we explore the
presence of relative abundance ratios gradients in our  sample of LDEGs 
To do so, we compare the metallicity gradients
obtained with different indices (combined with H$\beta$) and  make the 
assumption that the differences in the derived values are due 
to the different sensitivity of the Lick indices to changes in the chemical 
composition.
While avoiding a detailed
quantitative analysis  (due to the difficulty to perform this kind 
of analysis with the current stellar population models, see Paper II for details), 
the qualitative 
trends presented below do offer
invaluable information concerning the formation of these galaxies and the
timescales for star formation therein. 

\begin{figure*}
\resizebox{0.32\hsize}{!}{\includegraphics[angle=-90]{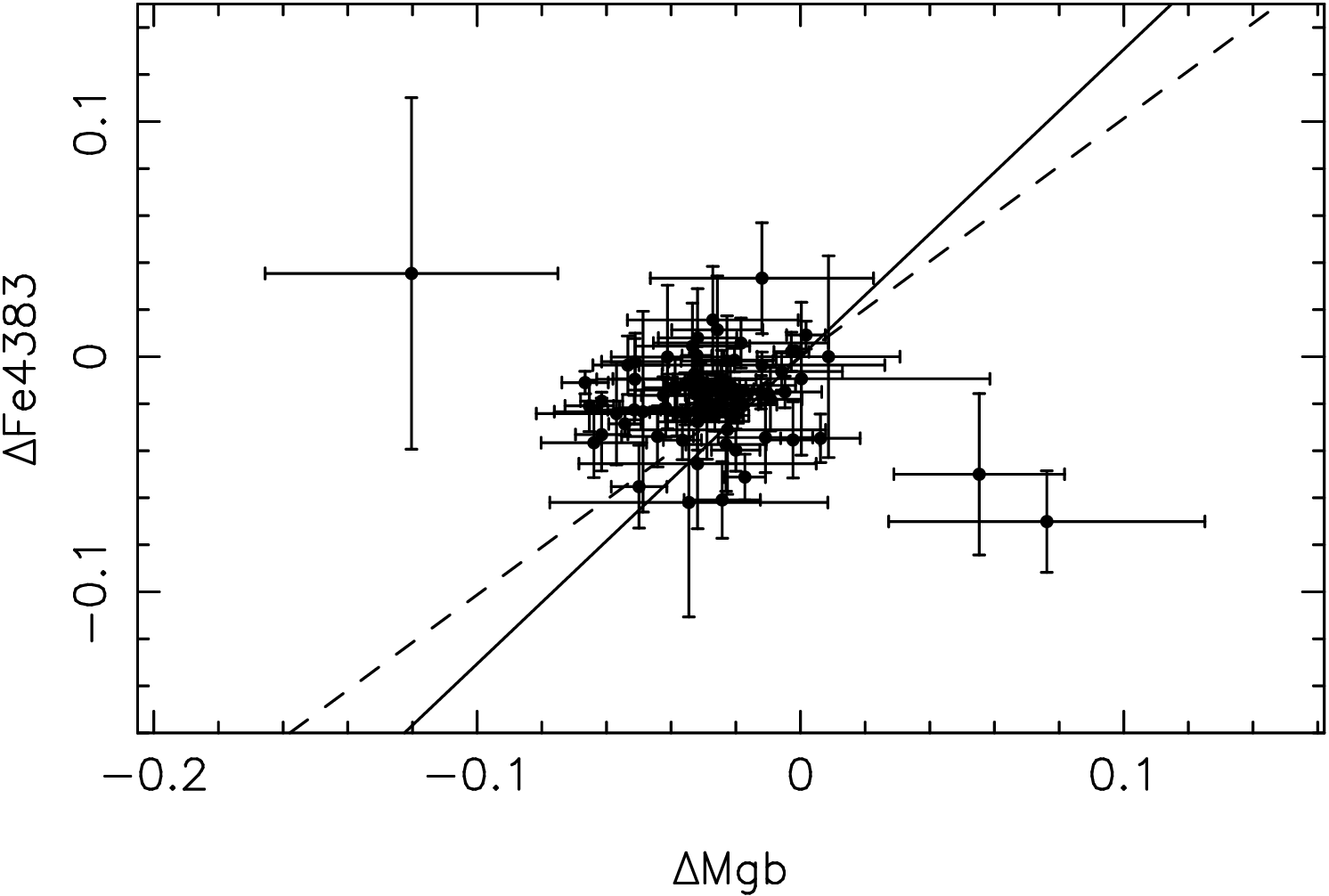}}
\hfill
\resizebox{0.32\hsize}{!}{\includegraphics[angle=-90]{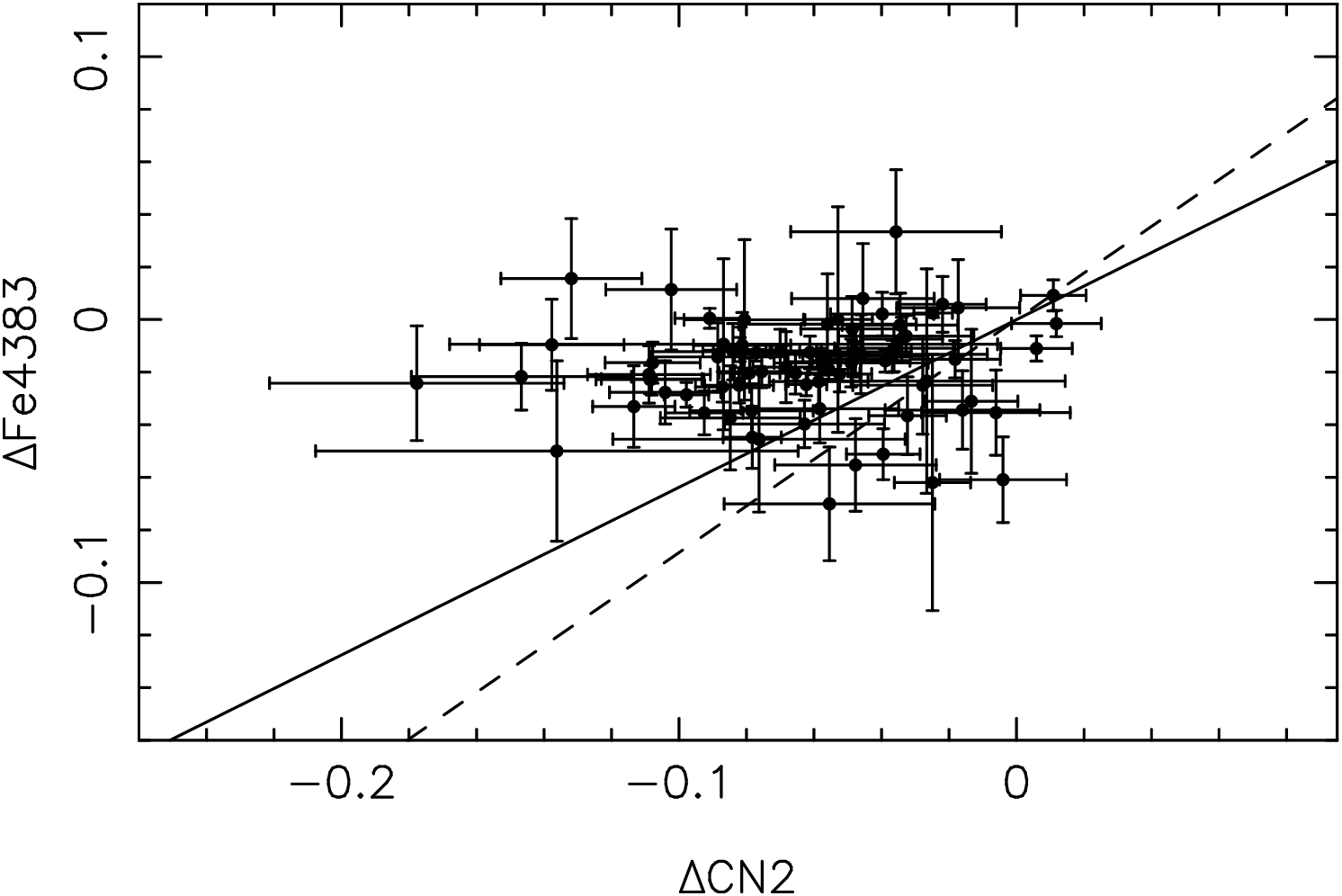}}
\hfill
\resizebox{0.32\hsize}{!}{\includegraphics[angle=-90]{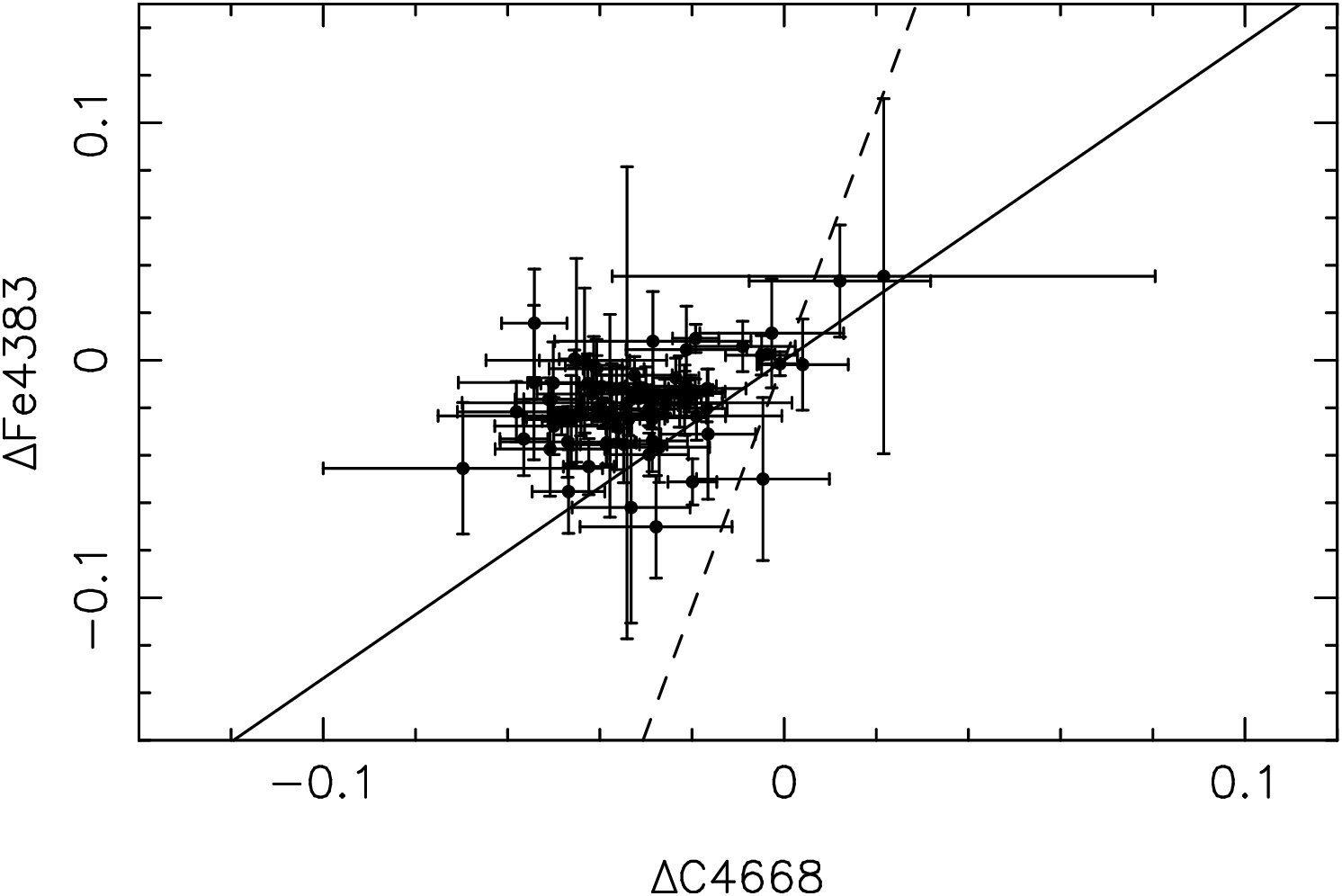}}
\caption{$\Delta$index--$\Delta$index diagrams in which we compare the Fe4383
gradients with the gradients of Mgb, CN$_2$ and C4668. Dashed lines indicate the
expected gradients if the only parameter varying with radius is the age
(assuming an invariant solar metallicity), while solid lines show the
expected trends if the only parameter changing with radius 
is the metallicity (assuming a constant age of 10 Gyr). 
\label{grad-grad}}
\end{figure*}

Fig.~\ref{grad-grad} shows three $\Delta$index--$\Delta$index diagrams in
which gradients of CN$_2$, Mgb and C4668 are compared against the
gradients of Fe4383. The lines corresponding to $\Delta$age=0 and
$\Delta$[M/H]=0 are over-plotted.  Due to the low sensitivity to age of 
all these indices, the diagrams appear highly degenerate -- i.e., the
iso-age and iso-metallicity lines are almost parallel. If the galaxies did
not have a gradient in the relative abundances, we would expect to find
all the points distributed along these lines.  However, they appear to be
systematically shifted toward the left of the diagram, although the
magnitude of these shifts is very different in the three plots.  In the
first panel ($\Delta$Mgb--$\Delta$Fe4383), the metallicity gradients
measured with Mgb are slightly steeper than the metallicity gradients
inferred from Fe4383, although the differences are admittedly small. 
In the second panel ($\Delta$CN$_2$--$\Delta$Fe4383)
the differences are more evident. The metallicity gradients obtained with
CN$_2$ are clearly steeper (in absolute value) than the ones inferred from
Fe4383. 
The last panel ($\Delta$C4668--$\Delta$Fe4383) is an
intermediate case between the first two. The metallicity gradient obtained
with C4668 is slightly steeper than the one estimated from Fe4383.
Although the trend is not as evident as in the second panel, the points
are visibly shifted toward the left of the constant age and metallicity
lines.  

\begin{figure*}
\resizebox{0.3\hsize}{!}{\includegraphics[angle=-90,bb=119 144 558 599]{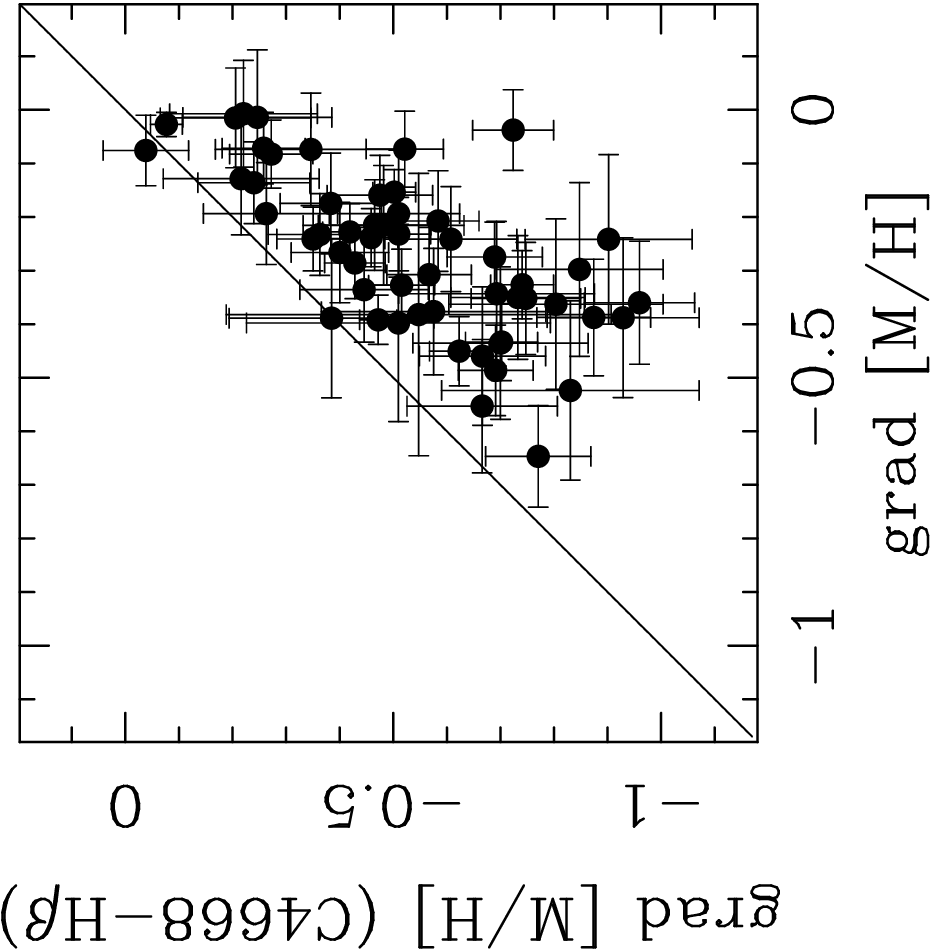}}  
\hfill
\resizebox{0.3\hsize}{!}{\includegraphics[angle=-90,bb=119 144 558 599]{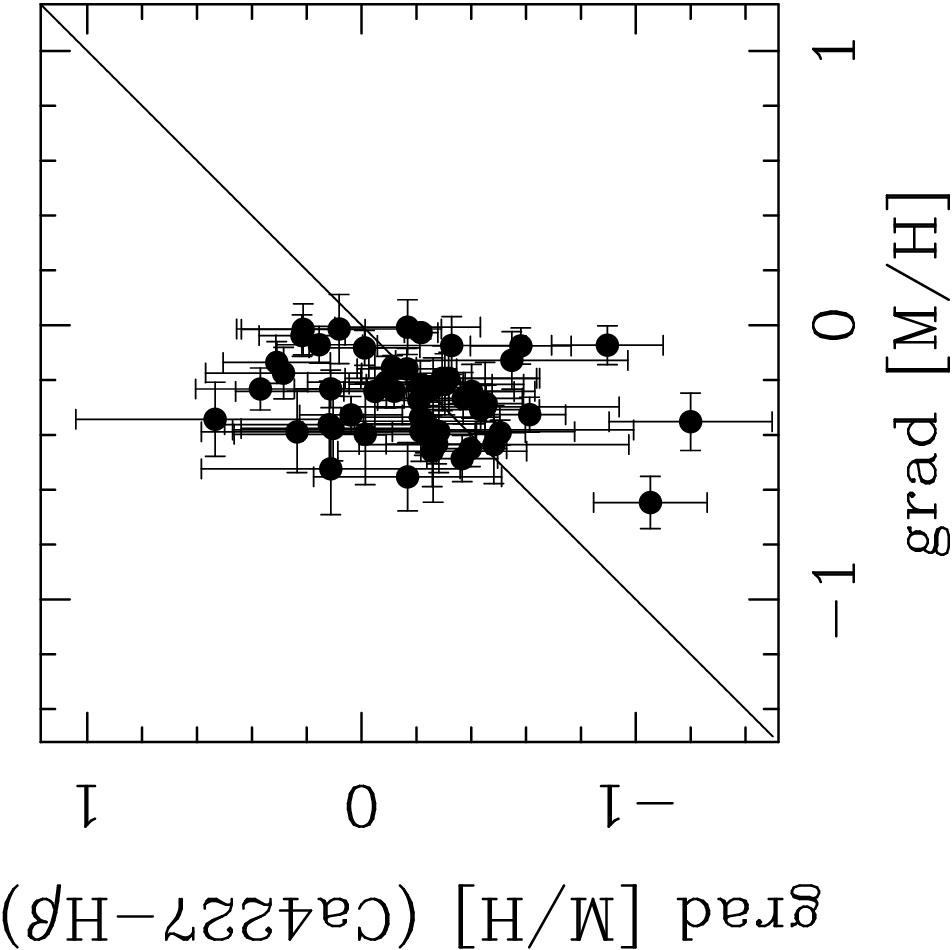}} 
\hfill
\resizebox{0.3\hsize}{!}{\includegraphics[angle=-90,bb=119 144 558 599]{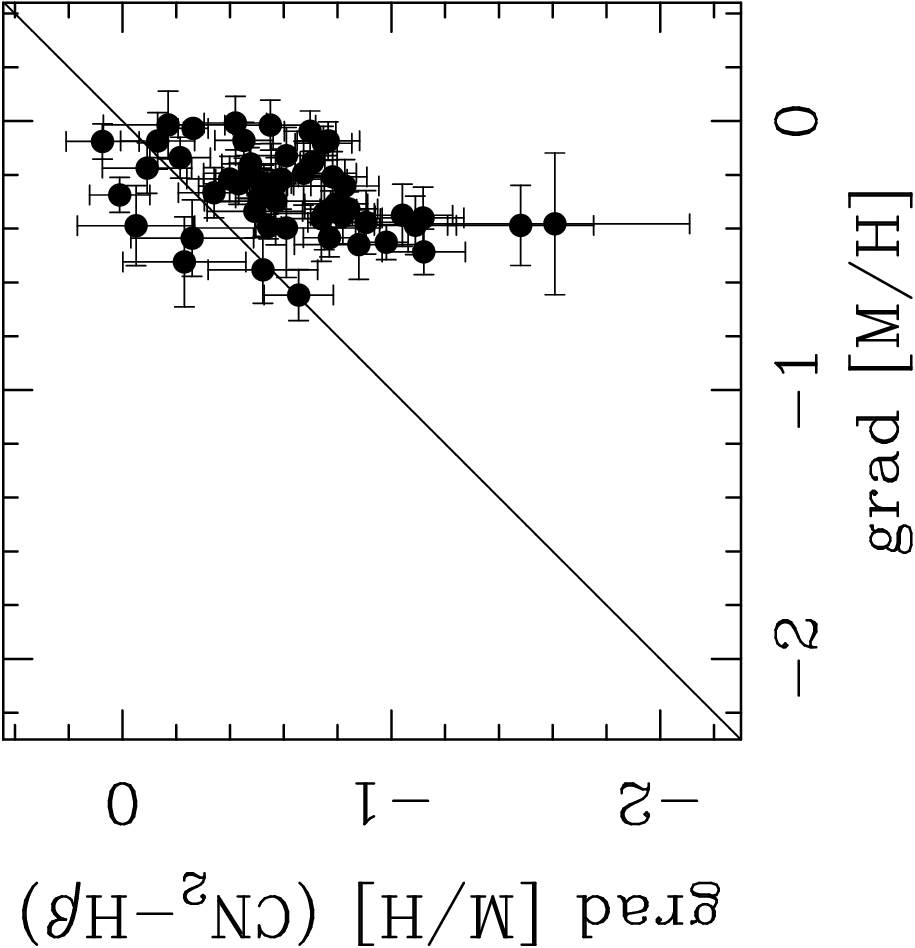}}   

\vspace{5mm}

\resizebox{0.3\hsize}{!}{\includegraphics[angle=-90,bb=119 144 558 599]{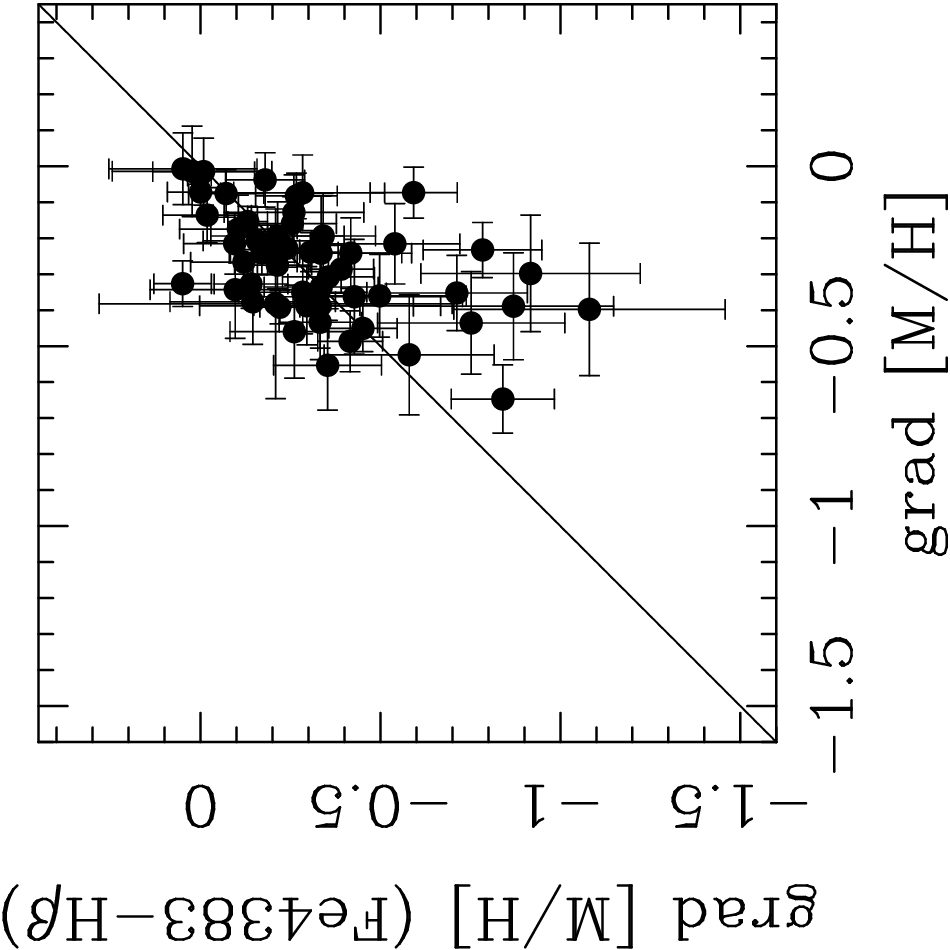}}
\hspace{5mm}
\resizebox{0.3\hsize}{!}{\includegraphics[angle=-90,bb=119 144 558 599]{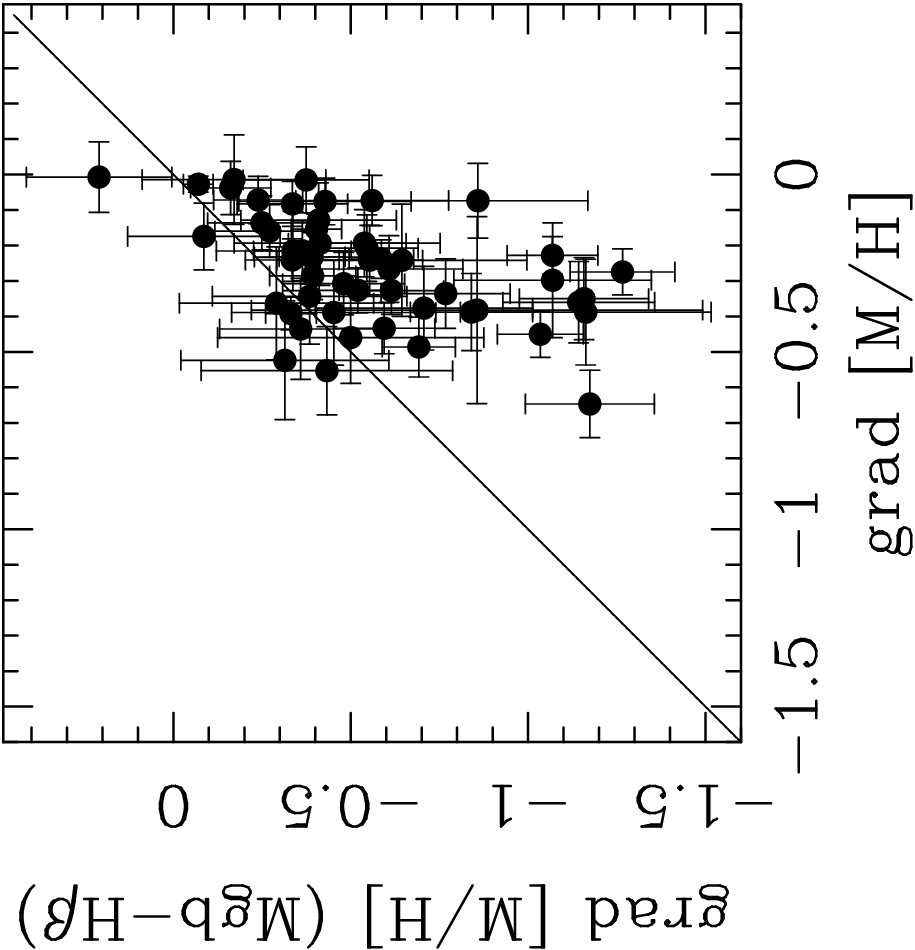}}    
\caption{Comparison of the metallicity gradient measured with different
indicators combined with H$\beta$, and the metallicity gradient calculated
combining ten different indicators (see 
text for details). \label{compara-gradientes}}
\end{figure*}

To analyse these differences in more detail, we calculated the metallicity
gradients in several index--index diagrams using the indices CN$_2$,
C4668, Fe4383, Mgb and Ca4227, combined with H$\beta$.  In
Fig.~\ref{compara-gradientes}, we compare these gradients with the values
obtained from ten different indicators, as described in
Sec.~\ref{sec-age.metallicity.gradients}.  Clearly, while the values
obtained with Ca4227 and Fe4383 are compatible with the gradients obtained
in Sec.~\ref{sec-age.metallicity.gradients}, the gradients calculated with
CN$_2$, C4668 and, perhaps, Mgb are steeper. We quantified these
difference by measuring the mean gradient using the various diagrams, the
results of which are summarised in Table~\ref{mean-gradients}.  The final
column of the table shows the probability that the mean gradients
calculated with the various indicators are the same as the gradients
calculated Section~\ref{sec-age.metallicity.gradients}.  The gradients
calculated using CN$_2$ and C4668 are almost twice as steep as the
gradients calculated with the average of ten indicators.  In the case of
Mgb, despite the mean gradient being almost twice as large steep as the
mean metallicity gradient, the result is less significant. In that case,
the initial hypothesis can be rejected with a significance level lower
than 0.025.

 We could argue that the differences in the metallicities derived
with different indicators are due to variations in the chemical 
composition 
as a function of galactocentric distance, as the sensitivity 
of different Lick indices to variations on the abundance
of different chemical 
elements is not the same
(Tripicco \& Bell 1995; Korn et al. 2004). But
the Lick indices show a dependence with gravity that, while not
large (Gorgas et al.\ 1993; Worthey et al.\ 1994), is also not null.  As
the behaviour of each line index under variations in the underlying IMF is
not identical, a change in the ratio of dwarf-to-giant stars with
galactocentric radius could result in the metallicity gradient inferred
from different indicators also being different. 

Cenarro et al.\ (2003) calibrated the CaT* (calcium triplet) index in the 
near-infrared and found
a relationship between [Fe/H], velocity dispersion, and the slope of the IMF.
One of the projections of this relationship is
\begin{equation}
\mu=2.41+2.78[{\rm M}/{\rm H}]-3.79[{\rm M}/{\rm H}]^2,
\end{equation}
where $\mu$ represents the slope of the IMF.  Using this equation we determined
the slope of the IMF from the average metallicities in the centre, and at a
distance of one effective radius, finding the following values:
\begin{itemize}
\item $\mu = 2.56$ in the central regions,
\item $\mu = 1.67$ at a distance of one effective radius from the galaxy centre.
\end{itemize}

To investigate the variation in the line-strength indices that such a variation
in the IMF, plus an average variation in metallicity as obtained in 
Section~\ref{sec.meangradients} would produce, we 
parameterised the indices (using the
V06 models) as a function of metallicity and $\mu$, obtaining:
\begin{eqnarray*}
 {\rm CN}_2   &
 \!\!\!\!\!\!\!\!=&\!\!\!\!\!\!\!\! 
 0.0374 - 0.0063 \mu + 0.1350 [{\rm M}/{\rm H}] + 0.0495 [{\rm M}/{\rm H}]^2\\
 {\rm Ca4227} &
 \!\!\!\!\!\!\!\!=&\!\!\!\!\!\!\!\! 
 1.3745 + 0.1600 \mu + 0.9547 [{\rm M}/{\rm H}] + 0.1566 [{\rm M}/{\rm H}]^2\\
 {\rm Fe4383} &
 \!\!\!\!\!\!\!\!=&\!\!\!\!\!\!\!\! 
 5.5963 + 0.0467 \mu + 4.2390 [{\rm M}/{\rm H}] + 0.8785 [{\rm M}/{\rm H}]^2\\
 {\rm C4668}  &
 \!\!\!\!\!\!\!\!=&\!\!\!\!\!\!\!\! 
 5.5930 - 0.2943 \mu + 6.4247 [{\rm M}/{\rm H}] + 1.8047 [{\rm M}/{\rm H}]^2\\
 {\rm Mgb}    &
 \!\!\!\!\!\!\!\!=&\!\!\!\!\!\!\!\! 
 3.7700 + 0.1303 \mu + 2.2567 [{\rm M}/{\rm H}] + 2.2567 [{\rm M}/{\rm H}]^2 \\
\end{eqnarray*}

A variation in the slope of the IMF, together with a variation in
metallicity equal to the mean metallicity gradient calculated in
Section~\ref{sec.meangradients}, would produce the following differences
in the selected indices:  $\Delta {\rm CN}_{2}=0.0278$~mag,
$\Delta$Ca4227$=0.0390$~mag, $\Delta$Fe4383$=0.0264$~mag,
$\Delta$C4668=0.0182~mag, and $\Delta$Mgb=0.0270~mag. The mean observed
variations for the LDEGs are: $\Delta$CN$_{2 \rm obs}= -0.0662\pm
0.0096$~mag, $\Delta$Ca4227$_{\rm obs}=-0.0157\pm 0.0097$~mag,
$\Delta$Fe4383$_{\rm obs}=-0.0174\pm 0.0064$~mag, $\Delta$C4668$_{\rm
obs}= -0.0314 \pm 0.0044$~mag, and $\Delta$Mgb$_{\rm obs}= -0.0301 \pm
0.0062$~mag.  As can be seen, a variation of the IMF of the form predicted
by Cenarro et al.\ (2003) cannot be responsible for the different
metallicity gradients obtained with the various indicators. In fact, this
variation of the IMF would produce positive gradients in the selected
indices. In this parameterisation we have not included the effect of age,
but the age gradients are not very strong in our sample (see
Section~\ref{sec.meangradients}) and the differences in the sensitivities
of the analysed indices to this parameter are not enough to produce the
observed differences. 

 More likely, the differences in the metallicity gradients obtained with
the different indices are due to changes in the relative abundance
gradients of distinct elements. However, it is difficult to quantify the
strength of those gradients with the present knowledge of line formation 
and the inherent limitations 
of stellar population synthesis models

\section{Correlation of the gradients with other parameters}
\label{section.correlation}

There are several physical processes that can produce a metallicity
gradient in early-type galaxies. The relationship between these gradients
and other global properties of the galaxies 
afford an opportunity to discriminate between the competing processes. 
Previous attempts have searched for correlations using colours and
empirical line-strength indices; in what follows,
we study the correlation between the gradients of the
derived simple stellar population (SSP) parameters and the global
properties of the host galaxies. 
 In this section, we only use the subsample of elliptical  galaxies, 
excluding the lenticulars (S0)  from 
the analysis. The reason for doing this is  that the correlations  between 
the gradients and other parameters 
are predicted by an specific mechanism of galaxy formation, and, although 
S0 and E galaxies seem to follow the same relations between the central 
properties and other parameters, very different mechanism have been proposed
for their formation. As the sample of S0 is small, we could not perform a 
comparative study using only S0 galaxies.

\subsection{Correlation of the metallicity gradient with the velocity
dispersion gradient}
\label{sec-gradsigma-gradmeta}

\begin{figure*}
\resizebox{0.32\hsize}{!}{\includegraphics[angle=-90,bb=119 20 558 723]{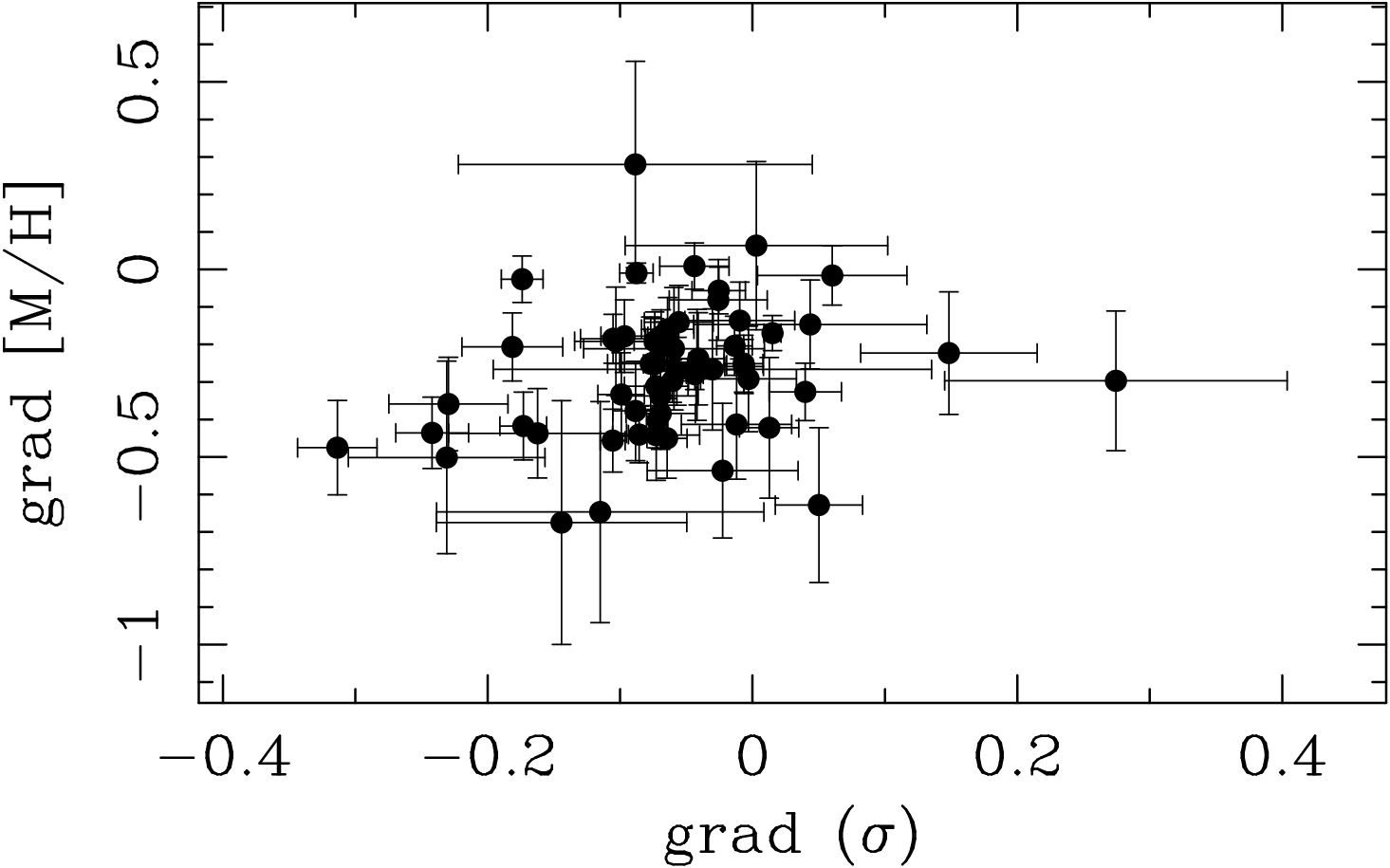}}
\hfill
\resizebox{0.32\hsize}{!}{\includegraphics[angle=-90,bb=119 20 558 723]{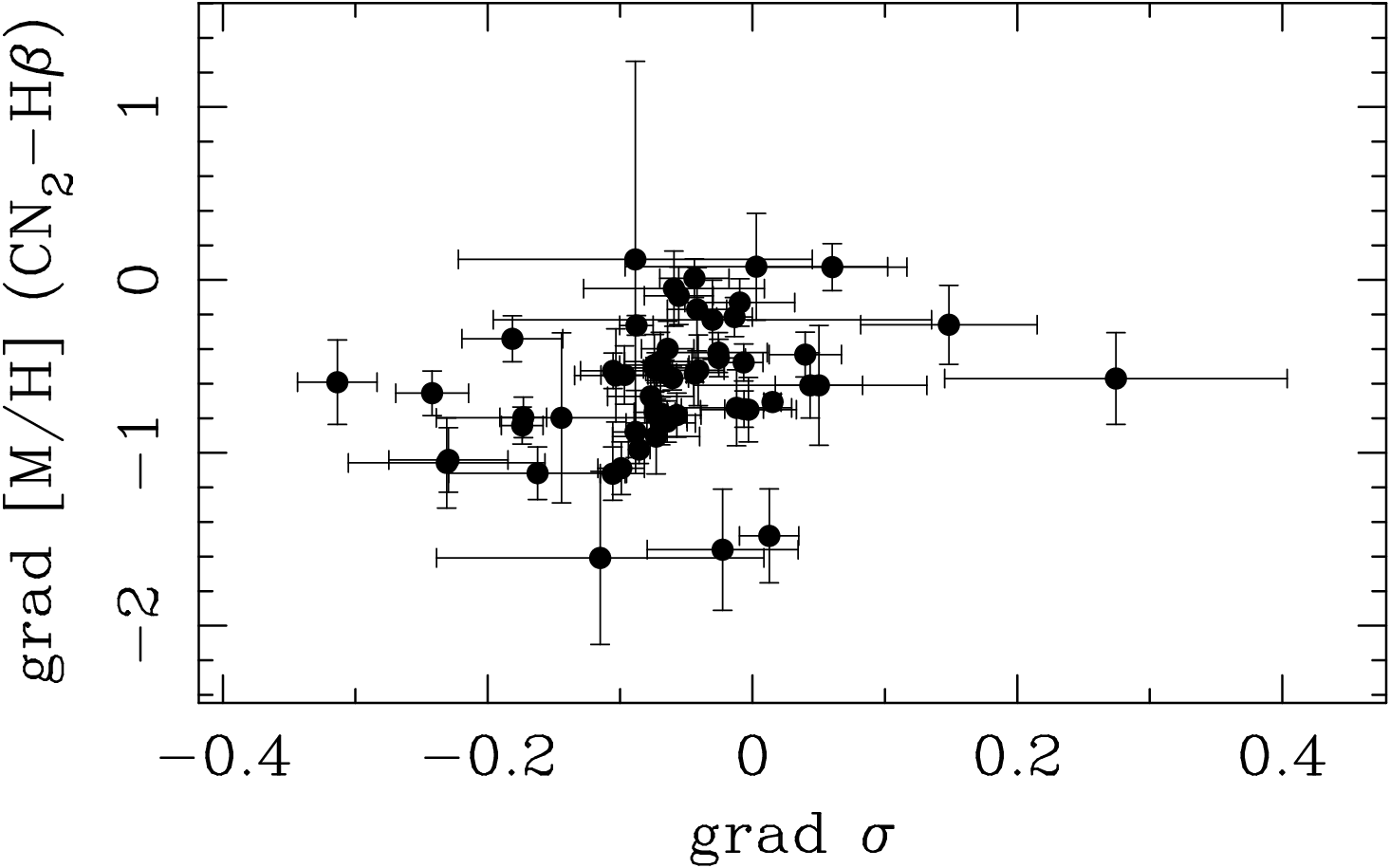}}
\hfill
\resizebox{0.32\hsize}{!}{\includegraphics[angle=-90,bb=119 20 558 723]{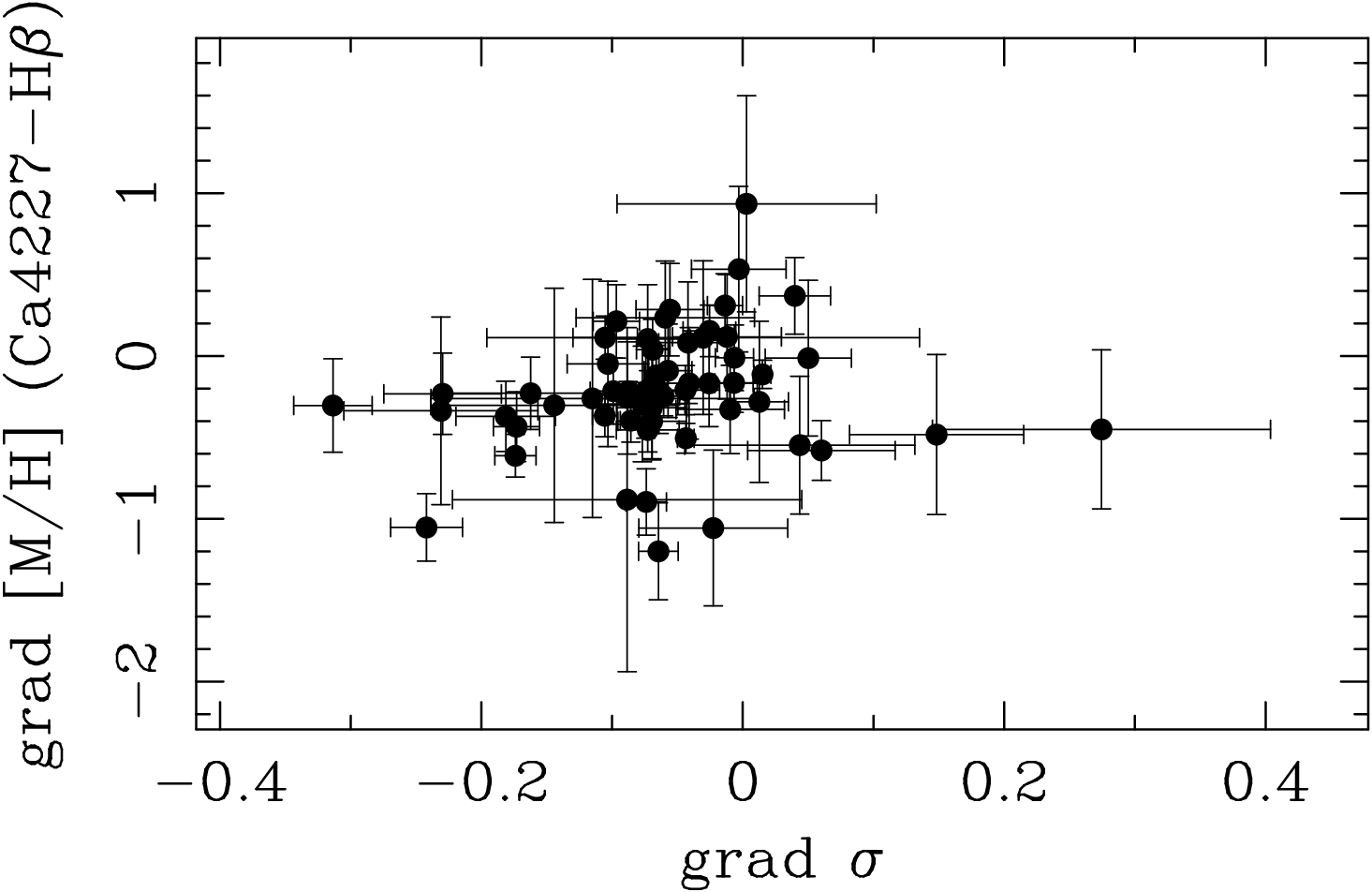}}

\vspace{4mm}

\resizebox{0.32\hsize}{!}{\includegraphics[angle=-90,bb=119 20 558 723]{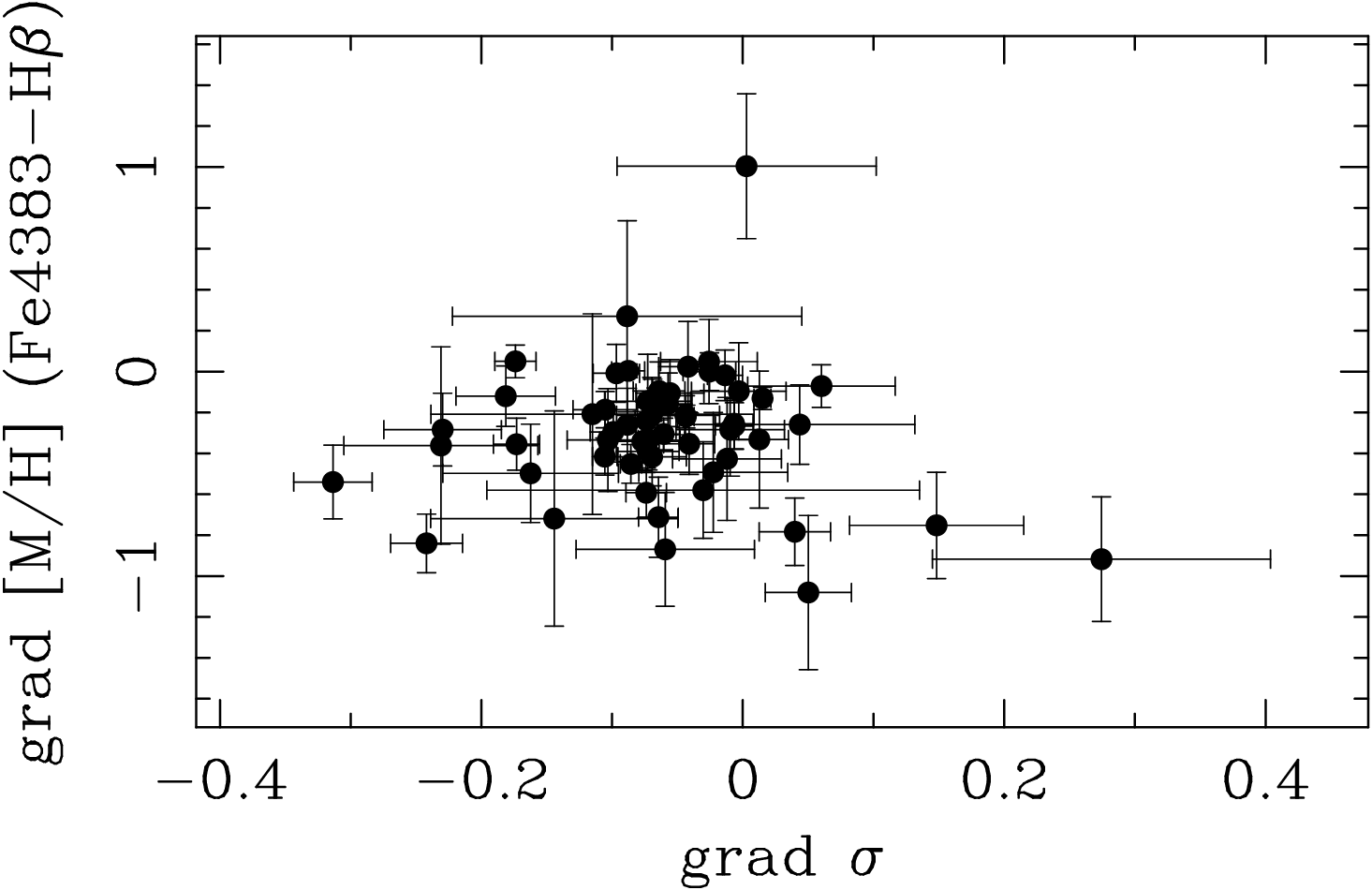}}
\hfill
\resizebox{0.32\hsize}{!}{\includegraphics[angle=-90,bb=119 20 558 723]{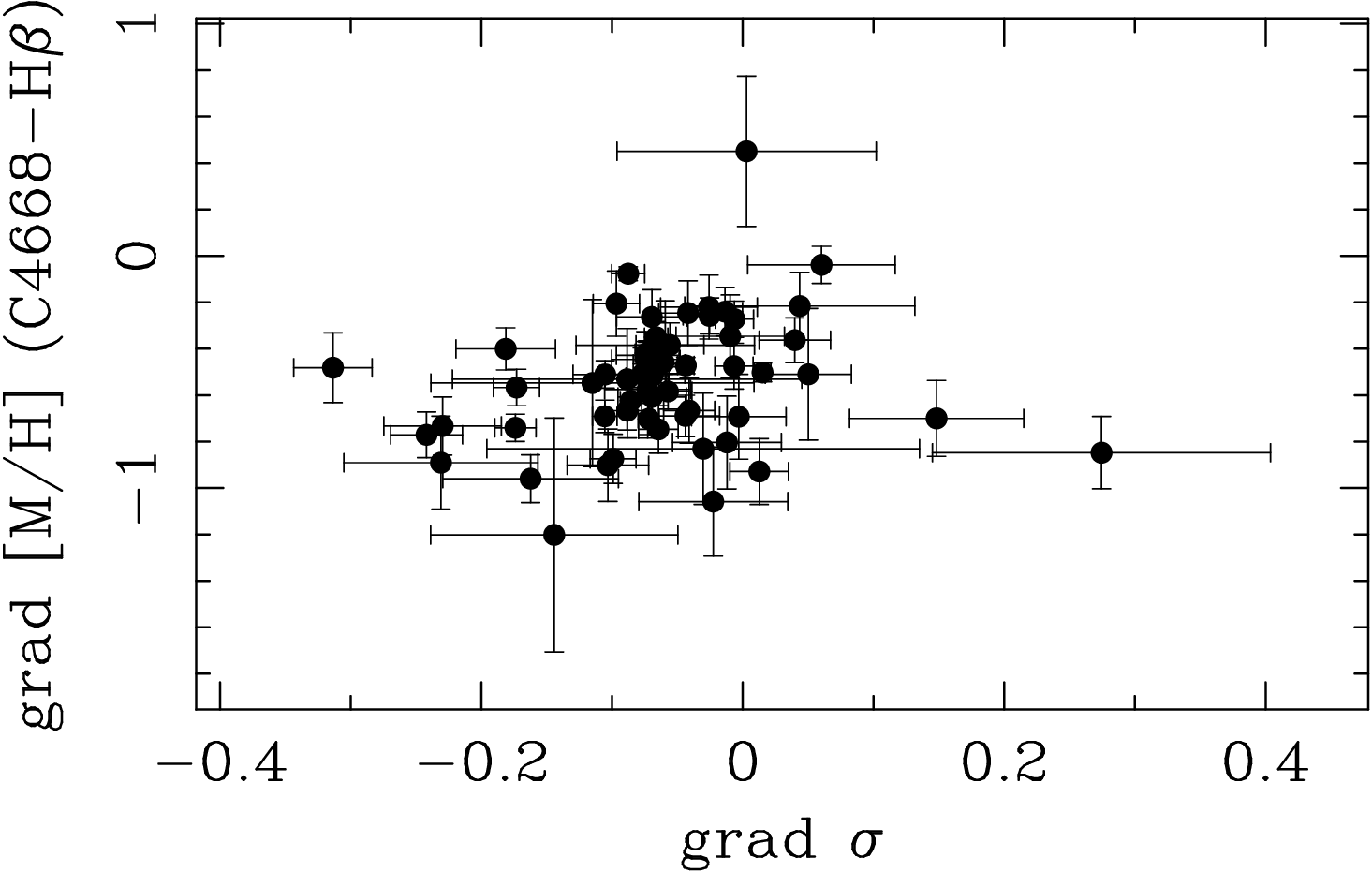}}
\hfill
\resizebox{0.32\hsize}{!}{\includegraphics[angle=-90,bb=119 20 558 723]{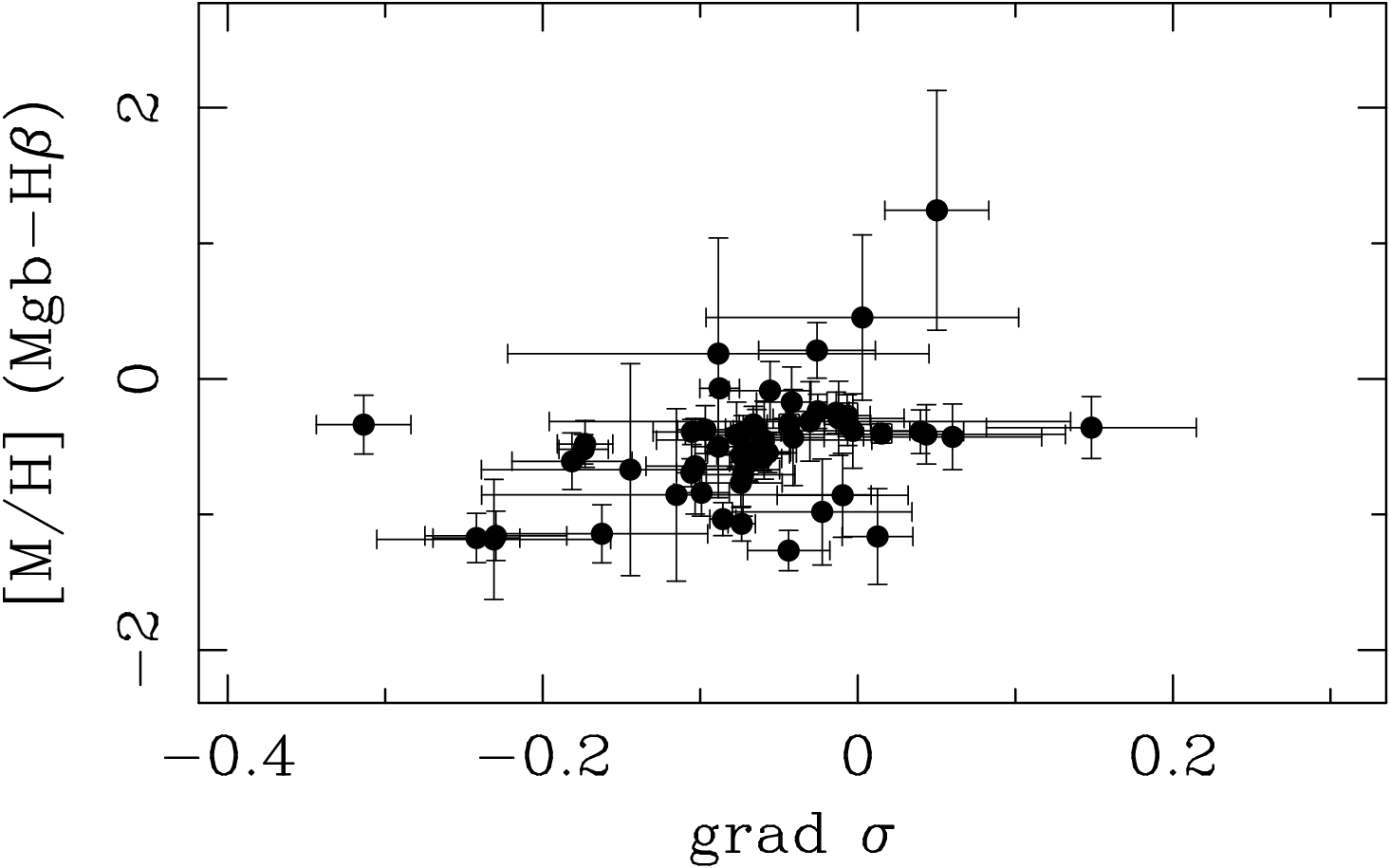}}
\caption{Relation between the metallicity gradients obtained in different
index--index diagrams (as indicated in the brackets) and the velocity
dispersion gradient for the sample of LDEGs. In the top left panel, the
metallicity gradient has been obtained using a combination of ten
indices, as described in Sec.~\ref{sec-age.metallicity.gradients}.
\label{age.meta.gradsigma}}
\end{figure*}

As noted in Section~1, Franx \& Illingworth (1990) found a correlation
between colour gradient and the gradient of {\it escape velocity}, which
they interpreted as a correlation between the local metallicity and the
local potential well depth in a sample of early-type galaxies. They
suggested that such a correlation was consistent with a galactic wind
origin, and inconsistent with a dissipative inward flow origin.  Others,
including Davies et al.\ (1993), Carollo \& Danziger (1994), and Mehlert
et al.\ (2003) find similar relations between the gradients of some
line-strength indices and the gradient of the potential
well.\footnote{Davies et al.\ (2003) calculated the local escape velocity
using the surface brightness and the kinematics of the galaxies (J\o
rgensen, Franx \& K\ae rgaard 1992). Carollo \& Danziger (1994) obtained
this parameter with asymmetric dynamical models which depended on the
total energy and the angular momentum along the symmetry axis. Mehlert et
al.\ (2003) used the gradient of velocity dispersion as a measure of the
local potential well depth.}. We now examine the likelihood for the
existence of this correlation within our sample of LDEGs, using the
inferred metallicity gradients and empirical velocity dispersion gradients
-- Fig.~\ref{age.meta.gradsigma} illustrates the observed trends. 

To study the degree of correlation we have performed two different tests:
a non-parametric Spearman rank order test and a $t$-test. The Spearman
test does not take into account the errors in the individual points, while
the $t$-test has the limitation of considering just a linear relation and
it assumes Gaussian probability distributions. A good estimate of the
degree of correlation between both magnitudes can be obtained by studying
the results of both tests. In the $t$-test, we check the hypothesis
``$b=0$'', where $b$ is the slope of the linear fit to the data. The
errors in the slope were calculated through Monte Carlo simulations, where
each point was perturbed in both axes assuming a Gaussian distribution
with standard deviation equal to the errors. For a significance level of
$\alpha=0.05$, a $t$ value higher than 1.96 indicates that a correlation
exists. The results of both tests are shown in
Table~\ref{table.gradsigma}. In the second row of the table (and left top
panel in Fig.~\ref{age.meta.gradsigma}) we show the correlation when the
metallicity is measured with the ten indicators described in
Sec.~\ref{sec-age.metallicity.gradients}. We do not find any correlation
between the metallicity gradients and the gradient of velocity dispersion.
The other panels in the figure show the metallicity gradients obtained in
different $\Delta$index--$\Delta$index diagrams in which we combined
H$\beta$ with various metallicity indicators (as indicated between the
brackets). Surprisingly, we find a marginally significant correlation
between the metallicity gradients and grad $\sigma$ when the metallicity
gradient is measured using Mgb and CN$_2$ indices, while we do not find
any correlation when the metallicity gradient is obtained using the other
indices.  If the correlation between these parameters is indicative of the
importance of galactic winds during the evollution of early-type galaxies,
the fact that the correlation only exists when the metallicity gradient is
calculated with some specific indices may indicate that this process
affects some chemical species more than others. This could happen if, for
example, the relative abundance patterns were different when the galactic
winds occurred. 

\begin{table*}
\caption{Linear fits and probability of no correlation between the metallicity
gradient and the gradient of velocity dispersion. $N$: number of galaxies in
the fit; $b+\sigma(b)$: slope and error of the linear fit (the errors have been
calculated by Monte Carlo simulations as described in the text); $t$:
$t$-statistic to verify the hypothesis ``$b=0$'' (a $t$ value higher than 1.96
allows the hypothesis to be rejected, 
with a significance level lower than 0.05);
Pnc: probability of no correlation in a non-parametric Spearman test.
\label{table.gradsigma}}
\centering
\begin{tabular}{@{}l rrrr@{}}
\hline\hline
                               & $N$ & \multicolumn{1}{c}{$b+\sigma(b)$} & 
                               $t$  & Pnc\\
\hline
grad age                       & 54  &$  1.727\pm 3.099$ & 0.56 & 0.885\\
grad [M/H]                     & 54  &$  0.175\pm 0.269$ & 0.65 & 0.238\\ 
grad [M/H] (CN$_2$--H$\beta$)  & 54  &$  1.134\pm 0.555$ & 2.04 & 0.002\\
grad [M/H] (Ca4227--H$\beta$)  & 54  &$  0.449\pm 0.686$ & 0.65 & 0.063\\
grad [M/H] (Fe4383--H$\beta$)  & 54  &$ -0.392\pm 0.442$ & 0.89 & 0.900\\
grad [M/H] (C4668--H$\beta$)   & 54  &$  0.451\pm 0.356$ & 1.27 & 0.062\\
grad [M/H] (Mgb--H$\beta$)     & 54  &$  0.876\pm 0.463$ & 2.11 & 0.001\\
\hline
\end{tabular}
\end{table*}

The dispersion in the relations is, in any case, very large. This large
scatter may be a consequence of using the velocity dispersion as an
indicator of the local potential and/or indicative that other processes
are driving the variation of the local metal content with radius.  Davies
et al.\ (1993), in fact, suggest that the local velocity dispersion is a
poor indicator of the local escape velocity, due to the (complicating)
presence of rotation and anisotropies. On the other hand, the existence of
kinematically decoupled cores in a large percentage of galaxies, the
presence of shells, dust lanes, and the observation of interacting
galaxies, seem to indicate that mergers and interactions are common
processes in the lives of galaxies. If these interactions have some
associated gas dissipation and/or star formation, it may produce
dispersion in the correlation between local metallicity and local escape
velocity.  In fact, other authors (Davies et al.\ 1993; Carollo \& Danziger 1994) have
found correlations between the Mg$_2$ and the velocity dispersion
gradients using the local escape velocity instead of the velocity
dispersion, finding also a large scatter in the relations.

We conclude that the local potential of early-type galaxies may play a
role in defining the metallicity gradient. However, the large scatter in
the derived correlation suggests that other processes may play a role in
modulating the final metal content.  These processes can be the 
consequence of  differences in the merger histories of galaxies as proposed by 
Kobayashi (2004).
The fact that the inferred
metallicity gradients differ depending upon the indicator adopted also
suggests that the gradients of the various chemical species have probably
been formed (and modified) through different physical mechanisms. 

\subsection{Correlation of the metallicity gradient with the central velocity
dispersion}

Dissipative collapse models of galaxy formation predict strong
correlations between the metallicity gradients and certain global
parameters, such as the luminosity, mass, and central velocity dispersion
(Larson 1974a; Carlberg 1984; Arimoto \& Yoshii 1987; Kawata 1999; Chiosi
\& Carraro 2002; Kawata \& Gibson 2003), in the sense that more massive
galaxies should possess steeper metallicity gradients.  Such a prediction
is driven primarily by the adopted mass-dependent feedback efficiency
within the models. Conversely, galaxy formation through hierarchical
clustering of small sub-units does not necessarily lead to a clear
prediction for any putative correlations between metallicity gradients and
global galactic properties -- merger history can readily play a part in
eroding any extant correlation (e.g. Kobayashi 2004). 

\begin{figure*}
 \resizebox{0.32\hsize}{!}{\includegraphics[angle=-90,bb=119 20 558 723]{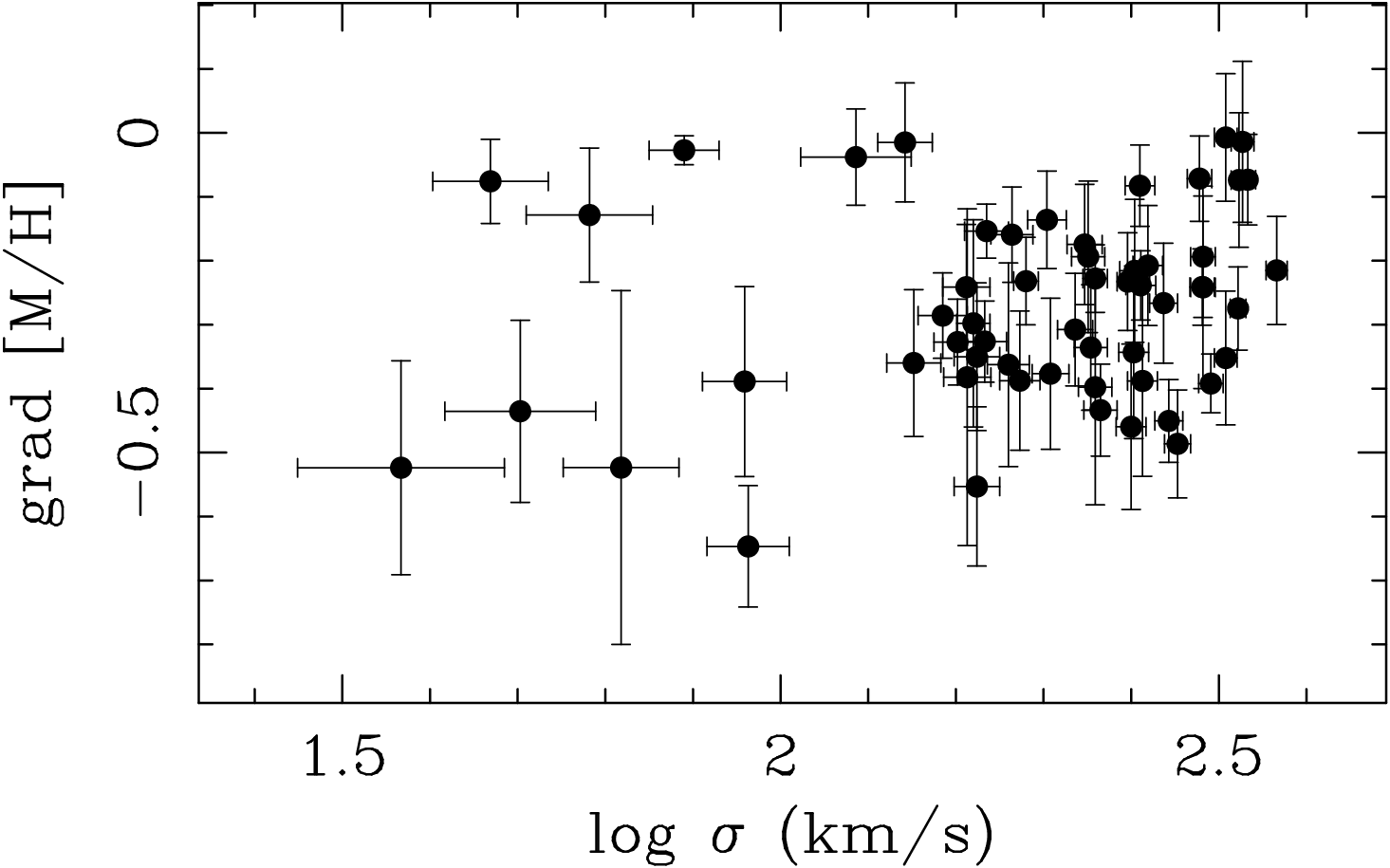}}
 \hfill
 \resizebox{0.32\hsize}{!}{\includegraphics[angle=-90,bb=119 20 558 723]{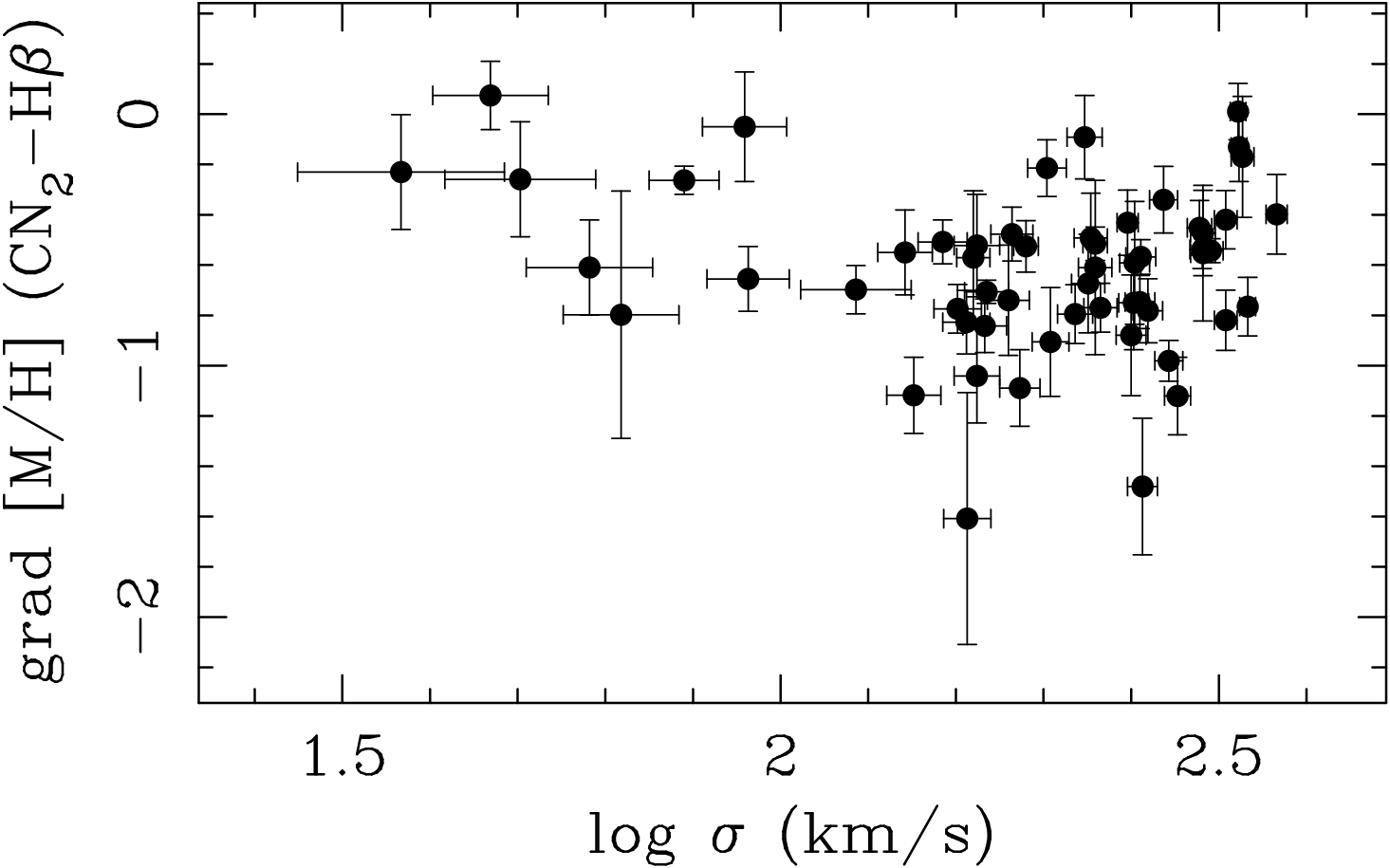}}
 \hfill
 \resizebox{0.32\hsize}{!}{\includegraphics[angle=-90,bb=119 20 558 723]{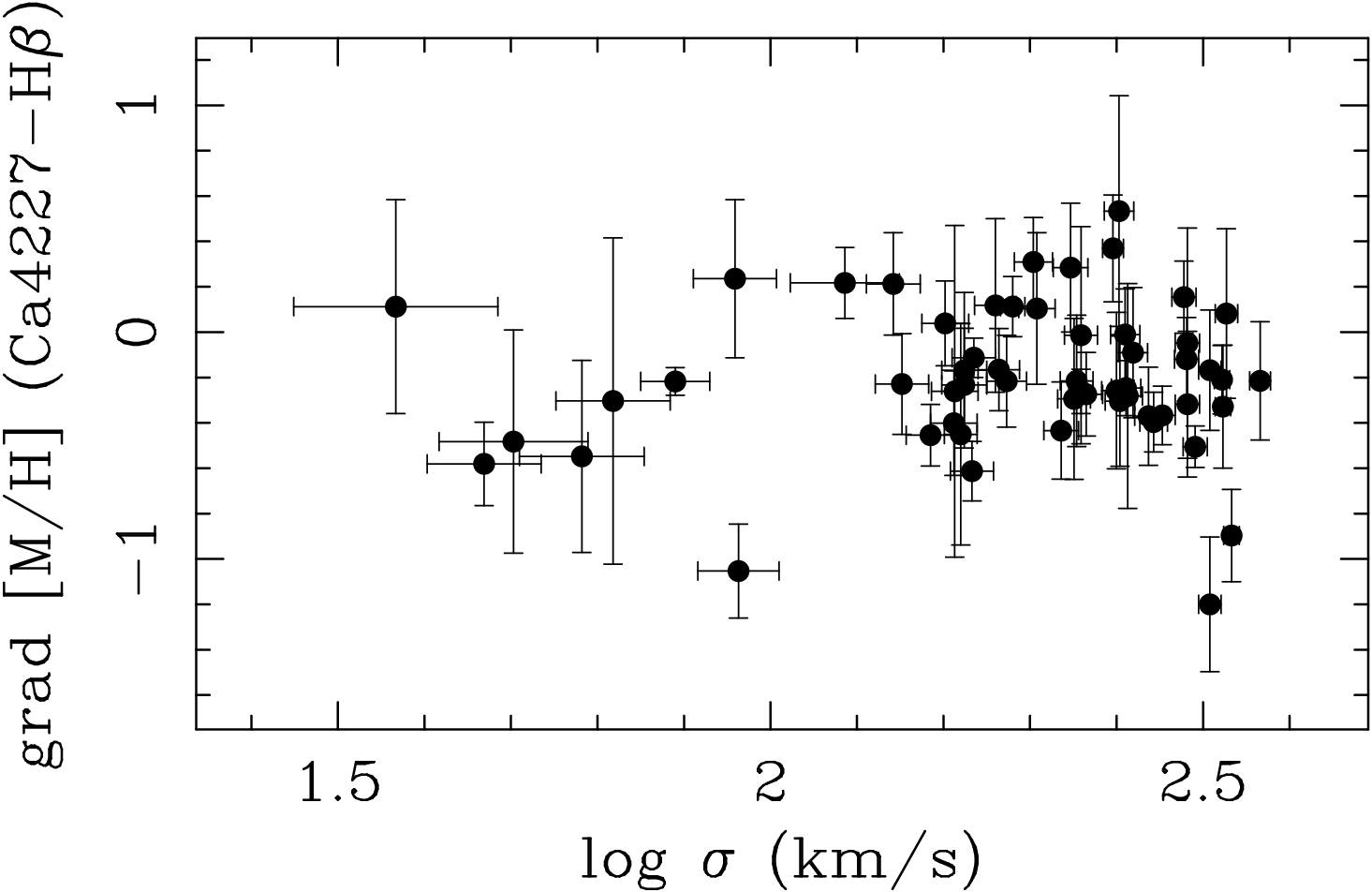}}

 \vspace{4mm}

 \resizebox{0.32\hsize}{!}{\includegraphics[angle=-90,bb=119 20 558 723]{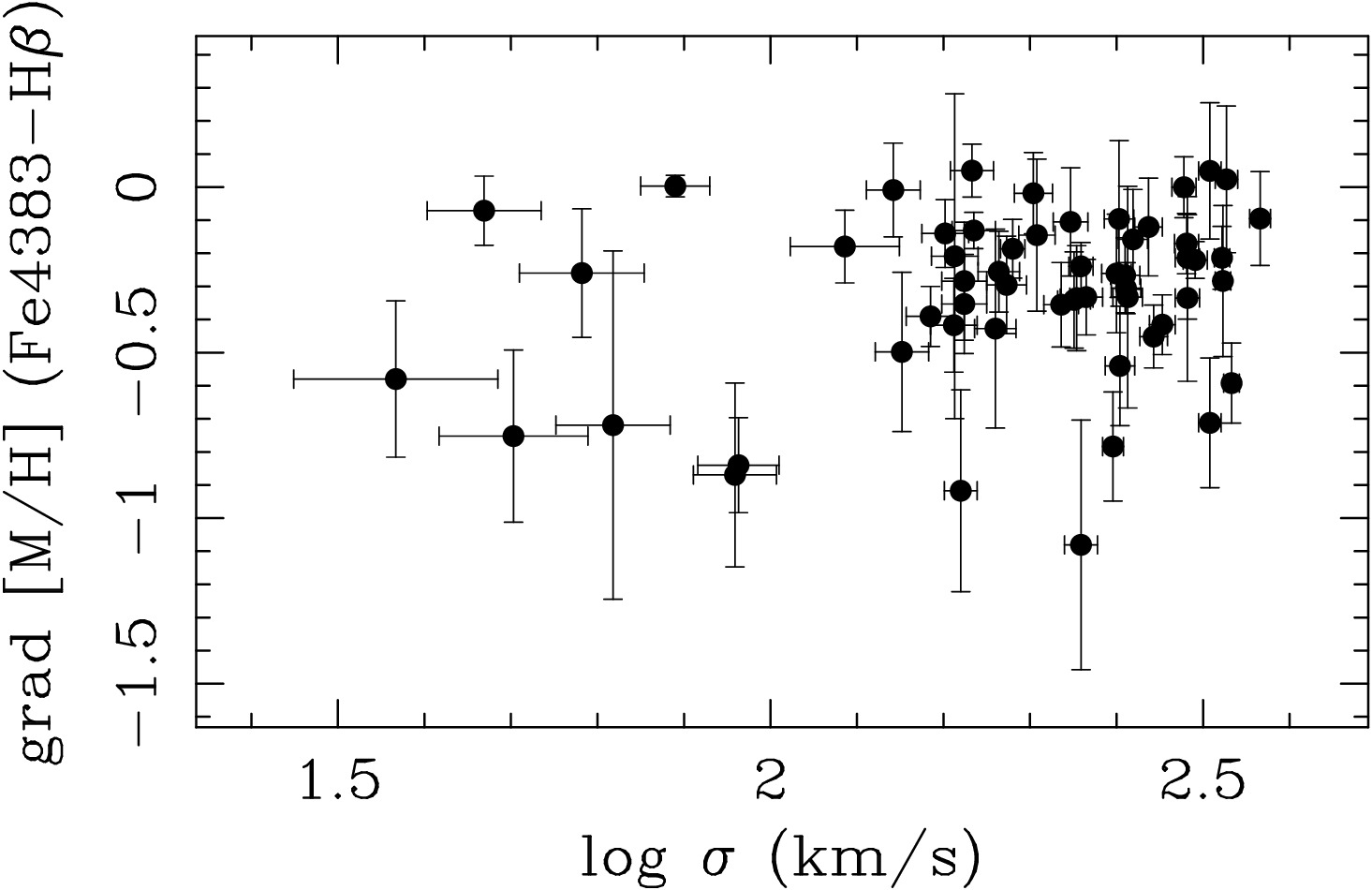}}
 \hfill
 \resizebox{0.32\hsize}{!}{\includegraphics[angle=-90,bb=119 20 558 723]{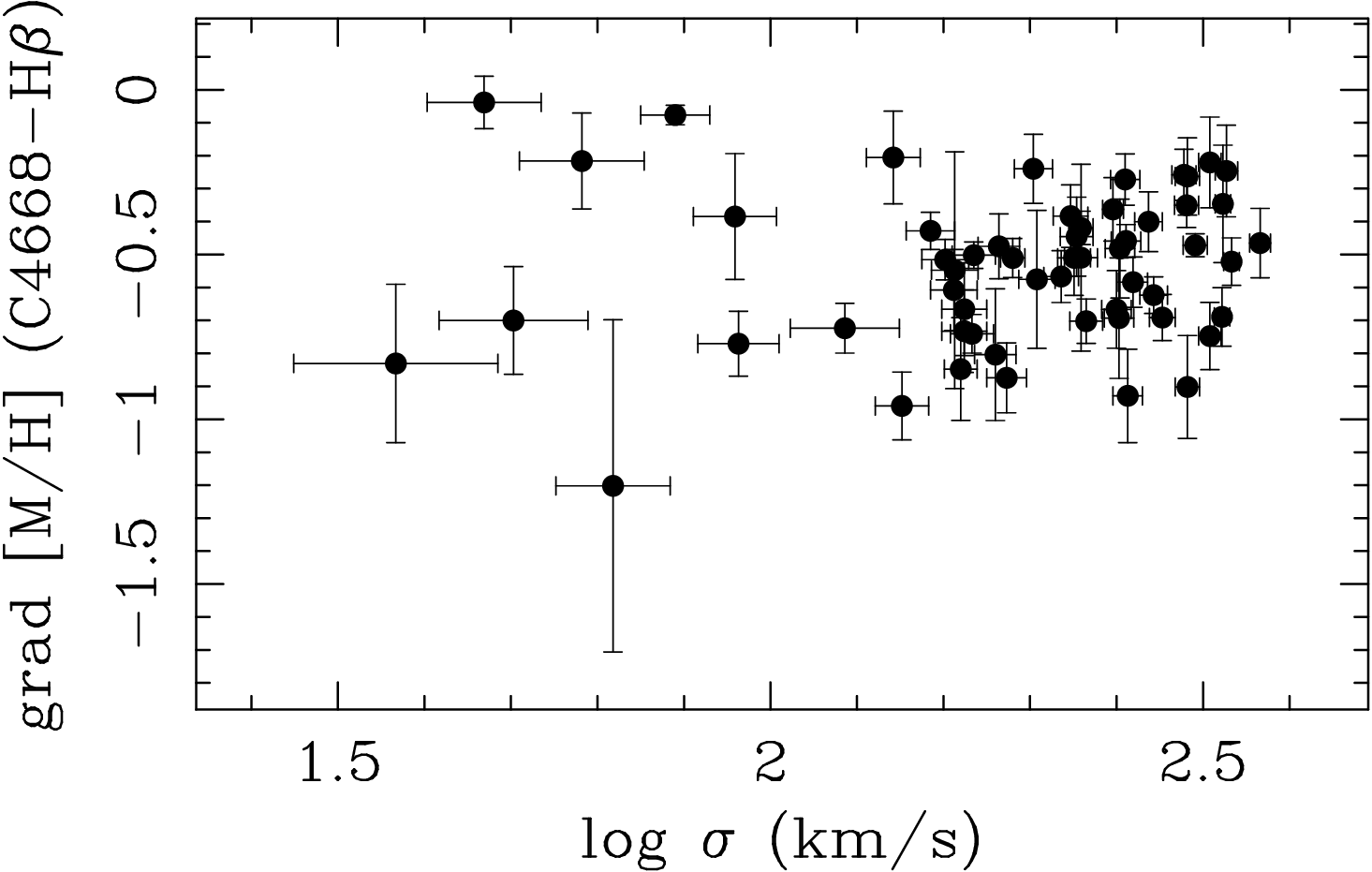}}
 \hfill
 \resizebox{0.32\hsize}{!}{\includegraphics[angle=-90,bb=119 20 558 723]{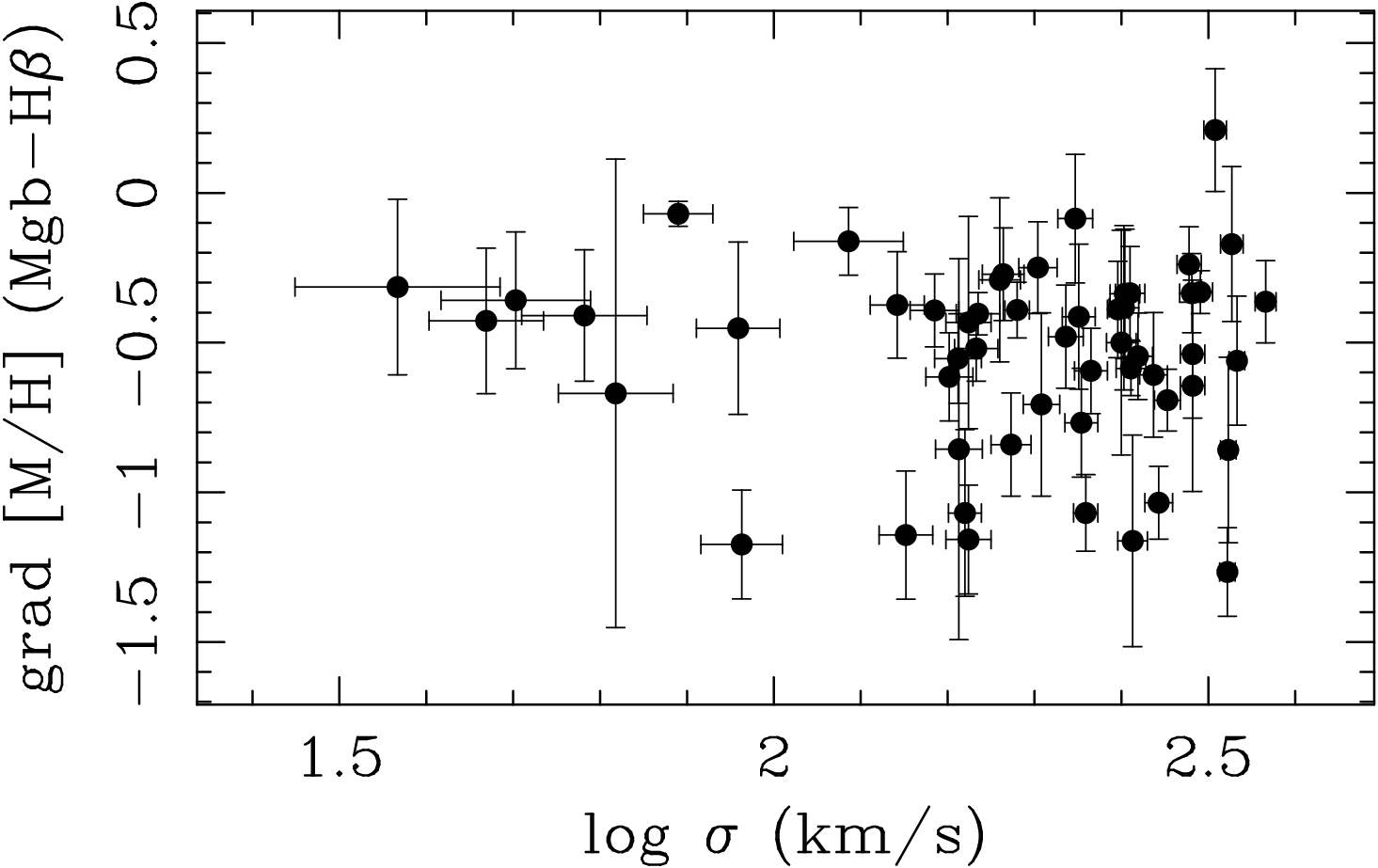}}
\caption{Metallicity gradients obtained in different index--index diagrams (as
indicated in brackets) versus the central velocity dispersion.
\label{grad.age.meta.logsigma}}
\end{figure*}

With the aim of exploring the possible correlation between metallicity
gradient and galactic mass, we compare in
Fig.~\ref{grad.age.meta.logsigma} the metallicity gradients (obtained as
described in Sec.~\ref{sec-age.metallicity.gradients}) with the central
velocity dispersion for the sample of LDEGs.  The other panels in the
figure show the relation between the metallicity gradient and the central
velocity dispersion when the metallicity gradient is measured using
CN$_2$, C4668, Fe4383, Mgb and Ca4227 combined with H$\beta$. 

As in Section~6.1, we studied the degree of correlation through a $t$-test
and a non-parametric Spearman test. The results of these tests are shown
in Table~\ref{table.corr.sigmacentral}.  We did not find any correlation
between the metallicity gradients and the central velocity dispersion,
regardless of the line-strength indices used to derive the former. 

\begin{table}
\caption{Correlation between the gradients of age and metallicity and the
central velocity dispersion. The brackets show the different indicators used
to calculate the metallicity gradients. The first and the second rows refer to
the values calculated using the ten indicators described in Section
\ref{sec-age.metallicity.gradients}.  $t$: t-parameter calculated as the
quotient of the slope and its error, computed through Monte Carlo simulations;
Pnc: probability of no correlation according to a non-parametric Spearman test.
\label{table.corr.sigmacentral}}
\centering
\begin{tabular}{@{}lrrr@{}}
\hline\hline
                              &N&  $t$ &  Pnc\\ 
\hline
grad age                      &54  &$-1.736$& 0.168\\
grad [M/H]                    &54  &  0.810 & 0.106\\
grad [M/H] (CN$_2$--H$\beta$) &54  &  1.075 & 0.789\\
grad [M/H] (Ca4227--H$\beta$) &54  &  0.251 & 0.986\\
grad [M/H] (Fe4383--H$\beta$) &54  &  1.303 & 0.244\\
grad [M/H] (C4668--H$\beta$)  &54  &  0.026 & 0.455\\
grad [M/H] (Mgb--H$\beta$)    &54  &  0.578 & 0.934\\   
\hline
\end{tabular}
\end{table}

In regards to putative correlations between line-strength indices and the
mass or the velocity dispersion, Peletier (1989), Peletier et al.\ (1990),
and G90, each found positive correlations between the Mg$_2$ gradient and
the central velocity dispersion, while Gonz\'alez (1993) and Davies \&
Sadler (1987) found the opposite trend.  Conversely, Davidge (1991, 1992)
and Davies et al.\ (1993) did not find any correlation between the
gradients of the indices Mg$_2$, $<$Fe$>$ or H$\beta$ and the central
velocity dispersion.  In this work, we have doubled the sample size
compared to any of the extant stuudies, and instead of analysing the
colours or raw line-strength gradients, we have employed the inferred
metallicity gradients. For the sample of LDEGs, we did not find any
correlation between these parameters and the central velocity dispersion. 

\subsection{Correlation of the gradients with the age of the central regions}
\label{gradvsage}

In Section~\ref{sec-gradsigma-gradmeta} we showed that, for the sample of
LDEGs, the metallicity gradients tend to correlate with the velocity
dispersion gradient when they are measured with {\it some} indicators (in
particular CN$_2$ and Mgb) but that this correlation disappears when the
metallicity gradient is measured with Fe4383, Ca4227, and C4668. 

\begin{figure*}
\centering
\resizebox{0.32\hsize}{!}{\includegraphics[angle=-90]{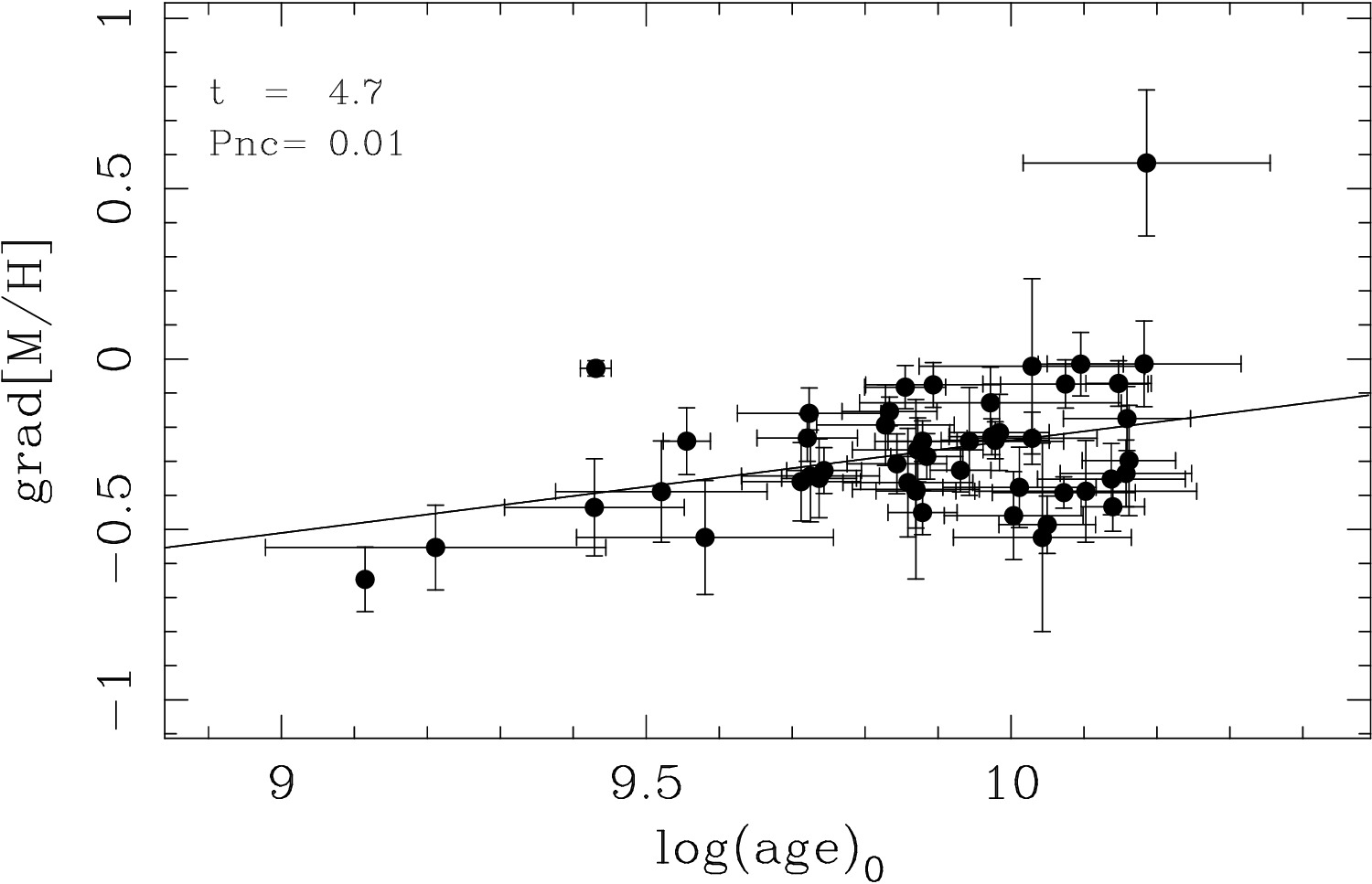}}
\hfill
\resizebox{0.32\hsize}{!}{\includegraphics[angle=-90]{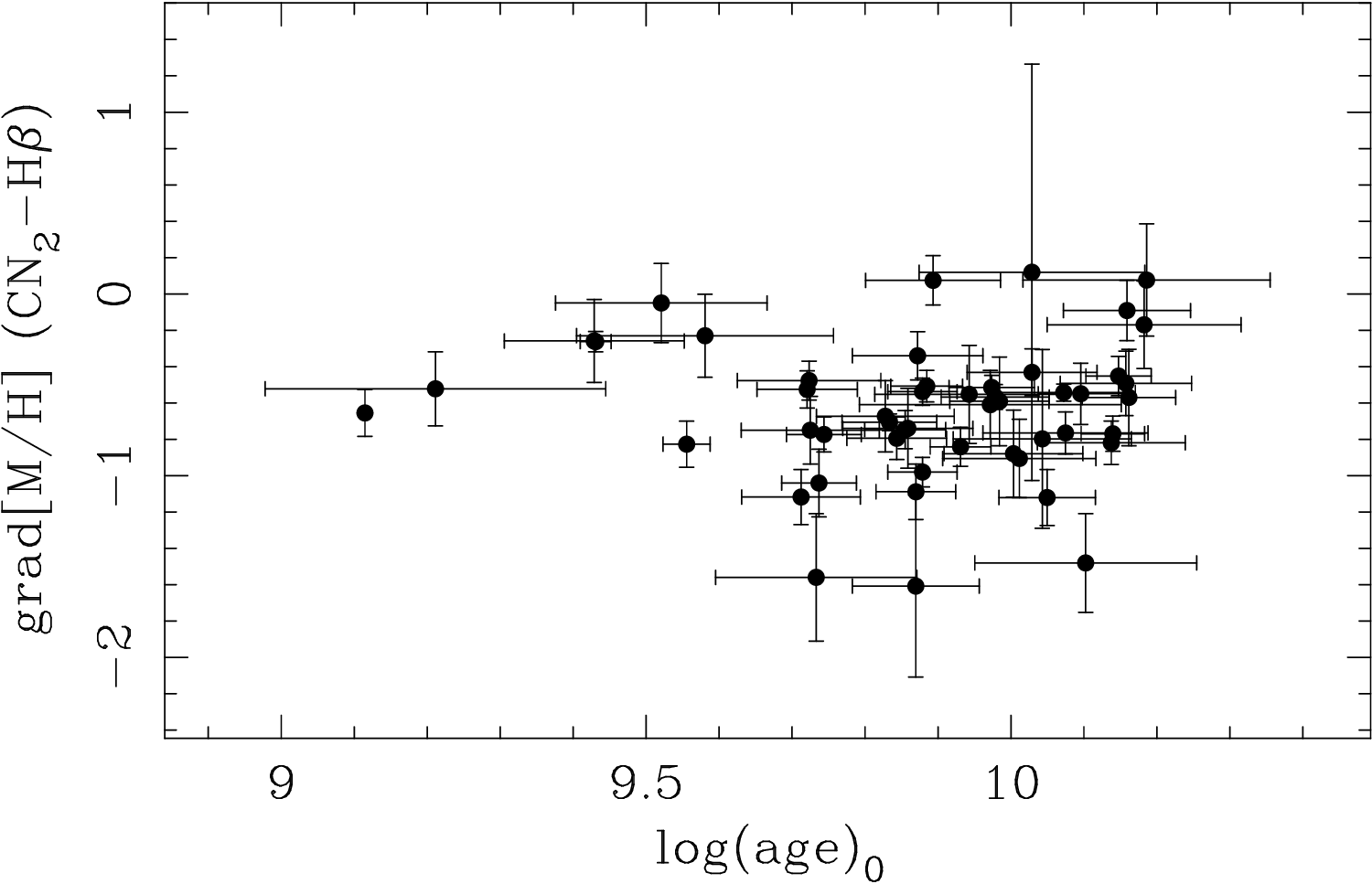}}
\hfill
\resizebox{0.32\hsize}{!}{\includegraphics[angle=-90]{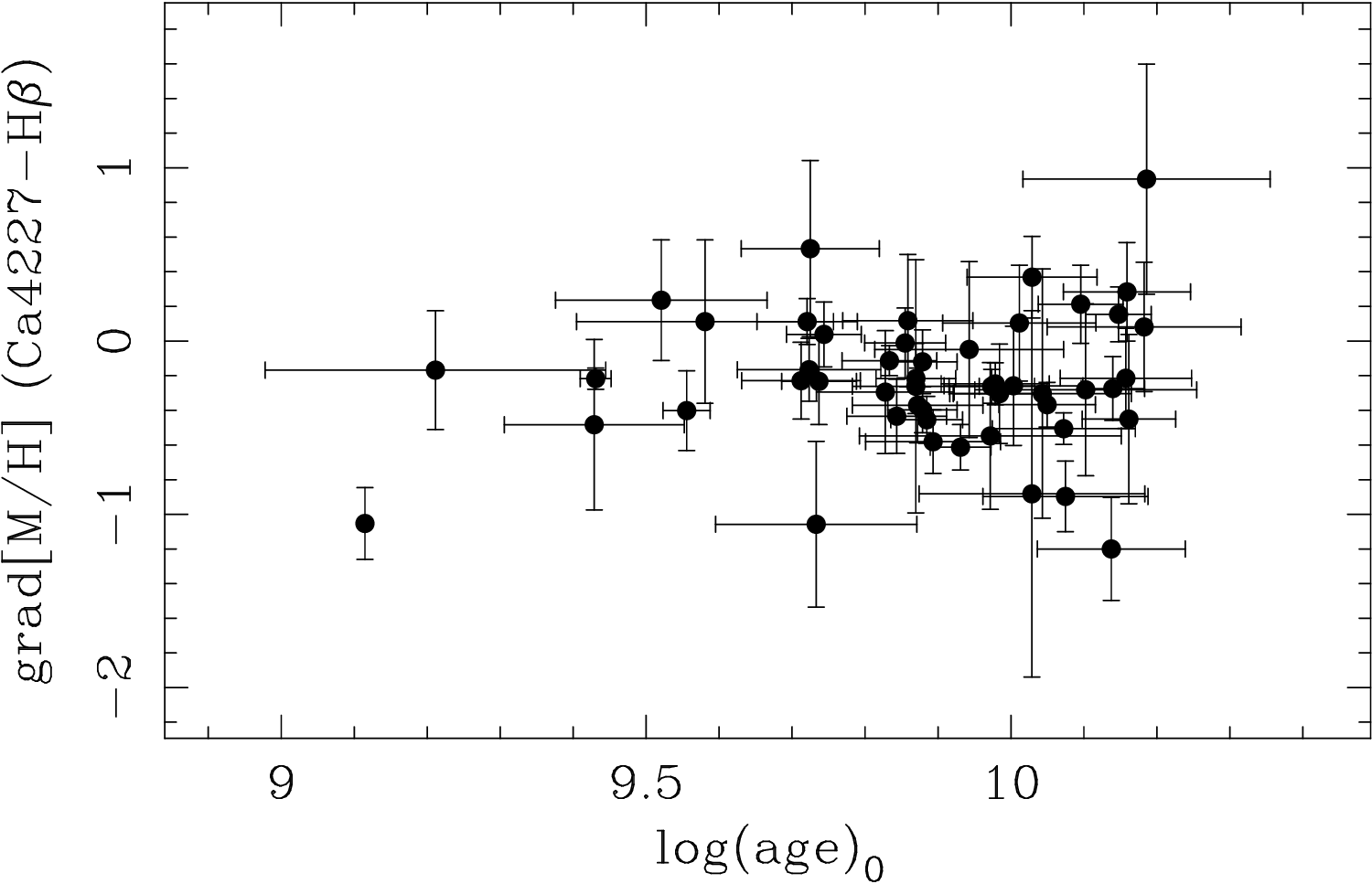}}

\vspace{4mm}

\resizebox{0.32\hsize}{!}{\includegraphics[angle=-90]{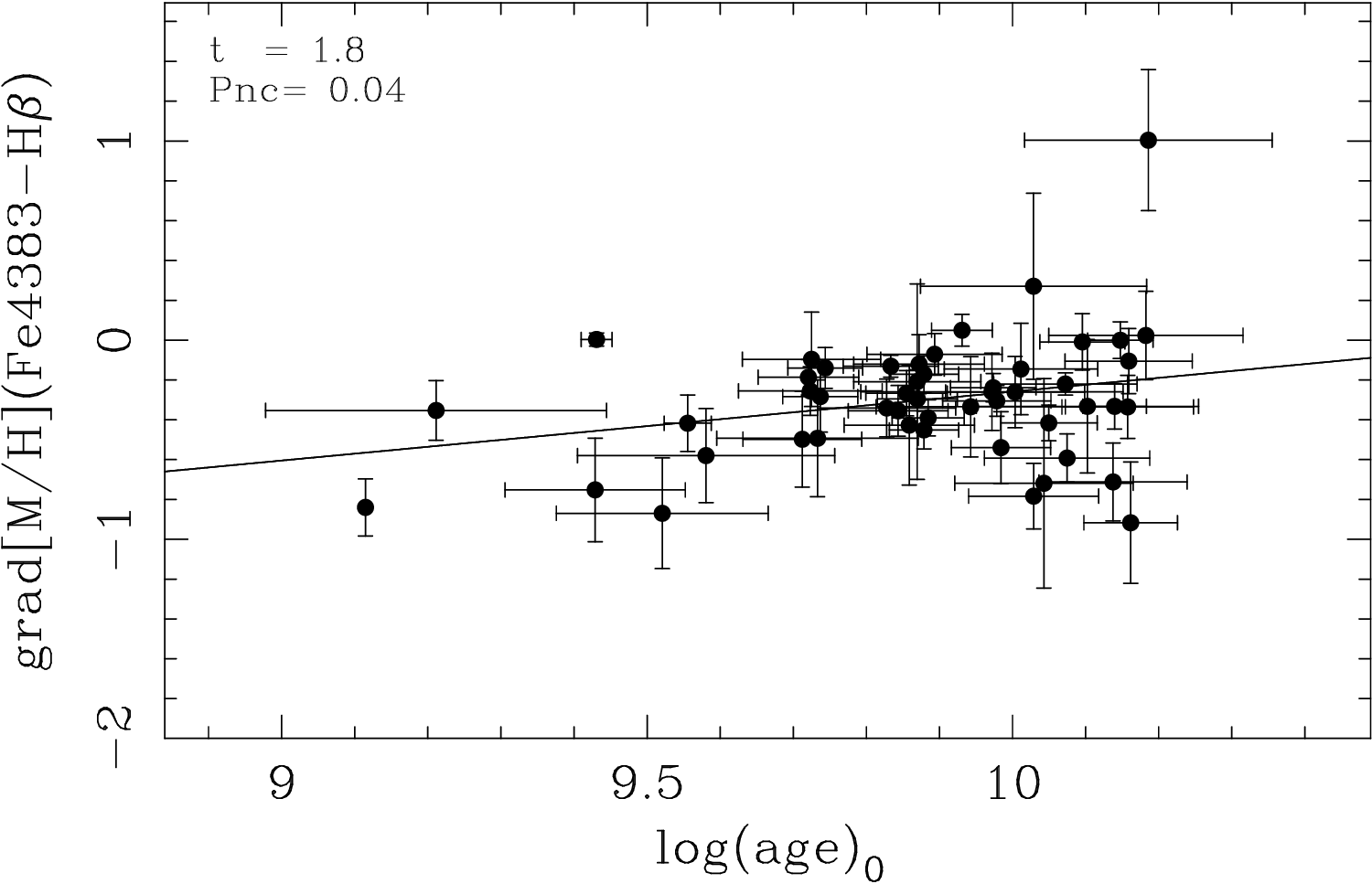}}
\hfill
\resizebox{0.32\hsize}{!}{\includegraphics[angle=-90]{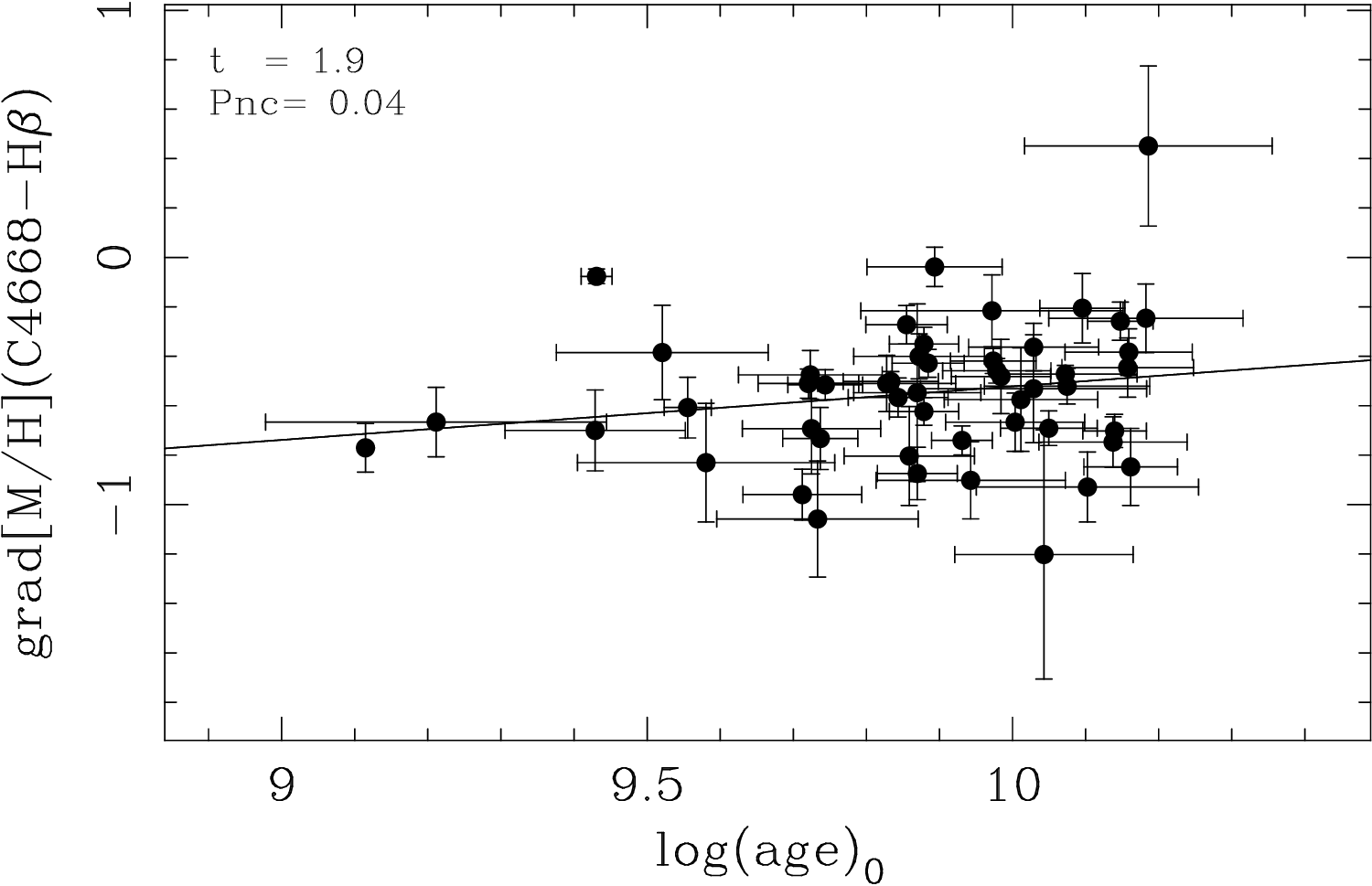}}
\hfill
\resizebox{0.32\hsize}{!}{\includegraphics[angle=-90]{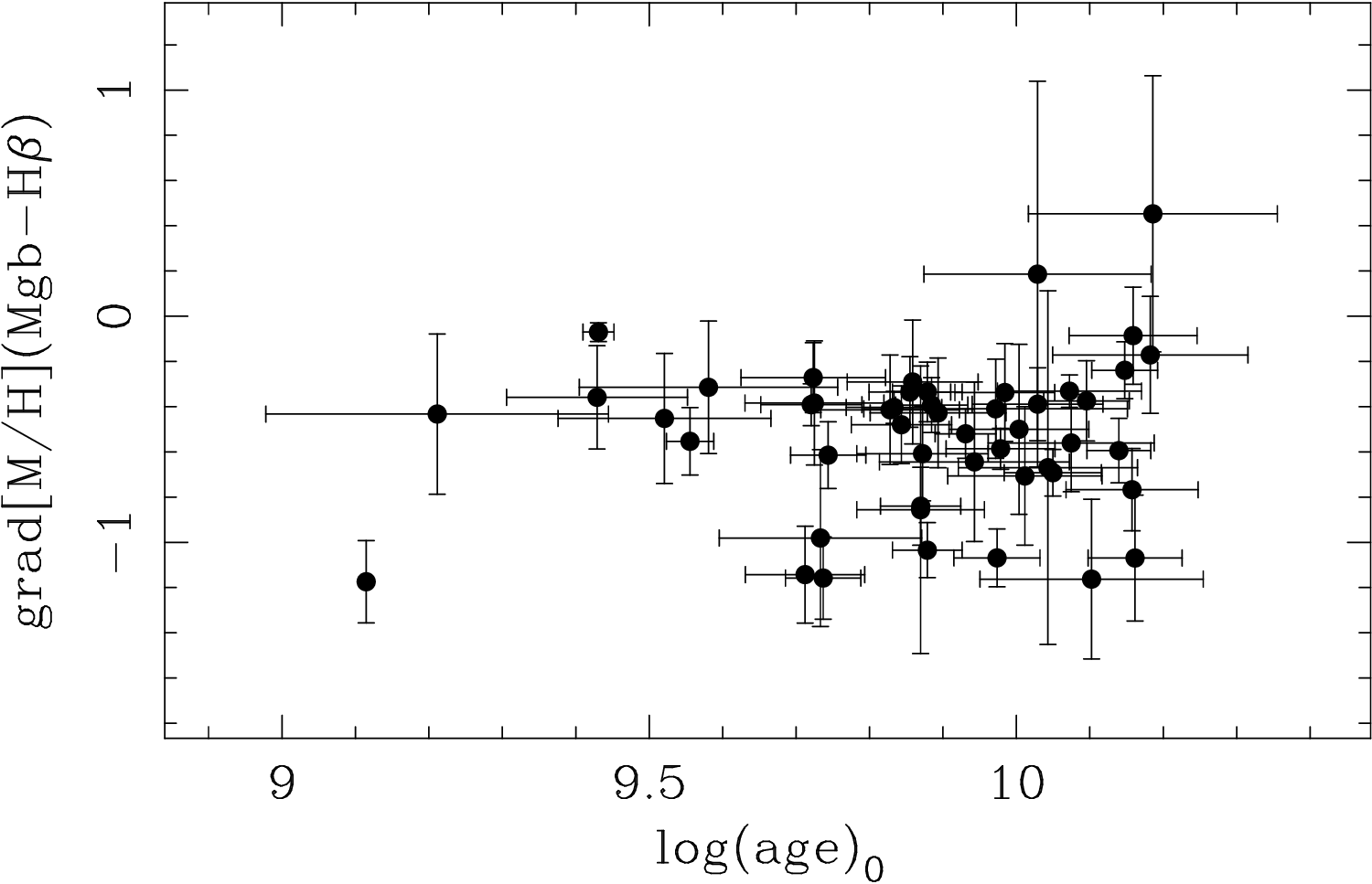}}
\caption{Relation of the metallicity gradient with the central age of the
galaxies. When the correlation is statistically significant a linear fit
weighted by the errors in both axes is also shown. In these cases, the $t$
parameter and the probability of no correlation (Pnc), as derived from a
non-parametric Spearman rank test, are also indicated in the top-left corner of
the panels. For a significance level of 0.05, a value of $t$ larger than 1.96
indicates the presence of correlation between both variables.
\label{metagradi.age}}
\end{figure*}

In Section~\ref{sec.meangradients} we showed that the mean age gradient in
our sample of LDEGs is not null, which might be indicative of the
occurrence of star formation events in the centre of the galaxies.
Furthermore, we showed in Paper~II than in these events, the relative
enrichment of some chemical species (the ones mainly released by low- and
intermediate-mass stars, like Fe) might be more important than others (the
ones mainly produced in Type~II supernovae, like Mg).  With the aim of
exploring if the occurrence of star formation processes in the centre of
the galaxies may be responsible for the abundance gradients in some
chemical species, we now study the correlation between the metallicity
gradients and the central ages obtained in Paper~II.
Fig.~\ref{metagradi.age} shows these correlations for different
measurements of the metallicity gradient. When the correlation is
statistically significant, a linear fit taking into account the errors in
the $x$- and $y$-directions is also plotted.  In these cases, the
probability of no correlation (Pnc) obtained in a non-parametric Spearman
test, and the $t$ parameter to verify the hypothesis $b=0$ (where $b$ is
the slope of the linear fit), are indicated in the panels.  As can be seen
in the different panels, while we do not find any significant correlation
when the metallicity gradients are measured with CN$_2$, Mgb, or Ca4227,
we {\it do} find a correlation between the metallicity gradients and the
central age of the galaxies when the former is measured using Fe4383 and
C4668.  We also find a significant correlation between the metallicity
gradient obtained with the ten indicators described in
Section~\ref{sec-age.metallicity.gradients} and the central age of the
galaxies.  We interpret these correlations in Section~6.4. 

\subsection{Interpretation of the Correlations for the LDEGs}
\label{discussion}

In the Sections~6.1-6.3, we studied the putative correlations between
metallicity gradient and other global properties of LDEGs.  Broadly
speaking, we found two distinct characteristics of the gradients depending
upon the specific indicator used to derive them: 

\begin{itemize}

\item {\it Metallicity gradients measured with CN$_2$ and Mgb}. These
correlate with the velocity dispersion gradient but they do not correlate
with the central velocity dispersion or with the central age of the
galaxies.  This behaviour may indicate that the gradients of some elements
(in particular Mg and N) are the result of processes such as galactic
winds which can develop after an intense burst of star formation in the
earliest phases of galaxy formation.  The possible subsequent star
formation in the galaxies has not modified substantially the variation of
these elements with radius.  This is in agreement with the results of
Papers~I and II where we argued that the relative enrichment of Fe in the
star formation processes which occurred in the centre of the galaxies had
to be more important than that of Mg. 

\item {\it Metallicity gradients measured with Fe4383 and C4668}. These
gradients do not correlate with the velocity dispersion gradient but they
do correlate with the central age of the galaxy. The correlation with the
central age may indicate that episodes of star formation in the centre of
the galaxies affects considerably the metallicity gradient for these
chemical species, increasing the central metallicity. 
 
This view agrees with the results of Papers~I and II. Note that in
galaxies that have suffered star formation episodes in their centres, the
[Mg/Fe] gradient flattens as a consequence of the Fe enrichment in the
central parts, which is supported by the relation between [Mg/Fe] and age
reported in Paper~I. This is also in agreement with the result presented
in Paper~II in which we found the existence of an age--metallicity
relation when the metallicity was measured with Fe4383, but not when this
parameter was measured with Mgb. 

Conversely, in S\'anchez-Bl\'azquez et al.\ (2003) we reported differences
in the N abundance between galaxies in different environments. If the star
formation processes in the centres of the galaxies do not affect
substantially the gradients in this element, then the differences must be
visible not only in the centres, but in the outer parts. 

We also note the behaviour of the metallicity gradient when inferred
from Ca4227. Specifically, the gradient does not correlate with any of the
obvious physical parameters -- i.e., neither with the central age nor with
the velocity dispersion gradient. 

\end{itemize}
 
\section{Global stellar population parameters}
\label{global}

Many of the results described in Papers I and II could be explained by
assuming the presence of a small percentage of young stars in the centre of
most of the galaxies, at least in the subsample of LDEGs.  We have also argued
that this could be the cause of the existence of non-null age gradients in our
sample of galaxies (see Sec.~\ref{sec.meangradients}). If this point of view is
correct, we would expect that the relations defined between the global
parameters of the galaxies (i.e. for the whole bodies of the galaxies) are
different than the one derived for the central regions.  To investigate this
possibility, we now compare for the subsample of LDEGs the
relation of these global values and the velocity dispersion with the relations
derived for the central values in Paper~II. The {\it global values} can be
obtained from the gradients, assuming a linear behaviour of the indices with
radius and evaluating the integral
\begin{equation}
I_{\rm global}=\frac{\displaystyle\int_{o}^{\infty} 
  I(r)\;2\pi I_{\rm e} \exp^{-7.67[(r/r_{\rm e})^{1/4}-1]}{\rm d}r}
{\displaystyle\int_{o}^{\infty}
  2\pi I_{\rm e} \exp^{-7.67[(r/r_{\rm e})^{1/4}-1]}{\rm d}r},
\end{equation}
where $I(r)=a+b\log (r/r_{\rm e})$ represents the index at a distance $r$ from
the galaxy as derived from the gradient, and $r_{\rm e}$ is the effective
radius.  To solve this integral, we have assumed that the spatial profile of
the galaxies can be approximated with a de~Vaucouleurs law.

\begin{figure}
\resizebox{1.0\hsize}{!}{\includegraphics[angle=0]{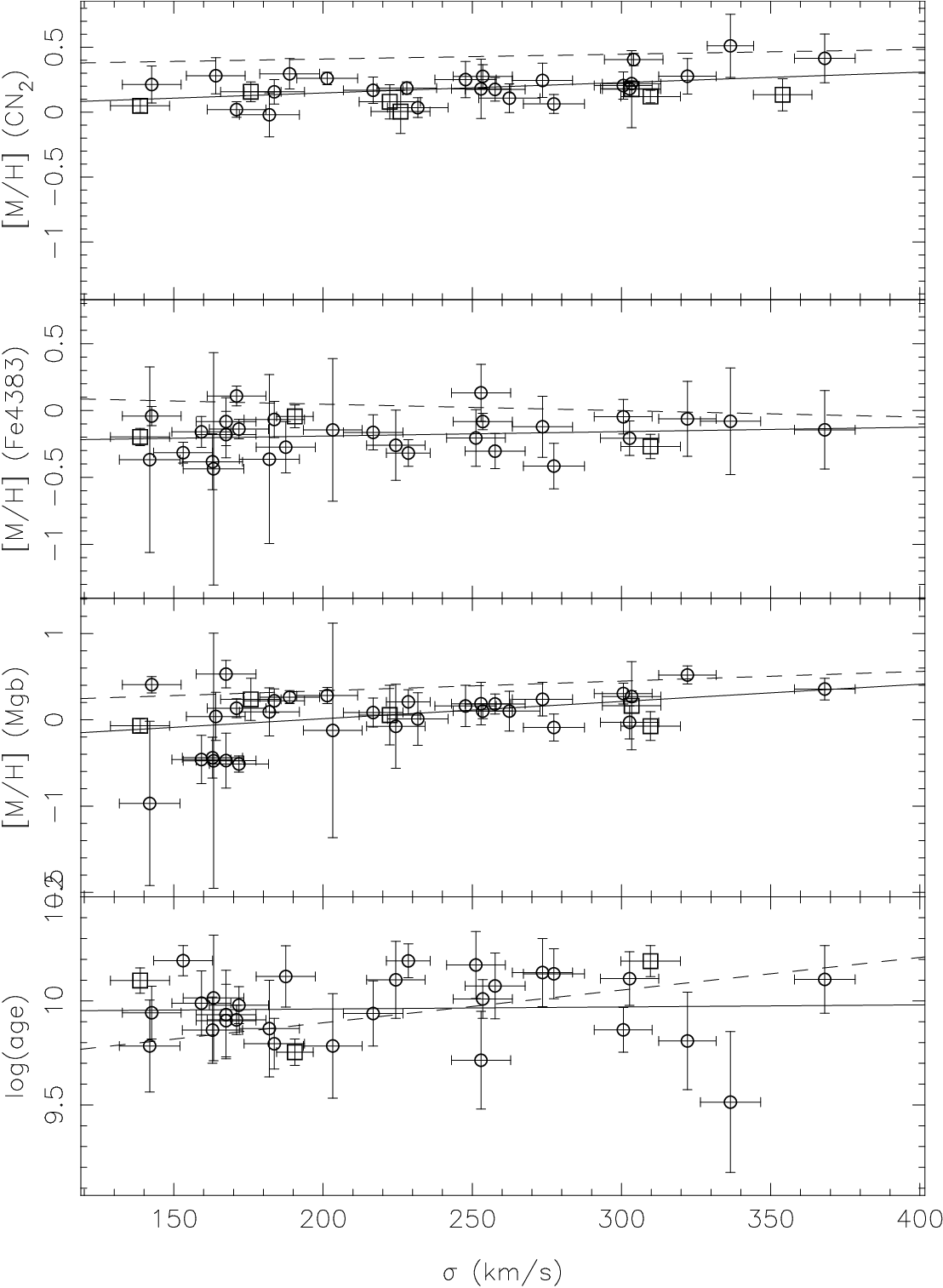}}
\caption{Relation between the global age and metallicity and the central
velocity dispersion for galaxies in low-density environments.  The solid line
represents a linear fit weighted with the errors for the global data. The
dashed line indicates the fit obtained for the central values reported in
Paper~II. Open circles represent E galaxies while the S0 are represented by squares.
\label{global.panelillo}}
\end{figure}

Fig.~\ref{global.panelillo} shows the relation between the global age and
global metallicities (as derived from different indicators) and the
central velocity dispersion. The solid line indicates a linear fit to the
data weighted with the errors in both parameters.  In order to compare
with the relations for the central regions, we have also plotted the
linear fits obtained for the central values derived in Paper~II (dashed
lines). Table~\ref{global.fits} summarises the parameters of the fits. We
carried out a $t$-test to verify the existence of significant differences
between the slopes of the trends defined by the central and the global
SSP-parameters.  A value of $t$ higher than 1.96 indicates that there
exist differences with a significance level lower than 0.05. 

\begin{table*}
\caption{Parameters of the linear fits, weighted with errors, of the age and
metallicities (measured in different index--index diagrams indicated between
brackets) and the central velocity dispersion. The table shows the results for
the central values ($a_c$ and $b_c$) and for the global values ($a_g$ and
$b_g$). The coefficients $a$ and $b$ represent the zero point and the slope of
the linear fit, respectively. The last column shows the $t$ parameter obtained
in a $t$-test to check the hypothesis $b_c=b_g$. A value of $t$ higher than
2.326 allows us to reject the hypothesis with a significance level lower than
0.01. \label{global.fits}}
\begin{tabular}{@{}l rrrrr@{}}
\hline
                        & \multicolumn{2}{c}{central} & \multicolumn{2}{c}{global} & \\
                        & \multicolumn{1}{c}{$a_c$}   &\multicolumn{1}{c}{$b_c$}&\multicolumn{1}{c}{$a_{g}$}&
\multicolumn{1}{c}{$b_{g}$}    & \multicolumn{1}{c}{$t$}     \\
\hline
$[$M/H$]$ (CN$_2$--H$\beta$) & $0.335\pm 0.081$ &$  0.00037\pm 0.00034$ &$ -0.012\pm 0.091$ &$ 0.00080\pm 0.00038$ & 0.86\\
$[$M/H$]$ (Fe4383--H$\beta$) & $0.148\pm 0.064$ &$ -0.00050\pm 0.00030$ &$ -0.275\pm 0.226$ &$ 0.00041\pm 0.00087$ & 0.98\\
$[$M/H$]$ (Mgb--H$\beta$)    & $0.113\pm 0.075$ &$  0.00112\pm 0.00031$ &$ -0.427\pm 0.286$ &$ 0.00212\pm 0.00107$ & 0.88\\
log age                 & $9.534\pm 0.075$ &$  0.00177\pm 0.00033$ &$  9.950\pm 0.115$ &$ 0.00007\pm 0.00049$ & 2.88\\
\hline
\end{tabular}
\end{table*}

The only case in which the slopes defined by the central and the global
values are significantly different is in the relation between the age and
the central velocity dispersion.  While the central age shows a
significant correlation with the central velocity dispersion (see
Paper~II), the global age as derived in this section does not correlate
with this parameter.  This result favours the idea that the young ages
that we found in Paper~II in a large percentage of galaxies are due to a
minor percentage of stars formed in the centre of the galaxies at a later
epoch than the bulk of the stars, as suggested by Trager et al.\ (2000b).
It also supports the suggestion that this minor percentage of young stars
in the centre of the galaxies is responsible for the age gradients
reported in Section~\ref{sec.meangradients}. 

On the other hand, the slopes of the relations between the metallicity and
the central velocity dispersion do not show a statistically significant
variation between the central and the global values, although there is a
tendency for the relations to be steeper for the global values. There are
also differences in the zero point as a consequence of the existence of
gradients, in disagreement with the earlier claims of Gonz\'alez \& Gorgas
(1996).  These authors found that the relation Mg$_2-\sigma$ is flatter
at one effective radii than in the center, concluding that the mass-metallicity
relation was much flatter at one effective radii than in the central parts 
of the galaxies. 
In light of our present series of papers, the
results of Gonz\'alez \& Gorgas could be explained if the differences
between the central and global relation of the age with $\sigma$
was causing these differences.

The lack of variation in the slope of the relation between the central and
the global metallicities measured with CN$_2$ and Mgb was expected, since
we  find a flat relation between the strength of the gradients
and the central velocity dispersion.
In the case of the metallicity measured with
Fe4383 we might have expected to see a steepening of the slope in the
global relation compared with the central one.  If the relative importance
of the star formation processes has been higher in the smaller galaxies
(as suggested by the age--$\sigma$ relation), the metallicity inferred
from Fe4383 in these galaxies should be also higher. The variation in the
slope obtained when comparing the central and the global relations goes in
this sense, but the differences are not statistically significant. We must
note that the errors in the global measurements are higher than for the
central values, which might explain the lack of statistical significance. 

\section{Differences in the gradients as a function of environment}
\label{sec.comagradientes}

If the environment in which galaxies reside has any influence over the
timescales of the star formation, the number of interactions, or the
dissipation of gas, we might expect to see an environmental dependence
upon the inferred age and metallicity gradients.  Fisher et al.\ (1995)
analysed a sample of bright cluster ellipticals, concluding that their
gradients and those of field ellipticals, were not significantly
different. Tamura \& Otha (2003) found, studying the photometry in the B
and R~bands as a function of radius for galaxies in Abell~2199,
correlations between the colour gradients and some global properties of
the galaxies, such as the luminosity and effective radius, which had not
been found in studies of field galaxies. These authors, however, only
found these correlations amongst the most luminous galaxies ($R<15$ mag
and with an effective \mbox{radius $>$3$^{\prime\prime}$}). We now we
examine the mean gradients of the SSP-parameters for the HDEGs and compare
these values with the ones obtained for the LDEGs. 

\subsection{Mean stellar population gradients in HDEGs}
\label{mean.gradients}

The second column of Table~\ref{table.mean.coma} lists the mean values of
the SSP-parameter gradients obtained for the HDEGs in different
index--index diagrams (indicated between brackets). As in
Section~\ref{sec.meangradients}, we quantified the probability that these
values are different from zero by chance.  The results of a $t$-test
indicate that, while the metallicity gradients obtained in the
CN$_2$--H$\beta$ and C4668--H$\beta$ diagrams are not compatible with
being null, the metallicity gradients obtained in the Fe4383--H$\beta$ and
Ca4227--H$\beta$ diagrams are compatible with zero, within the errors.
Furthermore, the mean age gradient for this subsample of galaxies is also
compatible with zero. We now analyse these results separately: 

\begin{table*}
\caption{Mean stellar population gradients and their errors for HDEGs.
$\sigma$: standard deviation about the mean; $N$: number of galaxies averaged;
$N_{\rm eff}$: effective number of points; $t$: $t$-statistic to verify the
hypothesis ``mean=0''; $\sigma_{\rm exp}$: standard deviation expected from the
errors;  $\alpha$: significance level to reject the hypothesis
``$\sigma$=$\sigma_{\rm exp}$'' in a $\chi^2$ test. Column~9 shows the
$t$-statistic to confirm if the gradients obtained with individual
indicators are
the same as the metallicity gradients obtained with a combination of ten
different indicators. As a reference, the $10^{\rm th}$ column shows the mean
gradients obtained for the LDEGs, from Section~\ref{sec.meangradients} (see
Table~\ref{mean-gradients}). The final column of the table shows 
the $t$-statistic, to verify if the mean gradients are the same for 
both samples of galaxies, LDEGs and HDEGs.
A high value of $t$ indicates significant differences.
\label{table.mean.coma}}
\centering
\begin{tabular}{@{}l rrrrrrrrrr@{}}
\hline\hline
                             &\multicolumn{1}{c}{mean}
                             &\multicolumn{1}{c}{$\sigma$} 
                             &\multicolumn{1}{c}{$N$} 
                             &\multicolumn{1}{c}{$N_{\rm eff}$} 
                             &\multicolumn{1}{c}{$t$} 
                             &\multicolumn{1}{c}{$\sigma_{\rm exp}$} 
                             &\multicolumn{1}{c}{$\alpha$} 
                             &\multicolumn{1}{c}{$t$} 
                             &\multicolumn{1}{c}{LDEGs}
                             &\multicolumn{1}{c@{}}{t$'$}\\
\hline
 grad log age                &$ 0.027\pm 0.067$ & 0.258  & 15 & 9.79 & 0.34 &
 0.138 &  0.0003 &      &  $0.082\pm 0.015 $& 0.8\\
 grad [M/H]                  &$-0.328\pm 0.064$ & 0.247  & 15 &11.85 & 5.59 &
 0.157 &  0.0017 &      & $-0.206\pm 0.019$& 1.8\\
 grad [M/H] (CN$_2$--H$\beta$)&$-0.643\pm 0.111$ & 0.432  & 15 &10.86 & 5.60 &
 0.236 &  0.0003 & 2.10 & $-0.582\pm 0.032$& 0.5 \\
 grad [M/H] (C4668--H$\beta$) &$-0.599\pm 0.092$ & 0.358  & 15 & 9.18 & 5.71 &
 0.185 & 4.2E$-06$ & 1.96 & $-0.459\pm 0.028$&1.4 \\
 grad [M/H] (Fe4383--H$\beta$)&$-0.168\pm 0.118$ & 0.456  & 15 & 8.95 & 1.47 &
 0.275 &  0.0014 & 0.95 & $-0.197\pm 0.025$&0.2\\
 grad [M/H] (Mgb--H$\beta$)   &$-0.537\pm 0.158$ & 0.611  & 15 &10.23 & 2.85 &
 0.337 & 4.2E$-05$ & 1.02 & $-0.407\pm 0.039$& 0.8\\
 grad [M/H] (Ca4227--H$\beta$)&$-0.168\pm 0.214$ & 0.831  & 15 &11.16 & 0.70 &
 0.434 & 1.7E$-06$ & 0.61 & $-0.238\pm 0.034$& 0.3\\           
\hline
\end{tabular}
\end{table*}

\begin{itemize}

\item {\it Age gradient.\/} If we assume that the most likely scenario to
explain the age gradient in early-type galaxies is the occurrence of star
formation processes in the centres of these systems, the lack of a mean
age gradient in the HDEGs indicates that these galaxies have undergone few
episodes of star formation in recent times, when compared with field
galaxies. If the star formation processes are triggered by the
interactions between galaxies, the differences could be explained due to
the lower probability of an interaction in the centre of the Coma cluster.
This is in agreement with the conclusions of Papers~I and II, where we
showed that the central stellar populations of the Coma cluster galaxies
possessed, on average, older ages than the LDEGs. 

\item {\it Metallicity gradients.\/} The mean metallicity gradient
obtained for this subsample of galaxies is $\Delta[$M/H$]/\log r
=-0.328\pm 0.064$, slightly steeper than the gradient obtained for the
LDEGs. From her simulations, Kobayashi (2004) predicts a metallicity
gradient for galaxies that have not suffered major mergers of $\Delta \log
Z/\Delta \log r \sim -0.3$. These predictions are in agreement with the
mean values obtained for the galaxies in the Coma cluster. On the other
hand, as for the LDEGs, the metallicity gradients are steeper when
obtained with some indicators (CN$_2$, C4668 and Mgb), although the
statistical significance of the differences is much lower than in the case
of the LDEGs (see final column of Table \ref{table.mean.coma}). This may
also be due to the larger errors in the determination of the gradients for
the galaxies in the Coma cluster. 

\end{itemize}

We have next checked to see if the scatter amongst the mean values was
compatible with the dispersion expected by the errors.  We carried out a
$\chi^2$ test of the hypothesis $\sigma=\sigma_{\rm exp}$ (where $\sigma$
is the observed scatter and $\sigma_{\rm exp}$ the scatter expected from
the errors), to see whether it could be rejected with a low significance
level ($\alpha$). Column~8 of Table~\ref{table.mean.coma} shows the
$\alpha$ values resulting from this test. We find a real scatter in both
the age and the metallicity gradients which cannot be explained by the
errors, in contrast with the findings of Mehlert et al.\ (2003). The
difference may be due to the fact that the sample used by Mehlert et~al.
spans a more limited range in velocity dispersion ($2.2<\log \sigma <
2.5$). 

\subsection{Correlation of the metallicity gradients in HDEGs with the central
velocity dispersion}
\label{corr.sigma.coma}

\begin{figure*}
\resizebox{0.35\hsize}{!}{\includegraphics[angle=-90,bb=119 20 558 723]{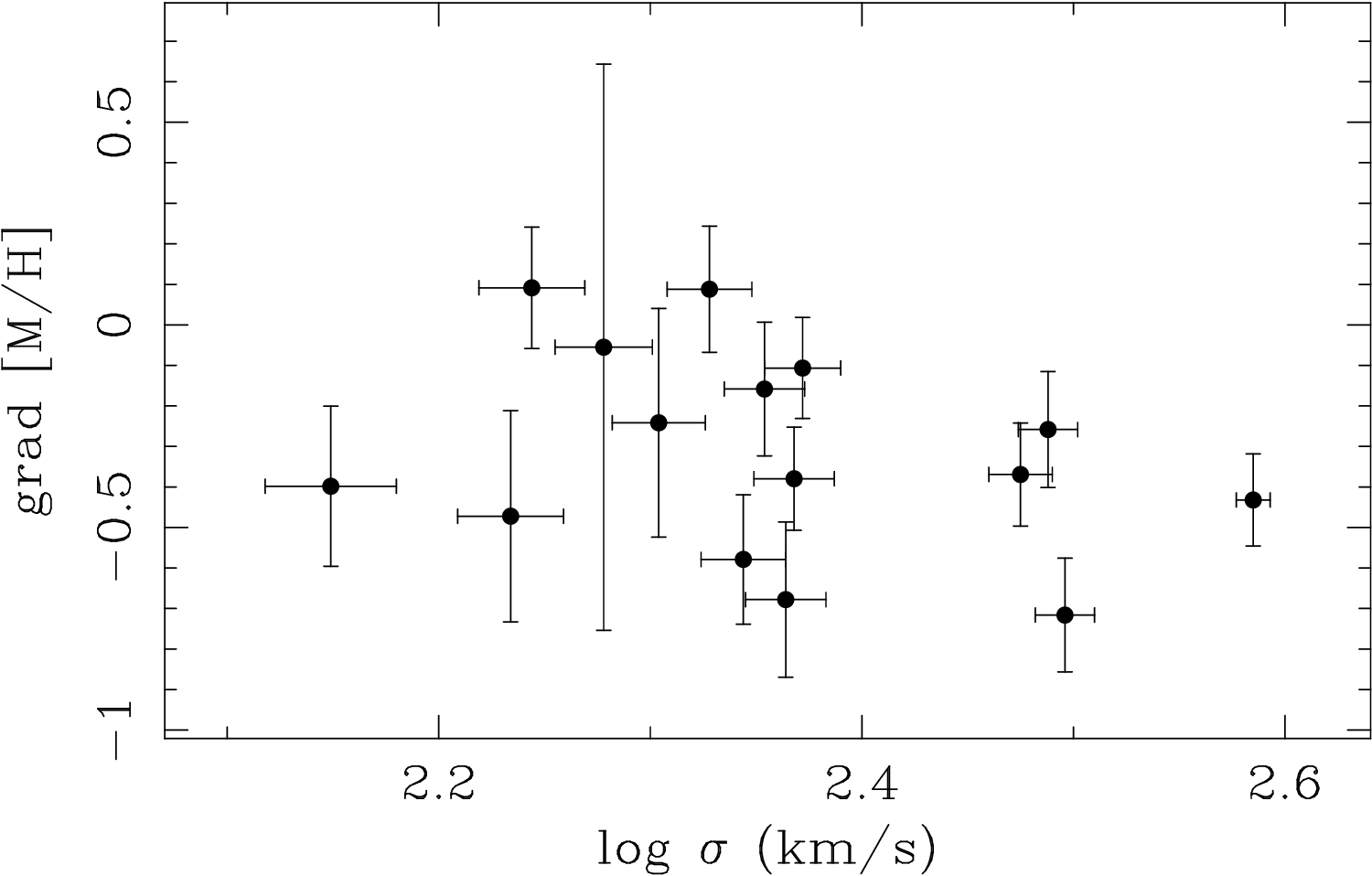}}
\hspace{10mm}
\resizebox{0.35\hsize}{!}{\includegraphics[angle=-90,bb=119 20 558 723]{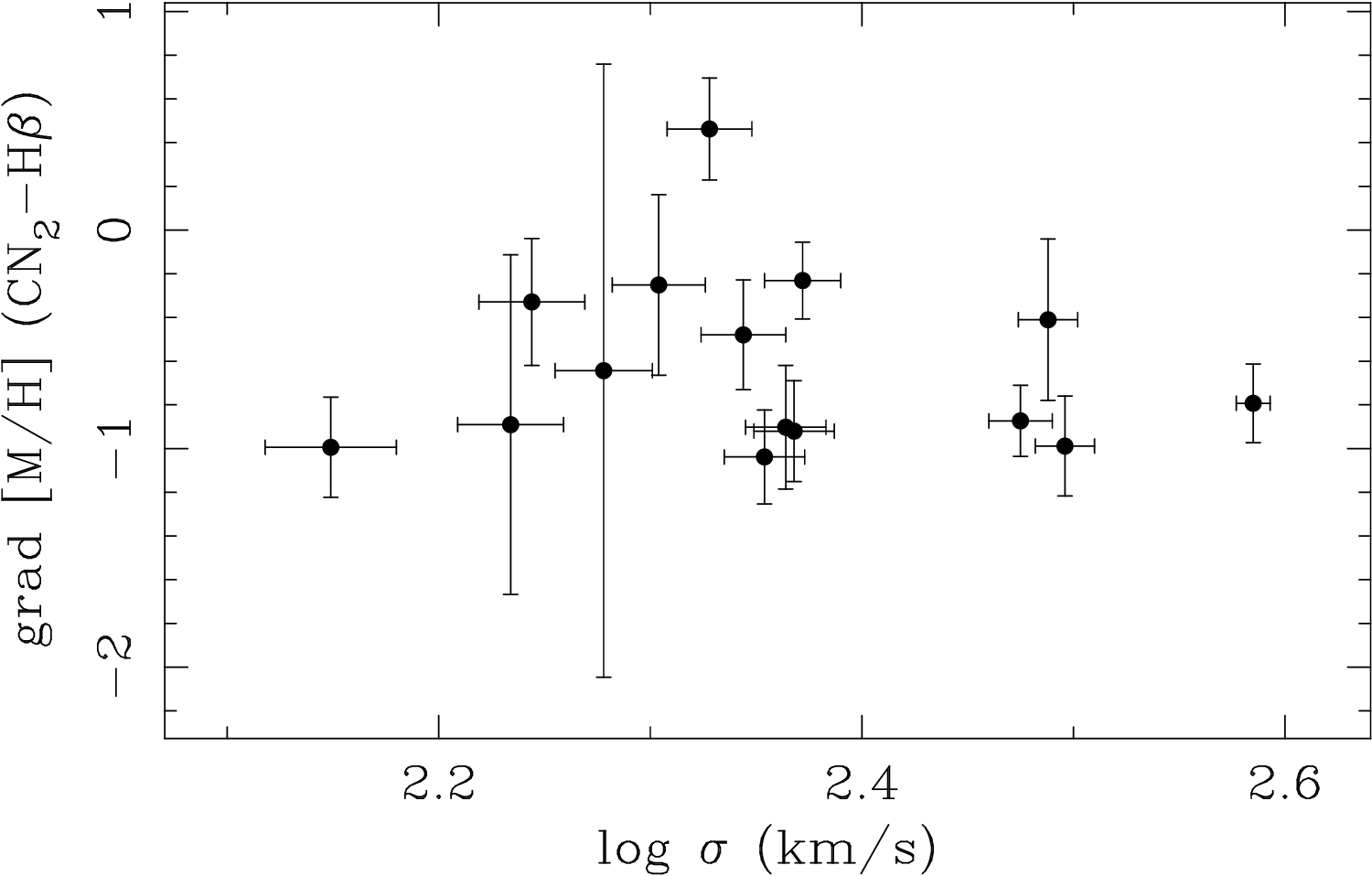}}

\vspace{5mm}

\resizebox{0.35\hsize}{!}{\includegraphics[angle=-90,bb=119 20 558 723]{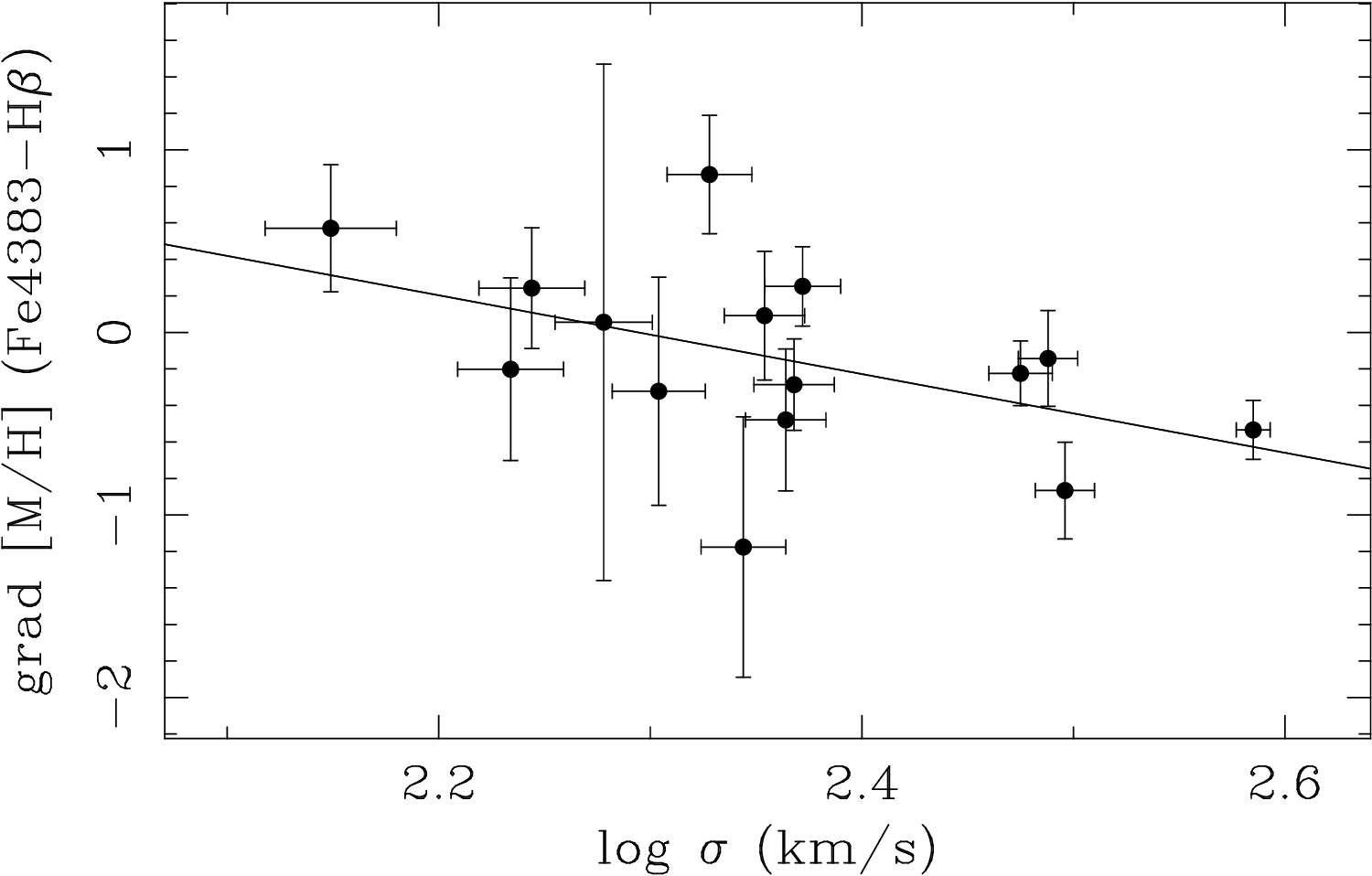}}
\hspace{10mm}
\resizebox{0.35\hsize}{!}{\includegraphics[angle=-90,bb=119 20 558 723]{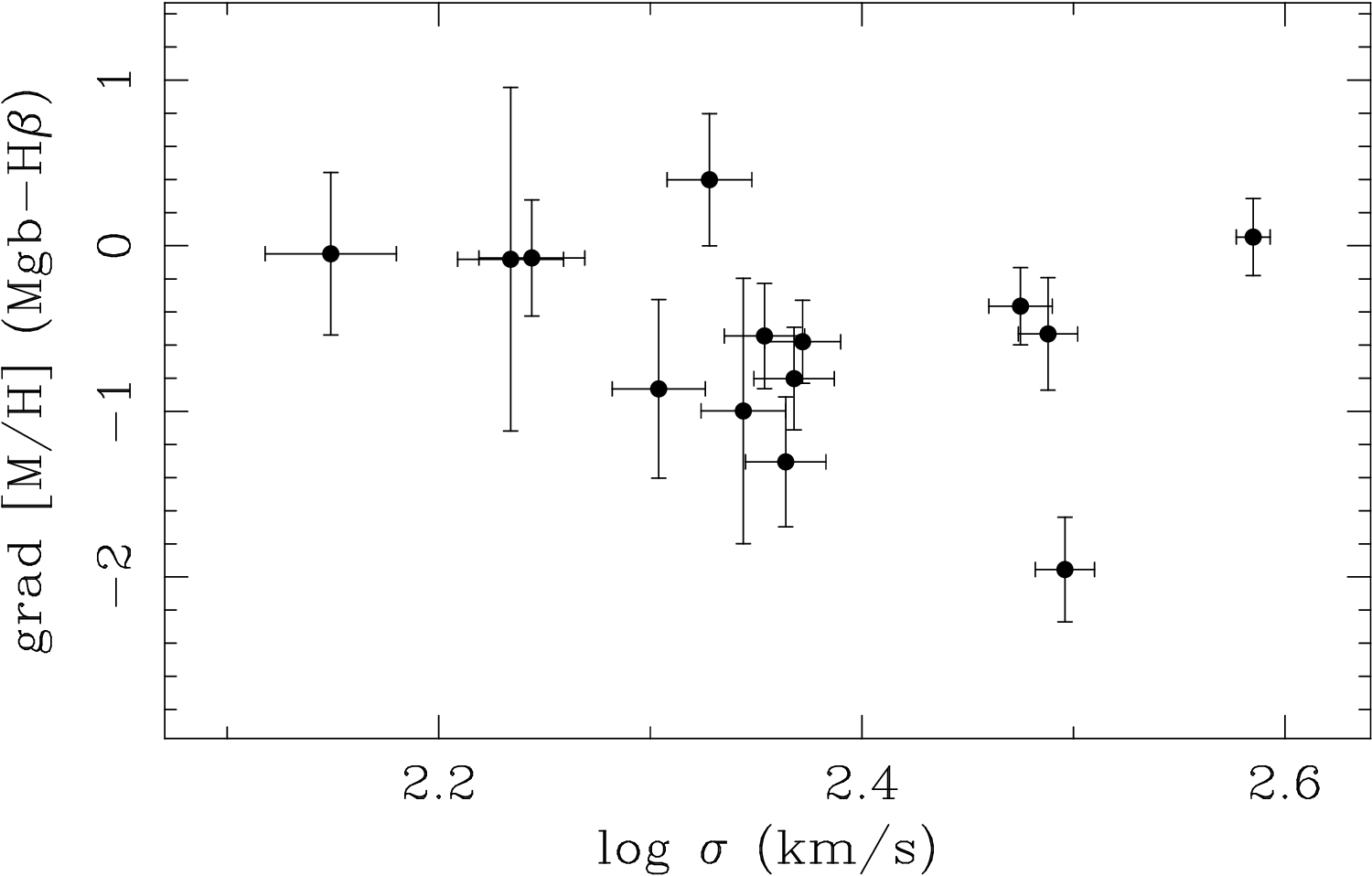}}

\vspace{5mm}

\begin{tabular}{@{}lrr@{}}
\hline
                              &\multicolumn{1}{c}{$t$}
                              &\multicolumn{1}{c@{}}{Pnc}\\
\hline                         
grad [M/H]                    &   1.16   &  0.13 \\  
grad [M/H] (CN$_2$--H$\beta$)  &   0.31   &  0.44 \\ 
grad [M/H] (Fe4383--H$\beta$)  &   2.31   &  0.03 \\
grad [M/H] (Mgb--H$\beta$)     &   1.40   &  0.18 \\
\hline
\end{tabular}
\caption{Metallicity gradients versus central velocity dispersion for HDEGs.
The different panels show the metallicity gradients obtained in different
index--index diagrams (indicated between the 
brackets). The first panel (top left)
compares the metallicity gradients obtained as described in Sec.
\ref{sec-age.metallicity.gradients}. In the case in which the correlation is
statistically significant, an error weighted linear fit is also shown. The
table at the bottom indicates the results of a $t$-test and a non-parametric
Spearman rank test (see text for details).
\label{grad.meta.losgiam}}
\end{figure*}

After Kobayashi (2004), the absence of correlation between the metallicity
gradient and the central velocity dispersion may be the consequence of
differences in the merger history of the galaxies. If the star formation
processes in recent epochs have been less frequent in the Coma cluster
galaxies, one might expect that any in situ correlation would have been
impacted upon less substantially. To explore if this is the case, in
Figure~\ref{grad.meta.losgiam} we present the metallicity gradients
measured in different $\Delta$index--$\Delta$index diagrams against the
central velocity dispersion for the elliptical galaxies of the Coma
cluster. The table at the bottom of the figure shows the results of a
$t$-test and a non-parametric Spearman rank test to check the degree of
correlation between both variables. In this case we find a correlation
between the metallicity gradient and the central velocity dispersion, but
only when the metallicity gradient is measured with the Fe4383 index.
While the correlation is not statistically significant when derived from
other indices, there does seem be a marginal trend in the sense that more
massive galaxies also show a steeper gradient.  To confirm this putative
mass trend and further constrain galaxy formation models, higher
signal-to-noise spectra must be obtained. 

\section{Conclusions}
\label{sec-conclusions}

We have carried out a study of the gradients in 23 different spectral
features for a sample of 82 early-type galaxies situated in different
environments. Our results can be summarised thusly: 

\begin{itemize} 

\item Using the new synthesis stellar population models of V06 we have
derived age and metallicity gradients for all the galaxies in the sample.
We have used a new method with employs ten different indicators in order
to reduce the scatter due to random errors. 

\item The mean age and metallicity gradients for the LDEGs are $\Delta
[$M/H$]/\log r=-0.205\pm 0.075$ and $\Delta \log$(age)/$\log r= 0.082\pm
0.032$, respectively. The mean age gradient is steeper and the mean
metallicity gradient flatter, than the predictions of dissipative collapse
models. On the other hand, the dispersion amongst the mean values is
larger than the dispersion expected by errors alone.  

\item We have studied the metallicity gradients derived from different
indicators, obtaining steeper gradients when using CN$_2$ and C4668 than
with Fe4383 and Ca4227.  Although it is beyond the scope of 
this paper  
to derive chemical abundances ratio gradients, we speculate that 
these aforementioned differences are reflecting the existence of 
radial variations in the 
relative abundances of some elements with respecto to iron.

\item We have studied the relation between the age and metallicity
gradients and the gradient of velocity dispersion, finding a correlation
when the metallicity gradient is calculated with CN$_2$ and Mgb. We do
not, however, find a correlation when the metallicity gradient is derived
from the Fe4383 and Ca4227 indices. The fact that the gradients of CN$_2$
and Mgb correlate with the gradient of the velocity dispersion may
indicate that the gradients in some elements (e.g. N and Mg) were shaped
early in the formation of galaxies, when the galactic winds presumably
dominate.  On the other hand, the lack of correlation when using Fe4383
and Ca4227 to measure the metallicity can indicate that other process,
such as secondary bursts of star formation, have had a stronger influence
in the final shape of these gradients. 
 
\item We have not found any correlation between the metallicity gradients
and the central velocity dispersion for the sample of galaxies in
low-density environments. A strong correlation between the gradients and
the mass of the galaxy is expected within dissipative collapse formation
scenarios. 
 
\item We have found a significant correlation between the metallicity
gradients and the central age for the LDEGs when the metallicity gradients
are measured with Fe4383 and C4668, in the sense that galaxies with a
younger central age also show a steeper metallicity gradient. 

\item The results quoted above suggest that the gradients of different
chemical species may have formed by different mechanisms. 

\item The mean age and metallicity gradients for the galaxies in the Coma
cluster are $\Delta \log ({\rm age})/\log r=0.027\pm$0.138 and $\Delta
{\rm [M/H]}/\log r= -0.328\pm 0.157$, respectively. The mean value of the
age gradient is compatible, to within the errors, with zero.  Both values
are also compatible with the predictions of Kobayashi (2004) for those
galaxies which have not undergone major mergers. However, the dispersion
amongst the mean values is higher than the scatter expected by the errors,
indicative of real differences between galaxies.

\item For galaxies in the Coma cluster, we have studied the correlation
between the metallicity gradients and the central velocity dispersion
finding a statistically significant correlation between both parameters
when the metallicity is measured with Fe4383. For the other indices, the
correlation is not statistically significant, but we observe a trend in
the sense that more massive galaxies tend to have a steeper metallicity
gradient. This trend is predicted by models of dissipative collapse, and
was not found for the LDEGs.  We should stress though that the quality of
the data for the HDEGs is lower than that for the LDEGS; higher quality
data for the Coma cluster galaxies are urgently needed.  In any case, in
this paper we have found systematic differences between the stellar
populations of galaxies in different environments which both confirms {\it
and} extends our conclusions from Papers~I and~II. 

\end{itemize}

\section*{Acknowledgments}
We would like to thank the anonymous referee for very helpful comments. 
PSB is most grateful to Brad Gibson for his continued support. 
The WHT is operated on the island of La Palma by the Royal Greenwich
Observatory at the Observatorio del Roque de los Muchachos of the Insituto
de Astrof\'{\i}sica de Canarias. The Calar Alto Observatory is operated
jointly by the Max-Planck Institute f\"{u}r Astronomie, Heidelberg, and
the Spanish Instituto de Astrof\'{\i}sica de Andaluc\'{\i}a (CSIC). This
work was partially supported by the Spanish research project AYA
2003-01840 and by the Australian Research Council. We are grateful to the
CATs for their generous allocations of telescope time.

\label{lastpage}

\begin{thebibliography}{99}

\bibitem[\protect\citeauthoryear{Arimoto \& Yoshii}{1987}]{AY87} Arimoto N.,
Yoshii Y., 1987, A\&A, 173, 23

\bibitem[\protect\citeauthoryear{Barnes \& Hernquist}{1991}]{BH91} Barnes J.E.,
Hernquist L.E., 1991, ApJ, 370, L65

\bibitem[\protect\citeauthoryear{Baugh, Cole, Frenk}{1996}]{BCF96} Baugh C.M.,
Cole S., Frenk C.S., 1996, MNRAS, 283, 1361 

\bibitem[\protect\citeauthoryear{Bender \& Surma}{1992}]{BS92} Bender R., Surma
P., 1992, A\&A, 258, 250

\bibitem[\protect\citeauthoryear{Binney}{1978}]{B78} Binney J., 1978, MNRAS,
183, 501

\bibitem[\protect\citeauthoryear{Boroson \& Thompson}{1991}]{BT91} Boroson
T.A., Thompson I.B., 1991, AJ, 101, 111

\bibitem[\protect\citeauthoryear{Bekki \& Shioya}{1999}]{BS99} Bekki K., Shioya Y., 1999, ApJ, 513, 108

\bibitem[\protect\citeauthoryear{Burstein}{1979}]{B79} Burstein D., 1979, ApJ, 232, 74

\bibitem[\protect\citeauthoryear{Cardiel et al}{1998a}]{CGA98} Cardiel N.,
Gorgas J., Arag\'on-Salamanca A., 1998a, MNRAS, 298, 977

\bibitem[\protect\citeauthoryear{Cardiel et al.}{1998b}]{CGC98} Cardiel N.,
Gorgas J., Cenarro J., Gonz\'alez J.J., 1998b, A\&AS, 127, 597

\bibitem[\protect\citeauthoryear{Carlberg}{1984}]{C84} Carlberg R.G., 1984,
ApJ, 286, 403

\bibitem[\protect\citeauthoryear{Carollo \& Danziger}{1994}]{CD94} Carollo
C.M., Danziger I.J., 1994, MNRAS, 270, 523

\bibitem[\protect\citeauthoryear{Carollo et al.}{1993}]{C93} Carollo C.M.,
Danziger I.J., Buson L., 1993, MNRAS, 265, 553

\bibitem[\protect\citeauthoryear{Cenarro et al.}{2003}]{C03} Cenarro A.J.,
Gorgas J., Vazdekis A., Cardiel N., Peletier R.F., 2003, MNRAS, 339, L12

\bibitem[\protect\citeauthoryear{Cohen}{1979}]{C79} Cohen J.G., 1979, ApJ, 228,
405

\bibitem[\protect\citeauthoryear{Chiosi \& Carraro}{2002}]{CC02} Chiosi C.,
Carraro G., 2002, MNRAS, 335, 335

\bibitem[\protect\citeauthoryear{Cole et al}{1994}]{CAF94} Cole S.,
Arag\'on-Salamanca A., Frenk C.S., Navarro J.F., Zepf S.E., 1994, MNRAS, 271,
781

\bibitem[\protect\citeauthoryear{Couture \& Hardy}{1988}]{CH88} Couture J.,
Hardy E., 1988, AJ, 96, 867

\bibitem[\protect\citeauthoryear{Davidge}{1991}]{D91} Davidge T.J., 1991, AJ, 102, 896
\bibitem[\protect\citeauthoryear{Davidge}{1992}]{DAV92} Davidge T.J., 1992, AJ,
103, 1512

\bibitem[\protect\citeauthoryear{Davies \& Sadler}{1987}]{DS87} Davies R.L.,
Sadler E.M., 1987, IAU symposium no 127, Structure and Dynamics of Elliptical
Galaxies, eds T. de Zeeuw (Dordrecth, Reidel), 441

\bibitem[\protect\citeauthoryear{Davies et al}{1993}]{DSP93} Davies R.L.,
Sadler E.M., Peletier R.F., 1993, MNRAS, 262, 650

\bibitem[\protect\citeauthoryear{Efstathiou \& Gorgas}{1985}]{EG85} Efstathiou
G., Gorgas J., 1985, MNRAS, 215, P37

\bibitem[\protect\citeauthoryear{Eggen et al}{1962}]{ELS62} Eggen O.J.,
Lynden-Bell D., Sandage A.R., 1962, ApJ, 136, 748

\bibitem[\protect\citeauthoryear{Faber et al.}{1985}]{FFB85} Faber S.M., Friel
E.D., Burstein D., Gaskell C.M., 1985, ApJS, 57, 711

\bibitem[\protect\citeauthoryear{Fisher, Franx, Illingworth}{1995}]{FFI95}
Fisher D., Franx M., Illingworth G., 1995, ApJ, 448, 119

\bibitem[\protect\citeauthoryear{Fisher, Franx \& Illingworth}{1996}]{FFI96}
Fisher D., Franx M., Illingworth G., 1996, ApJ, 459, 110

\bibitem[\protect\citeauthoryear{Franx \& Illingworth}{1990}]{FI90} Franx M.,
Illingworth, G., 1990, ApJ, 359, L41
\bibitem[\protect\citeauthoryear{Franx et al.}{1989}]{FIH89} Franx M., Illingworth G., Heckman T.,
1989, AJ, 98, 538

\bibitem[\protect\citeauthoryear{Gibson}{1997}]{G97}
Gibson B.K., 1997, MNRAS, 290, 471

\bibitem[\protect\citeauthoryear{Gonzalez}{1993}]{GON93} Gonz\'alez J.J., 1993,
PhD Thesis, University of California

\bibitem[\protect\citeauthoryear{Gonz\'alez \& Gorgas}{GG96}]{GG96} Gonz\'alez
J.J., Gorgas J., 1996, in  Fresh Views of Elliptical Galaxies, A., Buzzoni, A.
Renzini y A. Serrano (eds.), ASP Conf. Ser. Vol. 86., ASP, San Francisco, 225

\bibitem[\protect\citeauthoryear{Gorgas, Efstathiou \& Aragon}{1990}]{G90}
Gorgas J., Efstathiou G., Arag\'on-Salamanca A., 1990, MNRAS, 245, 217 (G90)

\bibitem[\protect\citeauthoryear{Gorgas et al.}{1993}]{GFB93} Gorgas J., Faber
S.M., Burstein D., Gonz\'alez J.J., Courteau S., Prosser C., 1993, ApJS, 86,
153

\bibitem[\protect\citeauthoryear{Gorgas et al.}{1997}]{GPG97} Gorgas J., Pedraz
S., Guzm\'an R., Cardiel N., Gonz\'alez J.J., 1997, ApJ, 481, L19

\bibitem[\protect\citeauthoryear{Joly \& Andrillat}{1973}]{JA73} Joly M.,
Andrillat Y., 1973, A\&A, 26, 95

\bibitem[\protect\citeauthoryear{J\o rgensen et al.}{1992}]{JFK92} J\o rgensen
I., Franx M., Kj\ae rgaard P., 1992, A\&AS, 95, 489

\bibitem[\protect\citeauthoryear{Kauffmann}{1996}]{KAU96} Kauffmann G., 1996,
MNRAS, 281, 487

\bibitem[\protect\citeauthoryear{Kauffmann \& Charlot}{1998}]{CK98} Kauffmann
G., Charlot S., 1998, MNRAS, 294, 705

\bibitem[\protect\citeauthoryear{Kawata}{1999}]{KAW99} Kawata D., 1999, PASJ,
51, 931

\bibitem[\protect\citeauthoryear{Kawata \& Gibson}{2003}]{KG03} Kawata D., Gibson B., 2003,
MNRAS, 346, 135

\bibitem[\protect\citeauthoryear{Kobayashi}{2004}]{K04} Kobayashi C., 2004,
MNRAS, 347, 740 

\bibitem[\protect\citeauthoryear{Kobayashi \& Arimoto}{1999}]{KA99} Kobayashi
C., Arimoto N., 1999, ApJ, 527, 573

\bibitem[\protect\citeauthoryear{Kormendy \& Djorgovsky}{1989}]{KD12} Kormendy
J., Djorgovski S., 1989, ARA\&A, 27, 235

\bibitem[\protect\citeauthoryear{Larson}{1974}]{LAR74a} Larson R.B., 1974a,
MNRAS, 166, 585

\bibitem[\protect\citeauthoryear{Larson}{1974}]{LAR74b} Larson R.B., 1974b,
MNRAS, 169, 229

\bibitem[\protect\citeauthoryear{Larson}{1975}]{LAR75} Larson R.B., 1975,
MNRAS, 173, 671 

\bibitem[\protect\citeauthoryear{Mathews \& Baker}{1971}]{MB71} Mathews W.G.,
Baker J.C., 1971, ApJ, 170, 241

\bibitem[\protect\citeauthoryear{Martinelli et al.}{1998}]{MMC98} Martinelli,
A., Matteucci F., Colafrancesco S., 1998, MNRAS, 298, 42

\bibitem[\protect\citeauthoryear{McClure}{1969}]{McC69} McClure R.D., 1969, AJ,
74, 50

\bibitem[\protect\citeauthoryear{Mehlert et al.}{2003}]{MTS03} Mehlert D.,
Thomas D., Saglia R.P., Bender R., Wegner G., 2003, A\&A, 407, 423

\bibitem[\protect\citeauthoryear{Mihos \& Hernquist}{1994}]{MH94} Mihos J.C.,
Hernquist L., 1994, ApJ, 427, 112

\bibitem[\protect\citeauthoryear{Mould}{1978}]{MOU78} Mould J.R., 1978, ApJ,
220, 434

\bibitem[\protect\citeauthoryear{Munn}{1992}]{MUN92} Munn J.A., 1992, ApJ, 399,
444

\bibitem[\protect\citeauthoryear{Oke \& Schwarzschild}{1975}]{OS75} Oke J.B.,
Schwarzschild M., 1975, ApJ, 198, 630

\bibitem[\protect\citeauthoryear{Peletier}{1989}]{PEL89} Peletier R.F., 1989,
PhD Thesis, University of Groningen

\bibitem[\protect\citeauthoryear{Peletier et al.}{1990}]{PDI90} Peletier R.F.,
Davies R.L., Illingworth G.D., Davis L.E., Cawson M., 1990, AJ, 100, 1091

\bibitem[\protect\citeauthoryear{S\'anchez-Bl\'azquez}{2004}]{SB04}
S\'anchez-Bl\'azquez P., 2004, PhD Thesis, Universidad Complutense de Madrid

\bibitem[\protect\citeauthoryear{S\'anchez-Bl\'azquez et al.}{2003}]{SGC03}
S\'anchez-Bl\'azquez P., Gorgas J., Cardiel N., Cenarro A.J., Gonz\'alez J.J.,
2003, ApJ, 2003, ApJ, 590, L91

\bibitem[\protect\citeauthoryear{S\'anchez-Bl\'azquez et al.}{2005a}]{SB05a}
S\'anchez-Bl\'azquez P., Gorgas J., Cardiel N., Gonz\'{a}lez J.J., 2005a,
MNRAS, in press (Paper I)

\bibitem[\protect\citeauthoryear{S\'anchez-Bl\'azquez et al.}{2005b}]{SB05b}
S\'anchez-Bl\'azquez P., Gorgas J., Cardiel N., Gonz\'{a}lez J.J., 2005b,
MNRAS, in press (Paper II)

\bibitem[\protect\citeauthoryear{S\'anchez-Bl\'azquez et al.}{2005c}]{SB05c}
S\'anchez-Bl\'azquez P., Peletier R.F., Jim\'{e}nez J., Cardiel N.,
Falc\'{o}n-Barroso J., Gorgas J., Selam S., Vazdekis A., 2005c, MNRAS,
submitted


\bibitem[\protect\citeauthoryear{Spinrad et al.}{1972}]{SST72} Spinrad H.,
Smith H.E., Taylor D.J., 1972, ApJ, 175, 649

\bibitem[\protect\citeauthoryear{Spinrad et al.}{1971}]{SGT71} Spinrad H., Gunn
J.E., Taylor B.J., McClure R.D., Young J.W., 1971, ApJ, 164, 11

\bibitem[\protect\citeauthoryear{Schweizer et al.}{1990}]{S90} Schweizer F.,
Seitzer P., Faber S.M., Burstein D., Dalle Ore, C.M., Gonz\'alez J.J., 1990,
ApJ, 364, L33

\bibitem[\protect\citeauthoryear{Tamura \& Ohta}{2003}]{TO03} Tamura N., Ohta
K., AJ, 126, 596

\bibitem[\protect\citeauthoryear{Thomas et al.}{2003}]{TMB02} Thomas D.,
Maraston C., Bender R., 2003, MNRAS, 339, 897
  
\bibitem[\protect\citeauthoryear{Thomsen \& Baum}{1989}]{TB89} Thomsen B., Baum
W.A., 1989, ApJ, 347, 214

\bibitem[\protect\citeauthoryear{Trager et al.}{1998}]{TWF98} Trager S.C.,
Worthey G., Faber S.M., Burstein D., Gonz\'alez J.J., 1998, ApJS, 116, 1

\bibitem[\protect\citeauthoryear{Trager et al.}{2000a}]{TFW00a} Trager S.C.,
Faber S.M., Worthey G., Gonz\'alez J.J., 2000a, AJ, 119, 1645

\bibitem[\protect\citeauthoryear{Trager et al.}{2000b}]{TFW00b} Trager S.C.,
Faber S.M., Worthey G., Gonz\'alez J.J., 2000b, AJ, 120, 165

\bibitem[\protect\citeauthoryear{van Albada}{1982}]{VAN82} van Albada T.S.,
1982, MNRAS, 201, 939

\bibitem[\protect\citeauthoryear{Vazdekis}{1999}]{VAZ99} Vazdekis A., 1999,
ApJ, 513, 224

\bibitem[\protect\citeauthoryear{Welch \& Forrester}{1972}]{WF72} Welch G.A.,
Forrester W.T., 1972, AJ, 77, 333

\bibitem[\protect\citeauthoryear{White}{1980}]{W80} White S.D.M., 1980, MNRAS, 191, 1
\bibitem[\protect\citeauthoryear{Worthey}{1994}]{WOR94} Worthey G., 1994, ApJS,
95, 107 

\bibitem[\protect\citeauthoryear{Worthey et al.}{1994}]{WFG94} Worthey G.,
Faber S.M., Gonz\'alez J.J., Burstein D., 1994, ApJS, 94, 687 


\end{thebibliography}
\end{document}